\documentclass[11pt]{article}
\usepackage{natbib}
 \bibpunct[, ]{(}{)}{,}{a}{}{,}%
\usepackage[utf8]{inputenc}
\usepackage{amssymb}
\usepackage{latexsym}
\usepackage{amsfonts}
\usepackage[english]{babel}
\usepackage{amsmath}
\usepackage{amsthm}
\usepackage{mathtools}
\usepackage[margin=1in]{geometry}
\usepackage{geometry}
\usepackage{longtable}
\usepackage{standalone}
\usepackage{graphicx}
\usepackage{color}
\usepackage{epstopdf}
\usepackage{graphicx}
\usepackage{grffile}
\usepackage[utf8]{inputenc}
\usepackage{amssymb}
\usepackage{latexsym}
\usepackage{amsfonts}
\usepackage[english]{babel}
\usepackage{amsmath}
\usepackage{amsthm}
\usepackage{mathtools}
\usepackage[margin=1in]{geometry}
\usepackage{geometry}
\usepackage{longtable}
\usepackage{standalone}
\usepackage{graphicx}
\usepackage{color}
\usepackage{epstopdf}
\usepackage{graphicx}
\usepackage{grffile}
\usepackage{caption}
\usepackage{lscape}
\usepackage{hyperref}
\usepackage{setspace}
\usepackage{tikz}
\usepackage{array}
\usepackage{caption}
\usepackage{lscape}
\usepackage{hyperref}
\usepackage{booktabs,caption}
\usepackage{threeparttable}
\usepackage{setspace}
\usepackage{tikz}
\usepackage{array}
\usepackage{layout}

\newtheorem{theorem}{Theorem}[section]
\newtheorem{lemma}{Lemma}[section]
\newtheorem{proposition}{Proposition}[section]
\newtheorem{corollary}{Corollary}
\newtheorem{assumption}{Assumption}

\newtheorem{remark}{Remark}
\newtheorem{definition}{Definition}
\usepackage{endnotes}
\let\footnote=\endnote

\usepackage{natbib}
 \bibpunct[, ]{(}{)}{,}{a}{}{,}%
\title{Credit Risk: Simple Closed Form Approximate Maximum Likelihood Estimator}

\author{Anand Deo, Sandeep Juneja\footnote{ This work was initiated when the second author
was working as an adjunct with CAFRAL (Centre for Advanced Financial Research and Learning), a research wing of Reserve Bank of India.} \\
{\small Tata Institute of Fundamental Research}}
\numberwithin{equation}{section}
\usepackage{natbib}
 \bibpunct[, ]{(}{)}{,}{a}{}{,}%
\title{Credit Risk: Simple Closed Form Approximate Maximum Likelihood Estimator}

\author{Anand Deo, Sandeep Juneja\footnote{ This work was initiated when the second author
was working as an adjunct with CAFRAL (Centre for Advanced Financial Research and Learning), a research wing of Reserve Bank of India.} \\
{\small Tata Institute of Fundamental Research}}
\numberwithin{equation}{section}

\begin{document}


\maketitle
\begin{abstract}
We consider discrete default intensity based and logit type  reduced form models for conditional default probabilities for corporate loans
where we develop simple closed form approximations to the maximum likelihood estimator (MLE) when the underlying covariates follow a stationary Gaussian process. In a practically reasonable asymptotic regime where the default probabilities are small, say $1-3\%$ annually, the number of firms and the time period of data available is reasonably large, we rigorously show that the proposed estimator behaves similarly or slightly worse than the MLE when the underlying model is correctly specified. For more realistic case of model misspecification, both estimators are seen to be equally good, or equally bad. Further, beyond a point, both are more-or-less insensitive to increase in data. These conclusions are validated on empirical and simulated data. The proposed approximations should also have applications outside finance, where logit-type models are used and probabilities of interest are small.\\
{\textit{{Keywords}} - credit risk, default probabilities, calibration, Logit models, Default intensity model, maximum likelihood estimator.\\}
\end{abstract}%

\thispagestyle{empty}

\section{Introduction}
\textbf{Overview: \,}Development and estimation of parametric  credit risk models to predict firm default probability is of great practical importance in financial credit risk and has generated considerable academic literature over the past fifty years (see, e.g., Altman 1968, Merton 1974, Ohlson 1980, Zmijewski 1984, Shumway 2001, Chava and Jarrow 2004, Giesecke and Goldberg 2004,  Gourieroux, Monfort, and Polimenis 2006 and  Giesecke, Longstaff, Schaefer, Strebulaev 2011). In addition, one looks to develop a regime that  accurately models the probabilities of joint defaults of many firms dynamically as a function of time - this becomes particularly useful in measuring dynamic evolution of portfolio credit risk exposure of financial institutions (see, e.g., Gagliardini and  Gourieroux 2005,
 Duffie, Saita and Wang 2007, Bharath and Shumway 2008, Duan, Sun and Wang 2012, Eymen, Giesecke, and Goldberg 2010 and references therein).
	
In this paper we revisit the well studied problem of estimating  the conditional default probability
of a firm as a function of firm specific  as well as macroeconomic covariates. Traditionally, dynamic evolution of risk is modelled using doubly stochastic point processes. Default intensity of each firm is modelled as a function of continuous time stochastic covariates.  Popular covariates include distance to default of each firm,  some firm-specific financial ratios as well as
 prevalent treasury rates and trailing stock index return. Dependence between obligors is captured by allowing the underlying covariates to evolve in a dependent manner.  Conditioned on the realised  default intensities, the firm default times are assumed to occur independently, each as the first event of a non-homogeneous Poisson process whose intensity corresponds to the realised default intensity of the firm.

In this paper, we work in a similar doubly stochastic framework, the difference being that we model the covariates as well as the corresponding obligor default intensities as discrete time stochastic processes. The benefits are that discrete time processes are usually simpler to analyse. Even continuous time models are typically first discretized both for parameter estimation, as well as for simulating sample paths, so this too makes analysis of discrete time models important (see, e.g., Duffie, Saita and Wang 2007,  Bharath and Shumway 2008, Duan, Sun and Wang 2012).

We consider a  popular
discretised default intensity based model where, for a firm surviving till time $t \in Z^+$,  the non-negative hazard rate intensity
within times $[t, t+1)$ is assumed to have the form $\exp(\beta^T V_t-\alpha)$ where  $V_t$ denote the vector of the underlying covariates at time $t$, and is typically assumed to be a parametric stochastic process.
 Parameters underlying  $V_t$ as well as parameters $(\beta, \alpha)$
are estimated from data using the popular maximum likelihood estimation (MLE) method. 
Observe that the above form of hazard rate intensity   implies that the  conditional probability of default  between times $t$ and $t+1$
	\begin{equation} \label{eqn:form1}
	= 1-\exp(- \exp(\beta^T V_t-\alpha))
	\end{equation}
	(see, e.g., Duffie et. al. 2007, Bharath and Shumway 2008, Duan et. al. 2012, Duan and Fulop 2013).

Logit type models are another class of popular doubly stochastic models to which our analysis is applicable. Here,  the conditional default probabilities have the form
	\begin{equation} \label{eqn:form2}
	\frac{\exp(\beta^{\intercal} V_t-\alpha)}{1+ \exp(\beta^{\intercal} V_t-\alpha)}
	\end{equation}
	(see, e.g., Shumway 2001, Chava and Jarrow 2004, Campbell, Hilscher and Szilagyi, 2008).

 One advantage of doubly stochastic framework is that under MLE the problem of estimating these two sets of parameters decouples (see Duffie et al 2007).
In this paper, we assume that stochastic process $V_t$ is given and focus on estimating the associated coefficients 
$(\beta, \alpha)$.
Standard methods to determine MLE for $(\beta, \alpha)$  involve solving a complex optimisation problem using non linear optimisation  or even sequential Monte Carlo methods (see Duan et al. 2012, Duan and Fulop 2013). These can be extremely time consuming given the huge data-sets that are often used for model estimation. These problems are also difficult since the defaults are rare events and sufficient amount of  data is needed to contain enough defaults to allow for accurate estimation. Further, these computational procedures provide little insight on the underlying factors that determine these parameters.

\textbf{\noindent {{ Main Contributions} : \bf 1. \, }}We develop {\em closed form} approximate expressions for estimated parameters, that are seen to be almost as accurate as the MLE in a variety of practically representative settings. These approximations rely on the observation that for 
small values the conditional default probabilities, both  (\ref{eqn:form1}) and
(\ref{eqn:form2}) can both be approximated by $\exp(\beta^{\intercal} V_t-\alpha)$. Further, for a Gaussian process $V_t$, expectation 
of $\exp(\beta^{\intercal} V_t-\alpha)$  and $V_t \exp(\beta^{\intercal} V_t-\alpha)$  have explicit closed form expressions.
 Note that the parametric form of conditional default probabilities 
is typically based on pragmatic considerations rather than any fundamental causal relationship. Thus, even if a covariate has a non-Gaussian marginal distribution, one may (as we do), look for functional transformations that are closer to the Gaussian distribution, and use the resulting time series as a covariate instead.  Empirically, on a large sample of U.S. Corporate data (henceforth referred to  as \textbf{C-DATA}), it can be seen that these transformations improve the predictive ability of the proposed estimators while not harming the predictive ability of the MLEs, for both intensity as well as logit models.

In our approximations for $\beta$ (see \eqref{eqn:interesting1}), {\em each parameter  is set to  a weighted sum of the corresponding covariate observed just before default occurrences}. This suggests that from the point of view of default prediction, accurate assessment  of  values of covariates just before default is key, while the  remaining data is less critical. As we see from our numerical experiments in Section 4, even if some of this remaining data is missing, the quality of the MLE is negligibly affected.  Interestingly, we show, theoretically as well as numerically, that when  the data is corrupted by additive zero mean noise, (as may happen in practice) the proposed estimator incurs a negligible bias compared to the  MLE. 
 Further, if the noise has a non-zero mean, both the proposed estimator and the MLE incur a similar bias. This suggests that in the presence of additive noise in the data, the proposed estimator is at least as accurate as the MLE. 

Another important advantage of the proposed estimator is that it provides a good starting point to  numerical MLE methods that considerably speeds up the computation. We observe that using the proposed estimator as an initial seed on  simulated data (for number of firms and time periods typically seen in practice), makes the MLE 5 to 7 times faster.   We also provide theoretical justification for these computational improvements  in Section 4.2 and \ref{EC:Supnum}.

\noindent {\bf 2. \,}  To gain further insights into the performance of the proposed estimator,  we embed  the estimation problem in a sequence of correctly specified statistical models indexed by the rarity of the underlying defaults. Specifically, we consider an asymptotic regime indexed by a rarity parameter $\gamma$, where as $\gamma\to 0$, default probability goes to zero and number of firms and time period of observation for each firm go to $\infty$. In this regime, we analyse the proposed estimator and the MLE, and develop precise bounds on errors in parameter estimation for each procedure. The resulting analysis  sheds further light into the accuracy of proposed approximations and the amount of data in terms of number of firms and the number of time periods needed for accurate estimation  as a function of the rarity of the underlying default probabilities. In a setting where the number of firms of data available is small, say about five thousand and the default probability is about 1\% per annum, we observe both analytically as well as numerically, that the proposed estimator and the MLE give the same order of magnitude of error. Even when the number of firms of data is large, we show numerically that the proposed estimator gives an error only marginally larger than the MLE.

\noindent {\bf 3. \,}
Often in practice, regularisation is used to provide robustness to parameter over-fitting. In the MLE setting, regularisation reduces to adding a penalty term 
to the log-likelihood function and then maximizing it to determine the {\it regularised MLE}. 
This adjustment can also be viewed in the Bayesian setting as reweighing the likelihood function with a prior distribution. We extend our approximation framework to these regularised 
MLE problems that are indexed by a prior distribution on the underlying parameters. 
We consider popular regularisations including

\begin{itemize}
    \item LASSO, obtained through a Laplace prior,  and
    \item Ridge, obtained through a Gaussian prior.
\end{itemize}
In our asymptotic regime, we show that  when ridge regularisation is used, the proposed approximation has a closed form.  Further, considering a fairly general class of priors, we show that the regularised MLE can be approximated by the solution to simple convex program. We also derive accuracy bounds. 

\noindent {\bf 4. \,}
In the existing credit risk literature, for the most part, the number of covariates used in predicting defaults are assumed to be few and well chosen. However, in some data driven settings, a large number of covariates may be used (see, e.g., Sirignano, Sadhwani and Giesecke, 2016). When the number of covariates is large, estimating the MLE becomes computationally demanding, particularly when default intensity based models are used. We test the performance of the proposed estimator when the number of covariates is large by allowing the number of covariates to increase in our asymptotic regime. We  identify explicitly the dependence of the error in estimation on the number of covariates. In practice, this suggests that the proposed approximations should work well for a reasonably large number (say about 20) of covariates, but not too large.  We also observe that the proposed approximations when used as an initial starting point, continue to significantly speed-up numerical algorithms, even when the number of covariates is large.

\noindent {\bf 5. \,}
Typically, the model that we assume for  generating default probabilities may be misspecified - it may lack some hidden co-variates  or it may be structurally  misrepresentative, or very likely, both (see Duffie, Eckner, Horel and Saita 2009). In such a setting, we show that the the misspecification error  dominates other errors so that the proposed estimator is equally accurate as the MLE. In particular, increasing data beyond a point, both in terms of the number of firms and the time periods considered, lead to virtually no improvement in the estimator quality. This may have ramifications both in search for more accurate models as well as on cost-benefit trade-offs in gathering data for model estimation. Dominance of model misspecification error is also suggested by the performance of the proposed estimators on C-DATA where they are seen to be about as accurate as ML methods in predicting defaults. 

\noindent {\bf 6. \,} As noted by Duffie et al. (2007) and Duan et al. (2012), typically firms exit not only due to default, but also due to mergers and acquisitions, etc.
We show that the proposed methodology adapts easily to this contingency (see \ref{EC:other1}). 
It is well known in corporate default literature that defaults tend to cluster displaying a contagion effect (see, e.g., Das, Duffie, Kapadia and Saita 2007, Azizpour, Giesecke and Schwenkler 2016). We observe that appropriately including contagion effect as a covariate  improves the empirical performance of the proposed estimator.\\
\textbf{{Outline}:} The remaining paper is as follows: In Section 2, we first arrive at the maximum likelihood estimators in  two popular regimes with conditional probabilities of the form (\ref{eqn:form1}) and (\ref{eqn:form2}). Further, we  identify the proposed estimator suggested by these equations under the assumption that covariates have a multivariate Gaussian distribution. In Section 3, we introduce the mathematical framework and  conduct an asymptotic  analysis of the proposed estimator. We also discuss the performance of the ML estimators under correct and misspecified models and the approximation to the MLE with regularisation. Numerical results based on simulation generated default data  are presented in Section 4. In Section 5, we compare the proposed estimator with MLE on C-DATA. In Section 6 we end with a brief conclusion. Due to a lack of space, proofs, some of the more detailed discussions, and some numerical results, are presented in the accompanying appendix.	
\section{Maximum Likelihood Estimator}\label{sec:MLE}
For ease of presentation, we first derive the proposed estimator in a simple setting. We assume  that the only form of exits are defaults, firms are homogeneous, and the conditional default probability has the discrete intensity form given by (\ref{eqn:form1}). Extensions to the cases where  firms also exit due to other reasons such as mergers and acquisitions (referred to as exits due to censoring in the literature), and where firms may belong to heterogeneous classes are provided in \ref{EC:Extensions}.
\subsection{Discrete default model}
Suppose that the data available  involves $m$ firms, observed over a discrete set of time periods  $\{0,1, \ldots, T\}$. Let $s_{i} \leq T-1$ denote the time  when firm $i$ came into existence,  with $s_{i}=0$ if it already existed at time $0$.  
   Let $\tilde{\tau}_{i}$ denote the default time for firm $i$. Specifically,  $\tilde{\tau}_i=t$ if the firm defaults between periods $t$ and $t+1$.  Else, if the firm does not default, i.e, $\tilde{\tau}_{i}\geq T$, set $\tilde{\tau}_i = \infty$,
   and set ${\tau}_{i} = \min(\tilde{\tau}_{i}, T-1)$.
Further,
\begin{itemize}
\item
for $d_1 \geq 1$, let  $(y_t \in \Re^{d_1}: t=0, 1, \ldots, T-1)$ denote the value of the common factors.
\item For $d_2 \geq 1$, let $(x_{i,t} \in \Re^{d_2}: s_{i} \leq t \leq \tilde{\tau}_{i})$ denote firm specific information available.
For each $i$ and $t$, let $v_{i,t} \in \Re^{d_1+d_2}$ denote the vector $(y_t, x_{i,t})$.
\item
Let $d_{i,t+1} =1$ if firm $i$ defaults between time $t$ and $t+1$. Else, $d_{i,t+1} =0$.
\item Let $\theta \in \Re^{d_1}$ and $\eta\in \Re^{d_2}$, respectively, denote the parametric sensitivities to global and idiosyncratic covariates respectively. We let $\beta \in \Re^{d_1+d_2}$ denote the tuple $(\theta,\eta)$.
\end{itemize}
Let $p(v_{i,t})$ denote the conditional probability that a firm  $i$, surviving at time $t$, defaults  between  time $t$ and $t+1$. This is assumed to be a function of $v_{i,t}$ given $v_{i,s}:s\leq t$. This probability also depends upon underlying parameters that need to be estimated from data. 

One popular model for credit risk is to view the default of a firm as the first arrival time of a Poisson process (see Lando 2009 or Duffie and Singleton 2012). In order to capture the temporal stochasticity in defaults, the non-negative intensity of this Poisson process is assumed to be an exponentially affine function of a diffusion. In implementation, one discretises this model by dividing the total observation period into equal time intervals $t\in\{1,\ldots, T\}$, and holding the intensity constant within a time interval. This leads to the discrete intensity model for conditional default probability (see Duffie et al. (2007), Duffie et al. (2009) or Duan et al. (2012)). Specifically, given all the covariate information up to time $t$, and that the firm has survived till $t$, the probability of it defaulting in $[t, t+1)$ is modelled as 
$p(v_{i,t}, \beta, \alpha) = 1-\exp(-e^{\beta^\intercal v_{i,t} -\alpha})$. In interest of space, details of the intensity model along with its discretisation are described in the electronic companion (see \ref{sec:ARMA-Int}).

For ${b} \in \Re^{d_1+d_2}$, $a \in \Re$, let  ${\cal L}({b},a)$ denote the likelihood function 
of seeing the default data $(d_{i,t}: s_{i} < t \leq \tau_{i})$ for each $i \leq m$. Then,
\begin{equation} \label{eqn:LR1}
{\cal L}({b},a) = \prod_{i \leq m} \prod_{t=s_{i}}^{\tau_{i}}
\left ( p(v_{i,t}, b,a)^{d_{i,t+1}}(1- p(v_{i,t},b,a))^{1-d_{i,t+1}} \right ).
\end{equation}
ML estimation of underlying parameters corresponds to finding parameters that maximise ${\cal L}({b},a)$, or equivalently, $\log {\cal L}({b},a)$. When the conditional default probabilities have the intensity structure, setting the partial derivatives with respect to each component of ${b}$ and  $a$
to zero, the following first order conditions are obtained, where $\hat{\beta}_M$ and $\hat{\alpha}_M$ are a solution to the MLE: 
\begin{equation}  \label{logit:001}
\sum_{i \leq m, t=s_{i}}^{ \tau_{i}}  \frac{v_{i,t} e^{\hat{\beta}^{\intercal}_M v_{i,t}  -\hat{\alpha}_M}}
{1-\exp(-e^{\hat{\beta}_M^{\intercal} v_{i,t}-\hat{\alpha}_M})}
   d_{i,t+1}
=  \sum_{i \leq m, t=s_{i}}^{ \tau_i}v_{i,t} e^{\hat{\beta}^{\intercal}_M v_{i,t}-\hat{\alpha}_M},
\end{equation}
and
\begin{equation}  \label{logit:003}
\sum_{i\leq m, t=s_{i}}^{ \tau_{i}}  \frac{e^{\hat{\beta}^{\intercal}_M v_{i,t}-\hat{\alpha}_M}}
{1-\exp(-e^{\hat{\beta}^{\intercal}_M v_{i,t}-\hat{\alpha}_M})}
   d_{i,t+1}
=  \sum_{i \leq m, t=s_{i}}^{ \tau_{i}}e^{\hat{\beta}^{\intercal}_M v_{i,t}-\hat{\alpha}_M}.
\end{equation}
Another popular model for defaults where the derived approximations are applicable is the logit model. Here the conditional default probability given covariates is $ p(v_{i,t}, \beta, \alpha) = \frac{\exp(\beta^\intercal v_{i,t}-\alpha)}{1+\exp(\beta^\intercal v_{i,t}-\alpha)}$. Similar equations for the logit model are easily derived:
\begin{equation*}  
\sum_{i\leq m, t=s_{i}}^{ \tau_{i}}v_{i,t} d_{i,t+1}
=
\sum_{i \leq m, t=s_{i}}^{ \tau_{i}}v_{i,t}
\frac{\exp(\hat{\beta}^{\intercal}_M v_{i,t}-\hat{\alpha}_M) }
{1+\exp(\hat{\beta}^{\intercal}_M v_{i,t}-\hat{\alpha}_M)},
\end{equation*}
and
\begin{equation*} 
\sum_{i \leq m, t=s_{i}}^{ \tau_{i}}d_{i,t+1}
=
\sum_{i \leq m, t=s_{i}}^{ \tau_{i}}
\frac{\exp(\hat{\beta}^{\intercal}_M v_{i,t}-\hat{\alpha}_M) }
{1+\exp(\hat{\beta}^{\intercal} v_{i,t}-\hat{\alpha}_M)}.
\end{equation*}
\subsection{Gaussian approximations}
Our approximation is based on two steps - a first order Taylor approximation of the default probability when it is small, and an  application of the law of large numbers to the MLE equations \eqref{logit:001} and \eqref{logit:003}. Let $\tau= \sum_{i \leq m} (\tau_{i}-s_{i})$ denote the firm-periods of data available. In our analysis  we assume that $\{y_t\}$ and
 $\{x_{i,t}:i \leq m\}$ for surviving firms, are realisations of a stationary process
$\{(Y_t, (X_{i,t}, i \leq m))\}$ (also denoted by $\{(V_{i,t}, i \leq m)\}$) observed at integer times $0 \leq t < T$, which is further assumed to be multivariate Gaussian.
As discussed in the introduction, the covariates originally may not be Gaussian, but we assume that they are suitably transformed to have a Gaussian marginal distribution. For instance, the logarithm of a covariate may be closer to a Gaussian distribution, and may instead be used as a covariate (see Section~\ref{sec:Numbers}). More specifically, the transformed variables form a stationary multivariate Gaussian process where each marginal has been normalised to have stationary mean zero and variance one. Using the law of large numbers and consistency of the MLE, we may  approximate RHS of (\ref{logit:003}) divided by $\tau$ by $E\left(\exp(\beta^{\intercal} V_{i,t} -\alpha)\right)$, which, as is well known, equals $\exp\left(\frac{1}{2} \beta^{\intercal} \Sigma \beta - \alpha\right)$,
where  $\Sigma$ denotes the correlation matrix
of $\{V_{i,t}\}$ and is assumed to be independent of $i$. Similarly, the RHS of (\ref{logit:001})
divided by $\tau$ may be approximated by
\[
E(V_{i,t} \exp(\beta^{\intercal} V_{i,t}-\alpha)) = \Sigma \beta   \exp\left(\frac{1}{2} \beta^{\intercal} \Sigma \beta - \alpha\right).
\]
While the default probability is small, the first order approximation $1-\exp(e^{\beta^\intercal v_{i,t}-\alpha}) \approx e^{\beta^\intercal v_{i,t}-\alpha}$ holds. Then the LHS of (\ref{logit:001}) and (\ref{logit:003}) approximately become $\tau^{-1}$ times 
\[\sum_{i\leq m}\sum_{t\leq \tau_i}v_{i,t}d_{i,t+1} \textrm{ and } \sum_{i\leq m}\sum_{t\leq \tau_i}d_{i,t+1}, \] 
respectively. Assume that $\Sigma$ is known and invertible. Then, the above discussion suggests that
\begin{equation}  \label{eqn:interesting1}
\beta \approx  \Sigma^{-1} \frac{\sum_{i \leq m, t=s_{i}}^{ \tau_{i}}v_{i,t} d_{i,t+1}}
 {\sum_{i \leq m, t=s_{i}}^{ \tau_i}d_{i,t+1}}.
\end{equation}

The RHS above, call it $\hat{\beta}$, is our proposed estimator for $\beta$. Motivated by (\ref{logit:003}), our estimator for $\alpha$ is set as
\begin{equation}  \label{eqn:interesting2}
\hat{\alpha} \triangleq  \log \left (     \frac{\sum_{i \leq m, t=s_{i}}^{ \tau_{i}} \exp(\hat{\beta}^{\intercal} v_{i,t}) }
{\sum_{i \leq m, t=s_{i}}^{ \tau_{i}}d_{i,t+1}}  \right ).
\end{equation}
The above discussion suggests, at least intuitively, that the proposed estimators, which may also be arrived at simply by approximately matching certain moments, are close to the MLE.
\begin{remark}\label{rem:Multiclass}\em
{Suppose that we assume that the firms can be classified into $K$ homogeneous classes
where the parameters $\beta$ are same across all classes while
the parameters $(\alpha_k: k \leq K)$ are allowed to be class dependent and they measure the riskiness of each class.
An easy extension of our approximation to this setting is to estimate
$\beta$ as above by  $\hat{\beta}$, assuming that the data comes from a single class.
The parameters $(\alpha_{k}: k \leq K)$  can be estimated as $\hat{\alpha}_{k} \triangleq  \log \left (     \frac{\sum_{i \leq m_k, t=s_{i,k}}^{ \tau_{i,k}} \exp(\hat{\beta}^{\intercal} v_{i,t}) }
{\sum_{i \leq m, t=s_{i,k}}^{ \tau_{i,k}}d_{i,k,t+1}}  \right )$, where the subscript $k$ attached to original notation denotes that the associated data $(s_{i,k}, \tau_{i,k}, v_{i,t},d_{i,t+1,k})$ is class specific.
}
\end{remark}

 \begin{remark} \em{
 Let  the weighted average of the covariates observed before defaults $\frac{\sum_{i \leq m, t=s_{i}}^{ \tau_i}v_{i,t} d_{i,t+1}}
 {\sum_{i \leq m, t=s_{i}}^{ \tau_i}d_{i,t+1}}$ be denoted by $\hat{w}$. Then, we have $\hat{\beta} = \Sigma^{-1} \hat{w}$.
 Now suppose that $\Sigma^{-1}$ is not known but is estimated from data by its empirical sample version, $\hat{\Sigma}^{-1}$. Then, a reasonable estimator of
 $\beta$ is $\hat{\Sigma}^{-1}\hat{w}$. This may be re-expressed as $\Sigma^{-1}\hat{w} + (\hat{\Sigma}^{-1}- \Sigma^{-1})\hat{w}$. In the asymptotic regime developed in Section 3, we show that the error in $(\hat{\Sigma}^{-1} - \Sigma^{-1})$ is in order of magnitude less than or equal to the error in the parameter estimation. Thus, conducting an analysis with $\Sigma$ assumed to be known does not affect our asymptotic results.}
\end{remark}
We next develop an asymptotic analysis that illuminates the sense in which the two sets of estimators are close to the true parameters.


\color{blue}

\color{black}

\section{Analysis of Proposed Estimators and MLE}\label{section:Main}
In this section we construct an asymptotic framework that sheds further light on, and helps compare  the proposed estimator and  the MLE - their dependence on the rarity of the probability of default, on the number of firms, and the number of time periods for which the data is available.  Further, we also arrive at the form of the proposed estimator when regularization is included
and again compare it with the regularized MLE.  We also study the effect of missing and corrupted data on both the estimators.  Finally, we analyze the performance of the two estimators when the underlying model is mis-specified. 

The analysis involved is technically demanding. To keep the exposition simple we restrict ourselves to a single class setting where the firms are statistically homogeneous and exit only due to defaults. In \ref{EC:Extensions}, we extend the asymptotic analysis to the case where firms are bucketed into K classes, and where exits due to censoring are allowed. 

Consider $m$ firms observed over a discrete set of time periods $t=0,1, 2, \ldots, T$. Again, to keep the notation simple, assume that  all firms are active at time zero. Recall  that  $Y_t \in \Re^{d_1}$ denotes the vector of common market information at time $t$, and  for firm $i \leq m$, $X_{i,t} \in \Re^{d_2}$ denotes a vector of company specific information at time $t$, and  $V_{i,t}= (Y_t, X_{i,t})$. Further,   $(V_{i,t}: i \leq m)_{ 0 \leq t \leq T}$ denotes a stationary Gaussian process restricted to integer times $0 \leq t \leq T$. Let $(V_{i}: i \leq m)$ denote the  random variables with the associated stationary distribution. These, as indicated earlier,  are all assumed to be normalized to have marginal mean zero and variance one. Let  $\Sigma\in\Re^{d\times d}$ denote the correlation matrix of $V_i$, where $d=d_1+d_2$. In this homogeneous set-up, we assume that this is same for all $i$. Let $D_{i,t+1}=1$ if firm $i \leq m$ that survives up to time $t$, defaults between times $t$ and  $t+1$. For each $t$, let ${\cal F}_t$ denote the sigma algebra generated by $(D_{i,s}, V_{i,s}: i \leq m, s \leq t)$.
\subsection{Asymptotic formulation}

Since the conditional default probabilities are typically very small, we analyze the estimation problem in a regime indexed by $\gamma$ as $\gamma \downarrow 0$. Specifically, we assume that
for each firm $i$, the conditional default probability $p(\gamma,V_{i,t}) $
of defaulting in period $(t,t+1]$ ($D_{i,t+1}=1$), conditioned on ${\cal F}_t$, and it surviving at time $t$, is small and is given by:
\begin{equation}\label{eqn:CDP}
   p(\gamma, V_{i,t}) =\exp(\beta^{\intercal} V_{i,t} - \alpha(\gamma))
(1+H(\gamma,V_{i,t})),
\end{equation}
where
$H(\gamma,V_{i,t}) \rightarrow 0$ as $\gamma \rightarrow 0$,  almost surely,
$\beta = (\theta, \eta)$ for  $\theta \in \Re^{d_1}$, $\eta \in \Re^{d_2}$
is independent of $\gamma$ and
\[
\alpha(\gamma) = \log (1/\gamma) - \log c,
\]
where $c>0$ is a constant. This ensures that the conditional default probabilities
are of order $\gamma$ as $\gamma \downarrow 0$. 
For presentation ease we have hidden the dependence $p(\gamma, V_{i,t})$ on $(\beta, \alpha(\gamma))$. We also assume that
\begin{equation} \label{eqn:prob_def_supp}
 |H(\gamma,V_{i})| \leq 
C \gamma \exp(\beta^{\intercal} V_i)
\end{equation}
a.s.  for a constant $C>0$.
\begin{remark} \em{ 
As mentioned in the introduction, logit and intensity based models are widely used to model conditional default probabilities. It can be seen that both of these obey (\ref{eqn:CDP}) and (\ref{eqn:prob_def_supp}). For example, in the discrete intensity model, the conditional default probabilities at any time $t$ have the form $1-   \exp \left ( - e^{\beta^{\intercal} V_{i,t} - \alpha(\gamma)}\right )$. Since $e^x(1- e^x /2) \leq 1- \exp(-e^x) \leq e^x$, it is easily seen that $C= \frac{c}{2}$ satisfies (\ref{eqn:prob_def_supp}). \\
When conditional default probabilities at any time $t$ have a logit representation
\[
 \frac{\exp(\beta^{\intercal} V_{i,t} - \alpha(\gamma))}
{1 + \exp(\beta^{\intercal} V_{i,t} -\alpha(\gamma)) },
\]
since for each $x$, $e^x(1- e^x) \leq  \frac{e^x}{1+e^x} \leq   e^x$, it is easily seen that
$C=  c$ satisfies (\ref{eqn:prob_def_supp}).} 
\end{remark}
Let $\tilde{p}(\gamma)$ denote the average default probability of a firm, that is, let $\tilde{p}(\gamma) = E(p(\gamma,V_{i,t}))$. The following observation is  easily seen:
\begin{equation} \label{eqn:00439}
\tilde{p}(\gamma) = c  \gamma  \exp(\frac{1}{2} \beta^{\intercal} \Sigma \beta) +O(\gamma^2).
\end{equation}

Further, in our asymptotic framework, the number of firms,  
 $m(\gamma) =\gamma^{-\delta}$ for $\delta >0$, and total number of periods
$T(\gamma) = \gamma^{- \zeta}$ for $\zeta \in (0,1)$ so they both increase as $\gamma \downarrow 0$. 

\begin{remark}\em
Similar asymptotic regimes, where default probabilities go to zero as the number of firms increase have been considered in literature on rare event analysis and efficient simulation of credit risk (see, e.g., Glasserman and Li 2004, Glasserman, Kang and Shahbuddin 2007, Bassamboo, Juneja and Zeevi 2008 and Spiliopolous and Sowers 2015). While in these works, and most literature on parameter estimation in credit risk (see, e.g., Duffie et al. 2007, Duan et al. 2012,   Sirignano and Giesecke 2018), a fixed time period is considered, in this paper, we also allow the time periods of observation to increase as $\gamma\downarrow 0$. The rationale behind this, as in  most asymptotic analyses, is that important structural insights are often better seen in an appropriate limiting regime. For instance, in Theorem~\ref{theorem:1}, we observe that for the estimator that we propose in \eqref{hatbeta} (motivated by (6) and the related discussion), if  for $m(\gamma) =\gamma^{-\delta}$, $\delta <1$, further increase in number of forms in the data, improves the mean square error of the estimator, while this is no longer true for $\delta>1$. We also observe how the mean square error decreases as $T(\gamma) = \gamma^{-\zeta}$ increases. Further, in Section 3.4, this asymptotic regime allows us to  derive rate of convergence of the MLE to the true parameters, as well as an associated central limit theorem where the covariance matrix has an explicit simple form. Both these insights would have been difficult to see if $T(\gamma)$ was assumed to be fixed as $\gamma\downarrow 0$.

The specific form for $T(\gamma)$ and $m(\gamma)$ is data driven - typical time periods may be in months, where the conditional default probabilities  are usually of order $10^{-3}$, so one may heuristically view $\gamma \sim  10^{-3}$.  We typically encounter data involving tens of thousands of firms, so $\delta \in [1,2)$, is reasonable although there may be cases where the data is limited and $\delta \in (0,1)$ is  a better representation. The time period can be in order of tens of years,
so $\zeta \in (0,1)$ appears reasonable.  Taking $\zeta \geq 1 $ would correspond to data of the order of hundreds of years. Procuring reliable data over such a long interval may be difficult. Further, the assumption that the covariates are stationary over such long intervals is increasingly untenable. 
\end{remark}

\subsection{Proposed estimators}
Guided by the discussion in Section 2, we  develop our parameter estimation methodology. Consider the following sequences of random variables indexed by $\gamma$,
\begin{align} \label{eqno:101}
\hat{D}_{\gamma} =\frac{1}{\gamma T(\gamma) m(\gamma)}\sum_{t=0}^{T(\gamma)-1}
\sum_{i=1}^{{m}(\gamma)}D_{i,t+1} \textrm{ \ \ and \ \ }   \hat{V}_{\gamma} =\frac{1}{\gamma T(\gamma)m(\gamma)}\sum_{t=0}^{T(\gamma)-1}
\sum_{i=1}^{m(\gamma)}V_{i,t} D_{i,t+1}.
\end{align}
As suggested in (\ref{eqn:interesting1}), the proposed estimator for $\beta$ is
\begin{equation}\label{hatbeta}
    \hat{\beta}(\gamma) \triangleq \Sigma^{-1}   \times \frac{\hat{V}_{\gamma}}{\hat{D}_{\gamma}}.
\end{equation}
As before, let
\begin{equation}\label{eqn:defaulttime}
\tau_{i} = \min\{T(\gamma)-1, \min \{t\geq 0: D_{i,t+1}=1\}\}.
\end{equation}
Thus, if firm $i$ defaults between times $t$ and $t+1$, $\tau_i=t$. Observe  that $E(D_{i,t+1}\vert \mathcal{F}_{t}) =  p(\gamma, V_{i,t}) \mathbb{I}(\tau_{i} \geq t)=\exp(\beta^{\intercal}V_{i,t}-\alpha(\gamma))(1+H(\gamma,V_{i,t}))\mathbb{I}(\tau_{i} \geq t)$. Hence,
\begin{equation}\label{alphaexact}
\alpha(\gamma) = \log\left(\frac{\sum_{i=1}^{m(\gamma)}\sum_{t=0}^{T(\gamma)-1} E(\exp(\beta^{\intercal}V_{i,t})(1+H(\gamma,V_{i,t}))\mathbb{I}(\tau_{i}\geq t))}{\sum_{i=1}^{m(\gamma)}\sum_{t=0}^{T(\gamma)-1} ED_{i,t+1}}\right).
\end{equation}
 This motivates  our empirical estimator for $\alpha$,
\begin{equation}\label{alphahat}
\hat{\alpha}(\gamma) \triangleq \log\left( \frac{\sum_{i=1}^{m(\gamma)}\sum_{t=0}^{T(\gamma)-1}\exp(\hat{\beta}^{\intercal}(\gamma) V_{i,t})\mathbb{I}(\tau_{i} \geq t)}{\sum_{i=1}^{m(\gamma)}\sum_{t=0}^{T(\gamma)-1}D_{i,t+1}}\right),
\end{equation}
where $\hat{\beta}(\gamma)$ from (\ref{hatbeta}) is the estimator for $\beta$.

\textbf{Covariates as a stationary process with short range dependence:} We further assume that that the covariates $(V_{i,t}: t \geq 0)$ for each $i \leq m$ have a short range dependence over time. Specifically, we require that the correlation between $V_{i,t}$ and $V_{j,t+k}$ decays exponentially fast over time. Let  $\|A\|$ denote  the operator norm for the  matrix $A$, that is, $\|A\| = \inf_{r \in \Re^{+}} (r  \colon \|Ax\|_{2} \leq r\|x\|_{2} \ \textrm{ for all $x$})$, where $\|y\|_{2}$ denotes the Euclidean norm for any vector $y$. Also, let $N(0,C)$ denote a Gaussian random variable with mean 0 and covariance matrix $C$.
\begin{assumption}\label{Assumption:Geometric}
For each $i$ the covariates $(V_{i,t}: t \geq 0)$  are distributed as  $N(0, \Sigma)$. Further, there exists a $\rho \in (0,1)$ and some constant $K$ such that for all $(i,j)$,
\begin{equation}\label{eqn:weakdependence}
\|E(V_{i,t}V_{j,t+k}^{\intercal})  \| \leq K\rho^{k}.    
\end{equation}
\end{assumption}
This assumption is satisfied, for example by causal ARMA models used widely in econometric literature (see Hamilton 1994). See \ref{sec:ARMA-Int} for further details.

\subsection{Main results}
Theorem~\ref{theorem:1} specifies the order of the square error of the proposed estimators for $\beta$ and $\alpha(\gamma)$. A few definitions are needed first:\\ 
{Let $X_{\gamma}$ be an indexed set of random variables, and $a_\gamma$ be an indexed set of real numbers, $\gamma > 0$. Then, $X_{\gamma} = O_{\mathcal{P}}(a_{\gamma})$ if for any $\varepsilon>0$, there exists an $M>0$, such that for all $\gamma$ small enough, $P\left(\big\vert \frac{X_{\gamma}}{a_\gamma}\big\vert >M \right) <\varepsilon$.}
\begin{theorem} \label{theorem:1} Under Assumption 1, with $\zeta\in(0,1)$ and $\delta>0$:  
\begin{enumerate}
    \item \begin{equation} \label{betabound1}
\|\beta - \hat{\beta}(\gamma)\|_{2}^{2} =   O_{\mathcal{P}}(\gamma^{\zeta+\delta-1}) + O_{\mathcal{P}}(\gamma^{\zeta}),
\end{equation}
\item \begin{equation}\label{eqnlemma2}
|\alpha(\gamma) - \hat{\alpha}(\gamma)|^{2} = O_{\mathcal{P}}(\gamma^{\zeta + \delta-1}) + O_{\mathcal{P}}(\gamma^{\zeta}).
\end{equation}
\end{enumerate}

\end{theorem}

{
Theorem~\ref{theorem:1} makes an interesting observation related to the sensitivity of the proposed estimator to the  systematic risk (risk that does not diversify away with increase in number of firms). Recall that $\zeta \in (0,1)$. Observe the obvious fact that if
$\delta +\zeta <1$ then the error is unbounded. This is because asymptotically, no defaults are observed in the data.  Now consider two regimes:
\begin{enumerate}
\item $\delta \geq 1$. As mentioned earlier, in this case, further increase in $\delta$ does not reduce the rate at which the error goes to 0. Thus, having more than order $\gamma^{-1}$ firms  does not help in improving accuracy of the proposed estimator.
\item $\delta < 1$. In this case, increasing $\delta$ does reduce the
rate at which the error of estimation goes to 0. Thus, having more firms data is useful up to order $\gamma^{-1}$, thereafter its utility to the proposed estimator is marginal. 
 \end{enumerate}
\begin{remark}\label{rem:Cov-Noise}\em
Suppose that  $\Sigma$ is not known, but is estimated from data by $\hat\Sigma = \frac{1}{T(\gamma)m(\gamma)} \sum_{i,t} V_{i,t}V_{i,t}^{\intercal}$.
Then, using the multidimensional central limit theorem, it can be seen that   
 $\|\hat{\Sigma} - \Sigma\| =  O_{\mathcal{P}}\left(\gamma^{\frac{1}{2}\zeta}\right)$ (see Bickle and Levina, 2008. See also, \ref{EC:Supnum}).  In that case, our estimator for $\beta$ becomes $\bar{\beta}(\gamma) = \hat\Sigma^{-1} \frac{\hat{V}_\gamma}{\hat{D}_\gamma}$. Observe that here, the error in estimation of the covariance matrix is in order of magnitude, less than or equal to the error in the parameter estimation. From Theorem~\ref{theorem:1}, it follows that $\|\beta - \bar{\beta}(\gamma)\|_2=O_{\mathcal{P}}\left(\gamma^{\frac{1}{2}(\delta+\zeta-1)}\right) + O_{\mathcal{P}}\left(\gamma^{\frac{1}{2}\zeta}\right)$, and asymptotic rate of convergence of the modified estimator to the true parameters remains the same as before.
\end{remark}
\textbf{Proof Outline of Theorem~\ref{theorem:1}} To prove (\ref{betabound1}) in Theorem~\ref{theorem:1}, observe that
\begin{equation}\label{betaboundm}
\|\beta - \hat{\beta}(\gamma)\|_{2}^{2} = \|\hat{\beta}(\gamma) - \beta^*(\gamma)\|_2^{2} + \|\beta^*(\gamma) - \beta\|_2^{2} + 2(\hat{\beta}(\gamma) - \beta^*(\gamma))^{\intercal}(\beta^*(\gamma) - \beta),
\end{equation}
where $\beta^*(\gamma) =  \Sigma^{-1}\frac{E(\hat{V}_{\gamma})}{E(\hat{D}_{\gamma})}$. The first term of \eqref{betaboundm}, $\|\beta^*(\gamma) - \beta\|_2^{2}$ is the bias associated with the first order approximation of the default probability. This is bounded by Lemma~\ref{lemma:betabias}. The second term of \eqref{betaboundm}, $\|\beta - \hat{\beta}(\gamma)\|_{2}^{2}$ is the estimation noise. This is bounded using Lemma~\ref{lemma:keylem}. The cross term in \eqref{betaboundm} is handled using the Cauchy-Schwarz inequality.
\begin{lemma}\label{lemma:betabias}
\begin{equation} \label{bound:betastar}
\Big\|\frac{E\hat{V}_{\gamma}}{E\hat{D}_{\gamma}} - \Sigma\cdot\beta\Big\|_{2} = O(\gamma).
\end{equation}
\end{lemma}
From Lemma~\ref{lemma:betabias}, it follows that $  \|\beta^*(\gamma) - \beta\|_2= O(\gamma),$ since
\begin{align*}
\|\beta^*(\gamma) - \beta\|_2 & = \Big\|\Sigma^{-1}\left(\frac{E\hat{V}_{\gamma}}{E\hat{D}_{\gamma}} - \Sigma\cdot\beta\right)\Big\|_{2} \leq \|\Sigma^{-1}\|\Big\|\frac{E\hat{V}_{\gamma}}{E\hat{D}_{\gamma}} - \Sigma\cdot\beta\Big\|_{2}.
\end{align*}
Equation (\ref{bound:betastar}) is the result of the following component-wise bound which is proved in the \ref{EC:Proofs}:
\begin{lemma}\label{1Dbound}
\begin{equation}\label{eqn:1Dbound}
    \Big\vert\frac{E\hat{V}_{\gamma}^{(i)}}{E\hat{D}_{\gamma}} - (\Sigma\cdot\beta)_{i}\Big\vert= O(\gamma) \textrm{ } \forall i \in \{1,2, \cdots ,d\}.
\end{equation}
\end{lemma}
Then (\ref{bound:betastar}) follows as
\begin{align*}
    \Big\| \frac{E\hat{V}_{\gamma}}{E\hat{D}_{\gamma}} - \Sigma\cdot\beta\Big\|_{2} &\leq d \max_{i=\{1,\ldots,d\}} \Big\vert\frac{E\hat{V}_{\gamma}^{(i)}}{E\hat{D}_{\gamma}} - (\Sigma\cdot\beta)_{i}\Big\vert.
\end{align*}
To see the intuition behind Lemma~\ref{1Dbound}, observe that roughly speaking, 
\begin{equation}\label{eqn:ProofApprox}
    \frac{E\hat{V}_{\gamma}^{(1)}}{E\hat{D}_{\gamma}} \approx \frac{E V_{i}^{(1)}p(\gamma,V_i)}{Ep(\gamma,V_i)},
\end{equation}
where recall that $V_i$ is distributed according to the stationary distribution of the covariates.  Next, recall that $p(\gamma,V_i) = \exp(\beta^{\intercal}V_{i})(1+H(\gamma,V_i))$, where $|H(\gamma,V_{i})|\leq \gamma\exp(\beta^\intercal V_i-\alpha(\gamma))$. Then,  
\[
Ep(\gamma,V_i) = \exp\left(\frac{1}{2}\beta^\intercal\Sigma\beta-\alpha(\gamma)\right)+O(\gamma^2) \ \ \textrm{ and }\ \ EV_{i}^{(1)}p(\gamma,V_i)=(\Sigma\beta)_{1}\exp\left(\frac{1}{2}\beta^\intercal\Sigma\beta-\alpha(\gamma)\right)+O(\gamma^2).
\]
Plugging this into the approximation \eqref{eqn:ProofApprox}, we obtain Lemma~\ref{1Dbound}. 

Lemma~\ref{lemma:keylem} bounds the noise $\|\hat{\beta}(\gamma) - \beta^*(\gamma)\|_2^{2} $:
\begin{lemma} \label{lemma:keylem}
\begin{equation}
    \|\hat{\beta}(\gamma) - \beta^*(\gamma)\|_2^{2} = O_{\mathcal{P}}(\gamma^{\zeta +\delta -1}) + O_{\mathcal{P}}(\gamma^{\zeta}).\label{varerrm}
\end{equation}
\end{lemma}

To see the intuition behind Lemma~\ref{lemma:keylem}, consider the first order Taylor approximation of a differentiable function $f(\cdot)$ for $\mathbf{u}$ in a multidimensional Euclidean space,
\[
f(\mathbf{u})= f(\mathbf{a}) + \langle \nabla f(\mathbf{a}), (\mathbf{u-a}) \rangle + R(\mathbf{u,a}),
\]
where $R(\cdot)$ denotes the first order remainder term.
Applying this to $f(x,y) =\frac{x}{y}$, with $x=\hat{V}_{\gamma}^{(1)}$ and $y=\hat{D}_{\gamma}$, setting $\mathbf{a}=(E\hat{V}_{\gamma}^{(1)}, E\hat{D}_{\gamma})$,
\begin{align}\label{eqn:TSEbeta}
    \hat\Sigma\cdot\left(\hat\beta(\gamma) - \beta^{*}(\gamma)\right) &= \frac{\hat V_{\gamma}}{\hat D_{\gamma}} - \frac{E\hat V_{\gamma}}{E\hat D_{\gamma}} \nonumber\\
    &=\frac{1}{E\hat{D}_{\gamma}} (\hat{V}_{\gamma}^{(1)} -E\hat{V}_{\gamma}) - \frac{E\hat{V}_{\gamma}^{(1)}}{(E\hat{D}_{\gamma})^{2}}(\hat{D}_{\gamma} - E\hat{D}_{\gamma}) + R((\Hat{D}_{\gamma},\Hat{V}_{\gamma}^{(1)}), (E\Hat{D}_{\gamma},E\Hat{V}_{\gamma}^{(1)})),
\end{align}
where $R((\Hat{D}_{\gamma},\Hat{V}_{\gamma}^{(1)}), (E\Hat{D}_{\gamma},E\Hat{V}_{\gamma}^{(1)}))$ is the first order Taylor remainder. Consider the term $(\hat{D}_{\gamma} - E\hat{D}_{\gamma})$ in \eqref{eqn:TSEbeta}. Expand this as 
\begin{equation}\label{eqn:KeyeqnTSE}
    \frac{1}{\gamma T(\gamma)m(\gamma)}\sum_{t=0}^{T(\gamma)-1}
\sum_{i=1}^{{m}(\gamma)}(D_{i,t+1} - E (D_{i,t+1}|{\cal F}_t))+\frac{1}{\gamma T(\gamma)m(\gamma)}\sum_{t=0}^{T(\gamma)-1} \sum_{i=1}^{{m}(\gamma)}(E (D_{i,t+1}|{\cal F}_t)- ED_{i,t+1})
\end{equation}
and recall (\ref{hatbeta}).
To keep things simple, assume that $d_1=1, d_2=0$, $\Sigma=1$, and note that
$E\hat{D}_{\gamma}$ is  a constant plus an $O(\gamma)$ term.
Now, the square of \eqref{eqn:KeyeqnTSE} equals $(E\hat{D}_{\gamma})^{-4}$ times
\begin{equation}\label{eqn:split}
 E  \left [ \frac{1}{\gamma m(\gamma) T(\gamma)}
\sum_{t=0}^{T(\gamma)-1}\sum_{i=1}^{m(\gamma)}
\left (D_{i,t+1} - E(D_{i,t+1}|{\cal F}_t) \right )
+
\sum_{t=0}^{T(\gamma)-1}\sum_{i=1}^{m(\gamma)}
\left (E(D_{i,t+1}|{\cal F}_t)  -   E{D}_{i,t+1}\right )  \right ]^2.   
\end{equation}
 Since the cross terms vanish, (\ref{eqn:split}) equals the sum of
 \begin{equation} \label{eqn:intuit01}
  E \left [ \frac{1}{\gamma m(\gamma) T(\gamma)}
\sum_{t=0}^{T(\gamma)-1}\sum_{i=1}^{m(\gamma)}
\left (D_{i,t+1} - E(D_{i,t+1}|{\cal F}_t) \right ) \right ]^2
\end{equation}
and
 \begin{equation} \label{eqn:intuit02}
E \left [ \frac{1}{\gamma m(\gamma) T(\gamma)} \sum_{t=0}^{T(\gamma)-1}\sum_{i=1}^{m(\gamma)}
\left (E(D_{i,t+1}|{\cal F}_t)  - E{D}_{i,t+1}\right ) \right ]^2.
 \end{equation}

 To see that (\ref{eqn:intuit01}) is   $\Theta(\gamma^{\zeta + \delta-1})$, notice that it comprises of uncorrelated terms
 each of which has a second moment that is $\Theta(\gamma)$. Further, to see that  (\ref{eqn:intuit02}) is   $\Theta(\gamma^{\zeta})$,
 observe that $V_{i,t}=Y_t$ is independent of $i$.  Equation \eqref{eqn:intuit02} thus simplifies to equal
 \begin{equation} \label{eqn:intuit03}
c^2 E \left [ \frac{1}{T(\gamma)} \sum_{t=0}^{T(\gamma)-1}
\left ( e^{\beta Y_t  }  - E( e^{\beta Y_t  }) \right ) \right ]^2,
\end{equation}
plus smaller terms. By Assumption 1, $(Y_t: 0 \leq t \leq T(\gamma)-1)$  are short range dependent.  It follows that
(\ref{eqn:intuit03}) is   $\Theta(\gamma^{\zeta})$.
 } This gives  Lemma~\ref{lemma:keylem}.\\
Finally, it follows from Lemmas~\ref{lemma:betabias} and \ref{lemma:keylem} and an application of the Cauchy-Schwarz inequality, that $(\hat{\beta}(\gamma) - \beta^*(\gamma))^{\intercal}(\beta^*(\gamma) - \beta)=O_{\mathcal{P}}(\gamma)$. Since $\zeta \in (0,1)$, (\ref{betabound1}) follows.
\begin{remark}\em
Equation \eqref{eqnlemma2} is somewhat surprising, since $\alpha(\gamma)\to\infty$. However, this follows as  the error in estimation of $\alpha(\gamma)$ and $\beta$ depends on similar quantities. Recall that $\alpha(\gamma)$ and $\hat{\alpha}(\gamma)$ are given by \eqref{alphaexact} and \eqref{alphahat}, respectively. Further, since the default probability of a firm is $O(\gamma)$, and $T(\gamma)=\gamma^{-\zeta}$, for $\zeta<1$, $\tau_i\sim T(\gamma)$, that is $\frac{\tau_i}{T(\gamma)} \xrightarrow{\mathcal{P}} 1$. Adding and subtracting $\log \left(\sum_{i=1}^{m(\gamma)}\sum_{t=1}^{T(\gamma)} \exp({\beta}^{\intercal}V_{i,t})\right) $ and re-arranging, 
\begin{align}
\hat{\alpha}(\gamma) - \alpha(\gamma) = & \log\left(\frac{\sum_{i=1}^{m(\gamma)}\sum_{t=1}^{T(\gamma)} \exp({\beta}^{\intercal}V_{i,t})}{\sum_{i=1}^{m(\gamma)}\sum_{t=1}^{T(\gamma)} E\exp(\beta^{\intercal}V_{i,t})}\right) \label{eqn:alpha1-M}\\
& +\log\left(\frac{\sum_{i=1}^{m(\gamma)}\sum_{t=1}^{T(\gamma)} ED_{i,t+1}}{\sum_{i=1}^{m(\gamma)}\sum_{t=1}^{T(\gamma)} D_{i,t+1}}\right) \label{eqn:alpha2-M}\\
 &+\log\left(\frac{\sum_{i=1}^{m(\gamma)}\sum_{t=1}^{T(\gamma)} \exp(\hat{\beta}^{\intercal}(\gamma)V_{i,t})}{\sum_{i=1}^{m(\gamma)}\sum_{t=1}^{T(\gamma)} \exp(\beta^{\intercal}V_{i,t})}\right) \label{eqn:alpha3-M}
\end{align}
plus smaller order terms. It can be seen from the proof of Theorem 1 (see \eqref{eqn:KeyeqnTSE} onward), that the error in estimation of $\beta$ (to the first order) depends on \eqref{eqn:alpha1-M} and \eqref{eqn:alpha2-M}. Since $\|\hat{\beta}(\gamma) - \beta\|_2 = O_{\mathcal{P}}(\gamma^{\frac{1}{2}(\delta+\zeta-1)}) + O_{\mathcal{P}}(\gamma^{\frac{1}{2}\zeta})$, \eqref{eqn:alpha3-M} can be shown to equal $O_{\mathcal{P}}(\gamma^{\frac{1}{2}(\delta+\zeta-1)}) + O_{\mathcal{P}}(\gamma^{\frac{1}{2}\zeta})$ using a Taylor series expansion.  We refer the reader to \ref{EC:Proofs} for details.
\end{remark}
\noindent In Proposition~\ref{theorem:2}, we find the order of the square of relative error of the firm conditional default probability when the covariates have a stationary distribution. This is perhaps a better measure of error vis-a-vis square error in estimating parameters, since ultimately our interest is in the error made in forecasting conditional probabilities. Also, given that in our asymptotic regime, default probabilities are decreasing to zero, relative error is a more appropriate measure of estimation error vis-a-vis absolute error. Let $\hat{p}(\gamma,V_i)$ denote the value of conditional default probability computed using the estimated parameters $\hat\beta(\gamma)$ and $\hat{\alpha}(\gamma)$ (for example, if the default probability were of the intensity form, $\hat{p}(\gamma,V_i) = 1-\exp(\mathrm{e}^{-\hat{\beta}^\intercal(\gamma)V_{i}-\hat\alpha(\gamma)})$). Here, $V_i$ is drawn from the stationary distribution of the covariates. Proposition~\ref{theorem:2} shows that the relative error of estimation of default probabilities goes to $0$ at the same rate as the error in parameter estimation. 
\begin{proposition} \label{theorem:2}
Under Assumption 1, if $V_i$ below is independent of $\hat{\beta}(\gamma)$ and $\hat{\alpha}(\gamma)$, with $\zeta\in(0,1)$ and $\delta>0$,
\begin{equation}\label{thm1}
\left(\frac{p(\gamma, V_i) -\hat{p}(\gamma, V_i)}{p(\gamma, V_i)}\right)^{2} =
 O_{\mathcal{P}}(\gamma^{\zeta + \delta-1})+ O_{\mathcal{P}}(\gamma^{\zeta}). 
\end{equation}
\end{proposition} 

The following technical assumption on the remainder terms in \eqref{eqn:TSEbeta} ensures that the tail of the error is well behaved and strengthens the results of Theorem~\ref{theorem:1}. It also facilitates more precise performance comparisons with MLE:
\begin{assumption}\label{assumption:TSremainder}
Define $\mu = (\delta+\zeta-1) \land \zeta$. Then, for each $i\in\{1,\ldots,d\}$, $\gamma^{-\mu}R^{2}((\Hat{D}_{\gamma},\Hat{V}_{\gamma}^{(i)}), (E\Hat{D}_{\gamma},E\Hat{V}_{\gamma}^{(i)}))$ is uniformly integrable, that is as $c\to\infty$.
\begin{equation}\label{eqn:UIdef}
    \sup_{\gamma} E(\gamma^{-\mu}R^{2}((\Hat{D}_{\gamma},\Hat{V}_{\gamma}^{(i)}), (E\Hat{D}_{\gamma},E\Hat{V}_{\gamma}^{(i)}))\mathbf{I}(\gamma^{-\mu}R^{2}((\Hat{D}_{\gamma},\Hat{V}_{\gamma}^{(i)}), (E\Hat{D}_{\gamma},E\Hat{V}_{\gamma}^{(i)}))\geq c) \to 0,
\end{equation}
\end{assumption}
\begin{proposition}\label{Cor:thm1}
Under Assumptions \ref{Assumption:Geometric} and \ref{assumption:TSremainder}, there exist positive constants $K_1$ and $K_2$, such that
\begin{equation} \label{betabound-cor}
E \left(\|\beta - \hat{\beta}(\gamma)\|_{2}^{2}\right) =  K_{1}\gamma^{\zeta + \delta-1} + K_{2}\gamma^{\zeta}+  o(\gamma^{\zeta+\delta-1}) +o(\gamma^{\zeta}),
\end{equation}
where $K_{1}$ and $K_2$ depend only on the system parameters, $\Sigma$ and $\beta$.
\end{proposition}
{Assumption~\ref{assumption:TSremainder} is difficult to verify analytically. The simulation results in Section 4 suggest
that the assumption holds for the experiments conducted. }\\ \noindent\textbf{{Extension to high dimensions}:}
In Theorem~\ref{theorem:1} and Proposition~\ref{Cor:thm1}, we have assumed that the number of dimensions $d$ of the data is fixed, that is $V_{i,t}\in\Re^{d}$, independent of $\gamma$. We relax this in \ref{EC:HD} to allow $d(\gamma)\to \infty$ as $\gamma\to 0$, and derive the corresponding approximation to MLE. We further show that this approximation produces vanishing errors for a large $d(\gamma)$, as long as it is $o(\gamma^{-\frac{1}{2}(\delta+\zeta-1)})+o(\gamma^{-\frac{1}{2}\zeta})$.

\subsection{MLE in our asymptotic framework}
Suppose that the conditional probability of default for the firm $i$ at time $t$ given its survival (denoted by $p(\gamma,V_{i,t})$) is specified by \eqref{eqn:CDP}. To make the analysis simpler, we assume that $\alpha(\gamma)$ is known, and equal to $\log\frac{1}{\gamma}$. 
Let $\hat{\beta}_{\textrm{M}}(\gamma)$ be the MLE that maximises \eqref{eqn:LR1} Define the matrix
\begin{equation}\label{eqn:Lim-Cov}
    \Gamma=(\Sigma + \Sigma\beta\beta^{\intercal}\Sigma)^{-1}\exp(-\frac{1}{2}\beta^{\intercal}\Sigma\beta).
\end{equation}
Observe that since $\Sigma$ is positive definite, $\Gamma$ is well defined. 
\begin{theorem}\label{thm:CLT}
Let Assumption~\ref{Assumption:Geometric} hold. Further, let $p(\gamma,V_{i,t})$ be differentiable, and let $H(\gamma, V_{i,t})= h(\beta^\intercal V_{i,t} -\alpha(\gamma))$ for some scalar function $h(\cdot)$, such that $\frac{1}{\gamma}h(\beta^\intercal V_{i,t} -\alpha(\gamma))$ is integrable, and has integrable derivatives of three orders for all $\gamma$ small enough. Then, for $\delta>0$, $\zeta\in(0,1)$ and $\delta+\zeta>1$, the random vector $X_{\gamma} \triangleq \gamma^{-\frac{\delta+\zeta-1 }{2}}(\hat{{\beta}}_{\textrm{M}}(\gamma) - \beta)\xrightarrow{\mathcal{L}} N(0,\Gamma)$. \textrm{Therefore, } 
\begin{equation}\label{eqn:MLEOP}
    \|\hat{{\beta}}_{\textrm{M}}(\gamma) - \beta\|^{2}_{2} = O_{\mathcal{P}}(\gamma^{\delta+\zeta-1}).
\end{equation}
\end{theorem}
The conditions of Theorem~\ref{thm:CLT} ensure that the conditional default probability and its first three derivatives are exponential to the first order. It is easy to check that both the logit and intensity models satisfy these conditions.
\begin{remark}\em
A Central Limit Theorem (CLT) for the MLE is known (see Farhmeir and Kaufman 1985). However, their result is only proved in an asymptotic regime where the conditional default probabilities are independent of $\gamma$.
 We provide a proof that uses arguments from Hall and Hyde (1980), Helland (1982) and weak convergence results from Van der Vaart (1998) to extend the central limit theorem to the case where default probabilities vanish as $\gamma \to 0$. This may be useful in developing improved confidence intervals for the MLE when the default event is rare, however, we do not explore this aspect further in the paper.
\end{remark}
\textbf{Heuristic derivation of CLT:} We  provide a brief heuristic explanation for the fact that $X_{\gamma}$ above converges to a Gaussian random vector, in a simple single dimensional set-up with $d_1+d_2=1$. Consider the first order conditions for MLE in our asymptotic framework. Let 
 $\tau_i$ be as defined in (\ref{eqn:defaulttime}). Recall that $\tau_i \sim T(\gamma)$ as  $\gamma \rightarrow 0$.
Then, from the first order conditions for MLE, loosely speaking, we have
\begin{equation}  \label{logit:0011}
\sum_{i\leq m(\gamma), t \leq \tau_i} V_{i,t} D_{i,t+1}
= \left [ \sum_{i \leq m(\gamma), t \leq \tau_i}V_{i,t} \exp(\hat{\beta}_{M}(\gamma) V_{i,t}-{\alpha}(\gamma)) \right ](1+ O(\gamma)).
\end{equation}
Expanding the RHS in (\ref{logit:0011}) using Taylor's expansion around $\beta$, dividing both sides by
${\gamma \, m(\gamma) \, T(\gamma)}$,
and observing that along the set $\tau_i \geq t$, $E (D_{i,t+1} |{\cal F}_t)  = \exp(\beta V_{i,t}-{\alpha}(\gamma))(1+O(\gamma))$,
we get
\begin{equation}  \label{logit:00111}
\frac{1}{\gamma m(\gamma) T(\gamma)} \sum_{i\leq m(\gamma), t \leq \tau_i} V_{i,t} \left ( D_{i,t+1} - E (D_{i,t+1} |{\cal F}_t) \right )
= (\hat{\beta}_{M}(\gamma)- \beta) \frac{1}{\gamma m(\gamma) T(\gamma)}  \sum_{i\leq m(\gamma), t \leq \tau_i} V^2_{i,t} \exp(\beta V_{i,t}-{\alpha}(\gamma))
\end{equation}
plus remainder terms. Observe that $\frac{1}{m(\gamma) T(\gamma)} \sum_{i\leq m, t \leq \tau_i} V^2_{i,t} \exp(\beta^{\intercal} V_{i,t})$ converges to a constant  as $\gamma \rightarrow 0$.  Since, the terms $\left (D_{i,t+1} - E (D_{i,t+1} |{\cal F}_t) \right )$ are zero mean random variables uncorrelated
for each $t$ and $i$, with variance of order $\gamma$ (as $\gamma \rightarrow 0$), it is easy to see that 
for $\gamma^{-(\delta+\zeta-1)/2}$ times the LHS of (\ref{logit:00111}) converges to a Gaussian random variable. \\
Corollary~\ref{Cor:UCLT} strengthens Theorem~\ref{thm:CLT}, and is required to compare the performance of the proposed approximations with the MLE.
\begin{corollary}\label{Cor:UCLT}
Suppose that $(\|X_{\gamma}\|^{2}_{2}: \gamma>0)$ is uniformly integrable, that is, as $c\to\infty$, $\sup_{\gamma}E(\|X_{\gamma}\|^{2}_{2} \mathbb{I}(\|X_{\gamma}\|^{2}_{2} > c)) \to 0$. Then,
\begin{equation}\label{eqn:MLEvar}
   E\left( \|\hat{{\beta}}_{\textrm{M}}(\gamma) - \beta\|_{2}^{2}\right) = \mathrm{Tr}(\Gamma)\gamma^{\zeta+\delta-1} + o(\gamma^{\zeta+\delta-1}).
\end{equation}
\end{corollary}
To see Corollary~\ref{Cor:UCLT}, write $E(\|\hat\beta_M(\gamma)-\beta\|_2^2) = \sum_{i=1}^{d}E(\vert (\hat\beta_{M}(\gamma))_i - \beta_i\vert^2))$.
From the convergence in Theorem~\ref{thm:CLT}, with the uniform integrability assumption, letting $\mathrm{Tr}(\Gamma)$ denote the sum of diagonal elements of $\Gamma$, Corollary~\ref{Cor:UCLT} follows.\\
In Proposition~\ref{prop:varianceratio}, we compare the mean square error (mse) of the MLE $\hat{{\beta}}_{\textrm{M}}(\gamma)$ with the mse of the proposed estimator $\hat{{\beta}}(\gamma)$ more precisely again when $d_1+d_2=1$.
 First note that if $\delta >1$, by (\ref{betabound1}) and (\ref{eqn:MLEvar}), mse of $\hat{{\beta}}(\gamma)$ is $\Theta(\gamma^{\zeta})$ and is asymptotically larger than the
 mse of $\hat{{\beta}}_{\textrm{M}}(\gamma)$, which is $\Theta(\gamma^{\delta+\zeta-1})$.
 Hence, consider the case where $\delta<1$. Here, both have mse $\Theta(\gamma^{\delta+\zeta-1})$.
Proposition~\ref{prop:varianceratio} shows that when the number of firms of data is limited to less than a few thousands, the error of the proposed estimator is at most a constant factor times that of the MLE. This fact is corroborated by numerical experiments in Section 4. The proof is provided in \ref{A4}.
\begin{proposition}\label{prop:varianceratio}
Suppose that the collection $(\|X_{\gamma}\|^{2}, \gamma >0)$ is uniformly integrable. Then,
\begin{equation}
    \lim_{\gamma \to 0}\frac{E(\|\hat{\beta}_{M}(\gamma)-\beta\|^{2}_{2})}{E(\|\hat{\beta}(\gamma)-\beta\|^{2}_{2})} \to \frac{1}{1+\beta^{2}}.
\end{equation}
\end{proposition}

\subsection{MLE with Regularisation}\label{sec:MAP}
Consider the problem of maximising the likelihood function $\mathcal{L}(b,a)$ in \eqref{eqn:LR1}, with a reweighing distribution on $b$, call it $g(b)$, that is, consider maximising $\mathcal{L}(b,a)g(b)$. Recall that this is equivalent to maximising $\log \mathcal{L}(b,a) + \log g(b)$. With the average default probability, $\tilde{p}(\gamma)$, given by \eqref{eqn:00439}, let $\log g(b) = \lambda u(b)\tilde{p}(\gamma)T(\gamma)m(\gamma)$ for some $\lambda>0$, and consider the following normalised version of this problem:
\begin{equation}\label{eqn:MAP-Reg-1}
    \max_{(b,a)\in \Re^{d+1}} \left( \frac{1}{\tilde{p}(\gamma) T(\gamma)m(\gamma)}\sum_{i\leq m(\gamma),t\leq \tau_i(\gamma)} D_{i,t+1}\log p(V_{i,t},b,a) + (1-D_{i,t+1})\log(1-p(V_{i,t},b,a)) +\lambda u(b)\right).
\end{equation}
The optimisation problem in \eqref{eqn:MAP-Reg-1} is the regularised version of the MLE, where $u(b)$ acts as the regulariser. The normalisation is selected to keep the objective function in \eqref{eqn:MAP-Reg-1} stochastically bounded as $\gamma\to 0$. Suppose the conditions of Theorem~\ref{thm:CLT} hold. Then, recall that, the value of the conditional default probability evaluated at $(b,a)\in\Re^{d+1}$ is given by
\[
p(V_{i,t}, b, a) = q(b^\intercal V_{i,t} -a), 
\]
where $q(x) = \exp(x)(1+h(x))$ is a real valued function of one variable. Write $q^{\prime}(\cdot)$ for the derivative of $q(\cdot)$. From the chain rule, the gradient of the conditional default probability with respect to $b$ is given by $\nabla p(V_{i,t},b,a) = V_{i,t} q^{\prime}(b^\intercal V_{i,t} - a)$, and its derivative with respect to $a$ is given by $-q ^{\prime}(b^\intercal V_{i,t} - a)$. Applying the first order conditions to \eqref{eqn:MAP-Reg-1}, the solutions to the regularised MLE, given by $(\beta^*_R(\gamma),\alpha^*_R(\gamma))$, satisfy
\begin{align} 
     \frac{1}{\tilde{p}(\gamma) T(\gamma)m(\gamma)}\sum_{i \leq m(\gamma), t=1}^{ \tau_i(\gamma)}  \frac{V_{i,t} q^{\prime}(V_{i,t})}
{q(V_{i,t})(1-q(V_{i,t}))}
   D_{i,t+1}
&=  \frac{1}{\tilde{p}(\gamma) T(\gamma)m(\gamma)} \sum_{i \leq m(\gamma), t=1}^{ \tau_{i}(\gamma)}V_{i,t}\frac{q^{\prime}(V_{i,t})}{1-q(V_{i,t})}- \lambda\nabla u(\beta_R^*(\gamma))\label{eqn:MAP-FOC-1}\\
\frac{1}{\tilde{p}(\gamma) T(\gamma)m(\gamma)}\sum_{i \leq m(\gamma), t=1}^{ \tau_i(\gamma)}  \frac{ q^{\prime}(V_{i,t})}
{q(V_{i,t})(1-q(V_{i,t}))}
   D_{i,t+1}
&=  \frac{1}{\tilde{p}(\gamma) T(\gamma)m(\gamma)} \sum_{i \leq m(\gamma), t=1}^{ \tau_{i}(\gamma)}\frac{q^{\prime}(V_{i,t})}{1-q(V_{i,t})}\label{eqn:MAP-FOC-2},
\end{align}
where $q(V_{i,t})$ and $q^{\prime}(V_{i,t})$ respectively stand for $q({\beta_R^*}^{\intercal}(\gamma)V_{i,t}-\alpha_R^{*}(\gamma))$ and $q^{\prime}({\beta_R^*}^{\intercal}(\gamma)V_{i,t}-\alpha_R^{*}(\gamma))$, and where we suppress the $\beta_{R}^*(\gamma)$ and $\alpha^*_{R}(\gamma)$ from this notation for ease of presentation. Since defaults are rare, observe that as $\gamma\to 0$, \[\frac{ q^{\prime}(V_{i,t})}
{q(V_{i,t})(1-q(V_{i,t}))} = 1+o_{\mathcal{P}}(1).\] 
Hence, the LHS of \eqref{eqn:MAP-FOC-1} is approximately $\frac{1}{\tilde{p}(\gamma) T(\gamma)m(\gamma)}\sum_{i=1}^{m(\gamma)}\sum_{t=1}^{T(\gamma)-1}V_{i,t}D_{i,t+1}$, and the LHS of \eqref{eqn:MAP-FOC-2} is approximately $\frac{1}{\tilde{p}(\gamma) m(\gamma)T(\gamma)}\sum_{i=1}^{m(\gamma)}\sum_{t=1}^{T(\gamma)-1}D_{i,t+1}$.
Set 
\[
\hat V^R_\gamma\triangleq\frac{1}{\hat{p}(\gamma)T(\gamma)m(\gamma)} \sum_{i=1}^{m(\gamma)}\sum_{t=1}^{T(\gamma)-1}V_{i,t}D_{i,t+1} \textrm{ \ \ and \ \ } \hat D_{\gamma}^R \triangleq\frac{1}{\hat{p}(\gamma)T(\gamma)m(\gamma)} \sum_{i=1}^{m(\gamma)}\sum_{t=1}^{T(\gamma)-1}D_{i,t+1}.
\]
Here $\hat{p}(\gamma) = \frac{1}{T(\gamma)m(\gamma)} \sum_{i=1}^{m(\gamma)}\sum_{t=1}^{T(\gamma)-1} D_{i,t+1}$ is the empirically observed default probability. Using the law of large numbers on the RHS of \eqref{eqn:MAP-FOC-1} and \eqref{eqn:MAP-FOC-2}, observe that 
\begin{equation}\label{eqn:MAP-Approx-sol}
    \hat{V}^R_{\gamma} = \Sigma\beta_R^*\hat{D}^R_{\gamma} - \lambda\nabla u(\beta_R^*)
\end{equation}
plus $o_{\mathcal{P}}(1)$ terms. Also notice that by definition, $\hat{D}_{\gamma}^R=1$ a.s. Consider ridge regularisation, so that $u(b) = - \frac{1}{2}b^\intercal\mathbf{Z}b$. Substituting for $u(\cdot)$  in \eqref{eqn:MAP-Approx-sol}, we obtain  a closed form approximation to the regularised MLE 
\begin{equation}\label{eqn:approx-L2reg}
    \hat{\beta}_R(\gamma) = (\Sigma + \lambda \mathbf{Z})^{-1} \hat{V}_\gamma.
\end{equation}
In general, the approximation to the regularised MLE is derived by solving \eqref{eqn:MAP-Approx-sol}. This is equivalent to solving for the maximum of the function
\begin{equation}\label{eqn:Super-function}
    f(b) = b^\intercal \hat{V}_{\gamma} - \frac{1}{2}b^\intercal\Sigma b +\lambda u(b).
\end{equation}
Further suppose that $u(b)$ is concave in $b$ (this is true, for example if $g(\cdot)$ has the Laplace density resulting in the LASSO regularised MLE), then $f(\cdot)$ is a concave function of $b$, and the approximation to the regularised  MLE reduces to finding the maximiser of a simple concave program. Proposition~\ref{lemma:MAP-Reg} below gives the rates of convergence of this approximation to the solution of \eqref{eqn:MAP-Reg-1}:

\begin{proposition}\label{lemma:MAP-Reg}
If $u(\cdot)$ is concave, then $f(\cdot)$ has an almost surely unique maximiser, call it $\hat{\beta}_R(\gamma)$. Further, if $\beta_R^*(\gamma)$ denotes the maximiser of \eqref{eqn:MAP-Reg-1}, then, under the conditions of Theorem~\ref{thm:CLT}, $\|\hat{\beta}_R(\gamma)-{\beta}^*_R(\gamma)\|_2=O_\mathcal{P}(\gamma^{\frac{1}{2}(\delta+\zeta-1)})+O_{\mathcal{P}}(\gamma^{\frac{1}{2}\zeta})$. 
\end{proposition}
The key steps in the proof of Proposition~\ref{lemma:MAP-Reg} are similar to those in the proofs of Theorems~\ref{theorem:1} and \ref{thm:CLT}. An outline is provided in \ref{sec:E.C.RegProof}. Proposition~\ref{lemma:MAP-Reg} shows that the regularised MLE can be accurately approximated by the solution of a convex program. Since $u(\cdot)$ is known to the modeller, solving the approximate problem of maximising \eqref{eqn:Super-function} is an easy task. Contrast this to solving the problem \eqref{eqn:MAP-Reg-1}, which may be computationally intensive. Even when \eqref{eqn:MAP-Reg-1} corresponds to a convex problem (e.g., in the logit case), computing derivatives of the objective function at each stage takes order $m(\gamma)T(\gamma)$ computations. The proposed estimator meanwhile requires a one time effort of order $m(\gamma)$ to compute $\hat{V}_\gamma$. Further, Proposition~\ref{lemma:MAP-Reg} also suggests that the approximate solution to the regularised MLE acts as a good starting point for algorithms used to solve \eqref{eqn:MAP-FOC-1} and \eqref{eqn:MAP-FOC-2}.

\subsection{Performance in presence of corrupted data}\label{sec:Noisy-MLE}
Recall that the proposed estimator for $\beta$ depends only on the value of the covariates just before the times of the defaults. Since from Theorem~\ref{theorem:1}, the MLE is close to the proposed estimator, it follows  that the MLE may itself be less sensitive to the data observed at other times where the firms do not default in the next time period.  In particular, if such data is missing that should not affect the MLE much. Similarly, if some data just before defaults is missing, both the proposed estimator as well as the MLE should be significantly inaccurate. These observations are validated on simulation data in Section 4.

Further, note that for a variety of reasons, some of the   data may be corrupted by noise. Interestingly, in such settings, the proposed estimator may be relatively unaffected by the noise in the data. To see this in a simple set-up, suppose that the covariates $V_{i,t}$ satisfy Assumption 1, and the default probability is given by \eqref{eqn:CDP}. Further,  the calibrator observes the data with added corruption or noise, $V_{i,t}^{c}  = V_{i,t} + A_{i,t}$. Here $A_{i,t}$ is zero mean, independent of $V_{i,t}$ and $D_{i,t+1}$, such that $E\|A_{i,t}\|_2^{2}\leq M<\infty$ for all $(i,t)$. Now, consider the proposed estimator of $\beta$:
\begin{equation}\label{eqn:Approx-MLE-Corruption}
    \hat{\beta}_c(\gamma) =\Sigma^{-1}\frac{\sum_{i,t} V_{i,t}^cD_{i,t+1}}{\sum_{i,t}D_{i,t+1}}. 
\end{equation}
Observe that since the corruption $A_{i,t}$ is independent of $V_{i,t}$ and $D_{i,t+1}$, and has   zero mean, 
\[
E(D_{i,t+1} V_{i,t}^{c} \vert \mathcal{F}_t) = E(D_{i,t+1} V_{i,t} \vert \mathcal{F}_t) = V_{i,t} \exp(\beta^\intercal V_{i,t} -\alpha(\gamma))\mathbb{I}(\tau_i\geq t). 
\]
Hence, even in presence of additive noise, the law of large numbers based arguments used to arrive at the estimator \eqref{hatbeta} continue to  hold, and we achieve consistent parameter estimates. Proceeding as in the proof of Theorem~\ref{theorem:1}, it can be shown that $\|\hat\beta_c(\gamma) -\beta\|_2 =O_{\mathcal{P}}(\gamma^{\frac{1}{2}(\delta+\zeta-1)}) + O_{\mathcal{P}}(\gamma^{\frac{1}{2}\zeta})$. Thus, the proposed estimator produces a consistent estimator of the parameters, even in the presence of additive noise. Lemma~\ref{lemma:Bias-MLE} shows in an simple setting, that due to its non-linearity, the MLE in presence of data with additive zero mean noise produces asymptotically biased estimators, as $\gamma \rightarrow 0$. 
\begin{lemma}\label{lemma:Bias-MLE}
Consider a case where covariate data is corrupted by additive noise as described above. Let $\hat{\beta}_{c,M}(\gamma)$ denote the MLE for $\beta$ in presence of such  additive noise. Further, suppose that $V_{i,t}\sim N(0,\mathbf{I})$ and $A_{i,t}\sim N(0,c\mathbf{I})$ for some $c>0$.  Then, $\hat{\beta}_{c,M}(\gamma) \xrightarrow{\mathcal{P}} \frac{1}{1+c^2}\beta$.
\end{lemma}
A similar conclusion may be obtained about the regularised MLE and its approximation. Lastly, if the corruption is systematic, that is, the noise terms $A_{i,t}$ have non-zero mean, then both the proposed estimator and the MLE incur a similar asymptotic bias. This suggests that in presence of additive, independent noise in the data, the proposed estimator may be preferred to the MLE, particularly if the noise has zero mean. 

\begin{remark}\em
If the covariance matrix $\Sigma$ is also estimated from the noisy data, both the proposed estimator and the MLE incur a similar bias. However, recall that as discussed in Remark~\ref{rem:Cov-Noise}, estimation of the covariance matrix from data leads to an additional error in parameter estimation, of $O_{\mathcal{P}}\left(\gamma^{\frac{1}{2}\zeta}\right)$. Thus, if a reasonable fraction of the data well spread over time is uncorrupted, then that may be used to estimate $\Sigma$. Here, the error in estimation of $\Sigma$ will still be at most of the same order as the error in the parameter estimation. Hence, our conclusion above, that the proposed estimator is asymptotically unbiased as $\gamma\to 0$ continues to hold. This is validated through simulation experiments in Section 4.
\end{remark}

\subsection{Performance of MLE under model misspecification}
In this section, we demonstrate the effect of model misspecification on the MLE as well as the proposed estimator in a simple illustrative setting. We capture misspecification by assuming that the underlying model generating defaults has two Gaussian factors common to all firms, while the modeller assumes that only one of the two factors exists; the other is unknown to the modeller. Let $(Y_{1,t}, Y_{2,t}: 1 \leq t \leq T(\gamma))$ denote the time series corresponding to the two factors. Further assume that  $(Y_{1,t}, Y_{2,t})$ have a stationary distribution under which random variables $Y_{1,t}$ and $Y_{2,t}$ are assumed to have zero mean, variance 1 and correlation $\rho$ amongst them. Let $(\beta_1, \beta_2, \alpha(\gamma))$ denote the parameters of default generation, where as before $\alpha(\gamma) = \log(1/\gamma) - \log c$.  Suppose it is thought that only the first factor with time series $(Y_{1,t}: 1 \leq t \leq T(\gamma))$ impacts the conditional default probabilities of firms. That is, while the modeller believes the conditional default probability is 
\begin{equation}\label{eqn:CDP-Mis-spec}
    p(\gamma,Y_{1,t})=q(\beta_{1}Y_{1,t}-\alpha(\gamma)),
\end{equation}
the true conditional probability of default is given by \eqref{eqn:CDP}. Here, $q(\cdot)$ is as defined in Section~\ref{sec:MAP}. Let $\hat\beta_{1,M}(\gamma)$ and $\hat\alpha_M(\gamma)$ solve the mis-specified MLE. Applying the first order conditions, these are solutions to the equations:
\begin{align} 
     \frac{1}{\gamma T(\gamma)m(\gamma)}\sum_{i \leq m(\gamma), t=1}^{ \tau_i(\gamma)}  \frac{Y_{1,t} q^{\prime}(Y_{1,t})}
{q(Y_{1,t})(1-q(Y_{1,t}))}
   D_{i,t+1}
&=  \frac{1}{\gamma T(\gamma)m(\gamma)} \sum_{i \leq m(\gamma), t=1}^{ \tau_{i}(\gamma)}Y_{1,t}\frac{q^{\prime}(Y_{1,t})}{1-q(Y_{1,t})}\label{logit:0011aa}\\
\frac{1}{\gamma T(\gamma)m(\gamma)}\sum_{i \leq m(\gamma), t=1}^{ \tau_i(\gamma)}  \frac{ q^{\prime}(Y_{1,t})}
{q(Y_{1,t})(1-q(Y_{1,t}))}
   D_{i,t+1}
&=  \frac{1}{\gamma T(\gamma)m(\gamma)} \sum_{i \leq m(\gamma), t=1}^{ \tau_{i}(\gamma)}\frac{q^{\prime}(Y_{1,t})}{1-q(Y_{1,t})}\label{logit:0011ab},
\end{align}
where $q(Y_{1,t})$ and $q^{\prime}(Y_{1,t})$ respectively stand for $q(\hat{\beta}_{1,M}(\gamma)Y_{1,t}-\hat\alpha_{M}(\gamma))$ and $q^{\prime}(\hat{\beta}_{1,M}(\gamma)Y_{1,t}-\hat\alpha_{M}(\gamma))$.
\begin{theorem}\label{thm:Mis-spec}
Let the assumptions of Theorem~\ref{theorem:2} hold. Consider a two factor default model, under the above mis-specification framework. Then, $\hat\beta_{1,M}(\gamma) \xrightarrow{\mathcal{P}}\beta_1+\rho\beta_2$ and $\alpha(\gamma)-\hat\alpha_M(\gamma)\xrightarrow{\mathcal{P}} \frac{\beta_2^2(1-\rho^2)}{2}$.  Further, the proposed estimator under the same assumptions also converges to these values.
\end{theorem}

The intuition behind Theorem~\ref{thm:Mis-spec} is as follows: when $\gamma m(\gamma) T(\gamma) \rightarrow \infty$ as $\gamma\to 0$, from \eqref{eqn:CDP-Mis-spec}, the LHS in (\ref{logit:0011aa})
converges to
\[
c EY_{1,t} \exp(\beta_1 Y_{1,t} + \beta_2 Y_{2,t})
= c (\beta_1+ \rho \beta_2) \exp \left ( \frac{1}{2}(\beta_1^2 + 2 \rho \beta_1 \beta_2 + \beta_2^2) \right ),
\]
and the RHS in  (\ref{logit:0011aa}) is asymptotically similar (as $\gamma \rightarrow 0$) to
\[
\frac{1}{\gamma}  EY_{1,t} \exp(\hat{\beta}_{1,M}(
\gamma) Y_{1,t} - \hat{\alpha}_M(
\gamma)) = \frac{1}{\gamma} \hat{\beta}_{1,M}(
\gamma) \exp \left ( \frac{1}{2}\hat{\beta}_{1,M}^2(
\gamma)  - \hat{\alpha}_{M}(
\gamma) \right ).
\]
Similarly, the LHS in (\ref{logit:0011ab})
converges to $
c E\exp(\beta_1 Y_{1,t} + \beta_2 Y_{2,t})
=  c \exp \left ( \frac{1}{2}(\beta_1^2 + 2 \rho \beta_1 \beta_2 + \beta_2^2) \right )$,
and the RHS in  (\ref{logit:0011ab}) is asymptotically similar (as $\gamma \rightarrow 0$) to
$\frac{1}{\gamma}  E \exp(\hat{\beta}_{1,M}(
\gamma) Y_{1,t} - \hat{\alpha}_M(\gamma)) = \frac{1}{\gamma} \exp \left ( \frac{1}{2}\hat{\beta}_{1,M}^2(
\gamma)  - \hat{\alpha}_M(\gamma) \right )$. Equating for parameters, it is seen that $\hat{\beta}_{1,M}(
\gamma) \approx \beta_1 + \rho \beta_2$ and $\hat{\alpha}_{M}(
\gamma) - \alpha(\gamma) \approx \frac{\beta_2^2(1-\rho^2)}{2}$. 

 The upshot is that while the MLE and the proposed estimator converge to the same value under this model misspecification, they both are equally wrong in the limit. Thus, practitioner may as well use the simpler proposed estimator. These observations are validated by numerical experiments in Sections~\ref{sec:Numbers} and~\ref{sec:corp_data}.

\section{Simulation Experiments}\label{sec:Numbers}
\textbf{Overview:} In this section we use simulation to generate default data using the intensity model, where solving for the MLE of the underlying parameters is computationally demanding, and on this data compare the proposed estimator and the MLE.  

We first consider  the case where default probabilities are about 1\% per annum (twelve times that per each month). The default generating model comprises of common Gaussian  distributed factors as well as idiosyncratic Gaussian factors for each firm.  Our broad conclusions are that when the model is correctly specified, our estimator is close in accuracy to the MLE in RMSE (Root Mean Square Error) between the true and the estimated parameters, when the number of firms is around 7,000 or less and number of time periods of observation is kept at 200 months. Consistent with the theory, MLE performs relatively better  when the number of firms increases from 5,000 to 13,000, as well as when the number  of time periods increase from 200 to 800 months. We also consider the case where default probabilities are of order 3\% and observe that the MLE performs somewhat better than the proposed estimator, although in all cases, the RMSE of both the estimators is small (see Table \ref{table:logit_correct_small}). To illustrate the effectiveness of our estimator when the firms are heterogeneous, we implement a simulation with two classes of firms, and find that the proposed estimator performs almost as well as MLE (see \ref{EC:Supnum}). We implement the MLE with ridge regularisation, and find that the approximation computed using \eqref{eqn:approx-L2reg} gives an RMSE comparable to the true regualrised MLE (see Table \ref{Reg-full}). In these settings, we find that using the proposed estimator as an initial seed leads to a 5-7 times reduction in the computational effort to solve the MLE.

The proposed estimator also provides an intuition about the factors affecting parameter estimation. In practice, covariate data is often missing or corrupted by noise. We show using a simple simulation experiment, that both the proposed estimator and the MLE are somewhat insensitive to missing data at points where firms are not in default.  Our results validate that gathering  data accurately just before a firm defaults helps in improving the performance of both the estimators. When covariate data is corrupted by noise, we find that the proposed estimator is at least as accurate the MLE, in terms of the RMSE between the true and the estimated parameters.

Often, covariates follow a heavy tailed time series (e.g. ARCH and GARCH models). As mentioned in the introduction, here, for the proposed estimator, the covariates need to be transformed to have approximately Gaussian marginals before parameter estimation. Thus, comparing the proposed estimator and the MLE using RMSE may not be accurate. In practice, one of the key aims of predicting default probabilities is to accurately identify the more risky firms. With this in mind, we adapt the testing methodology proposed  by Duffie et al. (2007), discussed later, to compare the ability of the proposed estimator and the MLE to rank firms in order of risk. We find that the proposed estimator is and the MLE give very similar rankings of risky firms.

When the underlying model is misspecified, such as when one of the common factors is latent to the modeller, we observe that both the proposed estimator as well as MLE have more or less identical RMSE even for large number of firms and long time periods of data availability (few thousands of firms and hundreds of months of data), both when the default probabilities are 1\% and 3\% per annum (see Table \ref{table:logit_incorrect_small}). 

A model may also be misspecified due to non-stationarity. We test robustness to this in a simple setting, where the covariates have a periodic component. Such periodic non-stationarity is designed to capture cyclicity in business cycles. We observe that in this setting, the proposed estimator performs about as well as the MLE (see \ref{EC:Supnum}).

\subsection{Set-up and comparison tests} 
 We test the performance of the proposed estimator when the defaults are generated according to the discrete intensity model, that is,
\[
P(D_{i,t+1}=1\vert (\mathbf{Y}_t, \mathbf{X}_{i,t})) = 1-\exp\left( -\mathrm{e}^{\beta^\intercal (\mathbf{Y}_t, \mathbf{X}_{i,t}) - \alpha}\right),  \ \mathbf{Y}_t\in\Re^{d_1}, \mathbf{X}_{i,t} \in \Re^{d_2},
\]
and number of firms, time period of observation and number of covariates are selected to be 5000-13,000, 200-800 months and 12, respectively, to keep thing similar to a typical practical set-up (see Section~\ref{sec:corp_data}). The covariates are assumed to follow the evolution
\begin{align}
    Y_{t,k}  &= 0.3\cdot Y_{t-1,k} + N_{t,k}(0,1), \ k\in[d_1] \textrm{, and} \label{eqn:Size-large}\\
    X_{i,t,k} &= N_{i,t,k}(0,1) ,\ k\in[d_2]\nonumber, 
\end{align}
where $\{Y_{k,t}\}_{k=1}^{d_1}$ and $\{X_{k,i,t}\}_{k=1}^{d_2}$ are the common and idiosyncratic covariates and all random variables, $N_{i,t,k}(0,1)$, are assumed to be independent standard Gaussian.  To remain close to practice, to compute the proposed estimator, we empirically estimate the covariance matrix $\Sigma$ from data (we also observe that the results remain similar if we assume the covariance matrix to be known). Further, in these experiments we set the covariance matrix $\Sigma$ to a diagonal matrix. Our conclusions remain the same even if $\Sigma$ is not diagonal, that is, there is dependence among the covariates. All experiments are performed using R, and a 2.3 GHz processor with 4 GB RAM. We compare the performance of the proposed estimator to the MLE with respect to three tests:\\
\noindent{\textbf{Test 1.}} When the model is correctly specified, we compare the RMSE of the proposed estimator and the MLE with the true parameters. This is calculated as
the square root of the average of the square of the Euclidean distance between estimated and true parameters.\\
\noindent {\textbf{Test 2.}} As mentioned before, we adapt the testing methodology proposed  by Duffie et al. (2007) (also see Das et al. (2007), Duffie et al. (2009) and Duan and Fulop (2013)) to compare the ability of the proposed estimator and the MLE to identify risky firms. Their methodology is designed for empirical data where the underlying model is not known. In our simulation setting, since we know the underlying model, we exploit that to arrive at a more reasonable test. Specifically, 
\begin{enumerate}
\item we consider $N$ firms and use the intensity model to generate default data for a total of $2T$ months.
The first $T$ months of data is used for training, that is, for learning the underlying parameters 
$(\beta, \alpha)$ using both the proposed method as well as the MLE. 
\item Thereafter, at each time period $t \in [T+1, \ldots, 2T]$, we sort the firms in the descending order of their true conditional default probabilities  (tcdf) as well as estimated conditional default probabilities (ecdf)
using the proposed method as well as the MLE. We compare the top 10\% of risky firms based on tcdf with the ecdf using the two methods, by determining the percentage overlap between the tcdf and the two ecdf lists. This is repeated for all other deciles, 20\%, 30\% and so on, as well as for months $T+2, \ldots, 2T$. The average of the percentage overlap is then reported. 
\end{enumerate}

For the purpose of these experiments, we select $N=10,000$ and $T=100$. Here, a larger overlap in the higher deciles between ecdf and tcdf for a particular estimation method, suggests that the estimation method accurately identifies high risk firms in the portfolio.

\noindent\textbf{Test 3:} Another reasonable way to compare the degradation in the quality  of the proposed estimator with the MLE when the underlying model is correctly specified  may be through comparing the value of the log-likelihood function evaluated at the respective estimators. We do this comparison again for the intensity model when the covariates follow \eqref{eqn:Size-large}, number of fims $N=10,000$ and number of months of data $T=200$. Again 100 independent iterations are conducted. While by definition the `log-likelihood score' of the proposed estimator is less than that of the MLE, 
we find that the average numbers equal -7.012 and -6.913 respectively, so that the log-likelihood function evaluated at the proposed estimator is within 2\% of the maximum possible value.

\subsection{Results}
\noindent\textbf{1) Correctly Specified Model:} We first consider the case where the calibrator is aware of the all underlying factors as well as the form of the default probability, and estimates $(\beta, \alpha)$ from the generated default data. In that case,  we generate the default
data for various values of $m$ and $T$, and arrive at the estimators for the parameters
$(\beta, \alpha)$ using the  proposed method and the MLE. These experiments are  repeated 100 times
and the RMSEs are estimated.  These estimated values are referred to as RMSE($\beta_{prop}$) and
RMSE($\beta_{ML}$) under the two methods.
Similarly, the errors associated with estimators for
$\alpha$ are referred to as RMSE($\alpha_{prop}$) and
RMSE($\alpha_{ML}$).
The experiments are conducted in two sets:  In the first set, we let the number of firms $m$ vary from 5,000 to 13,000. and fix number of time periods $T=200$.  In the second set, $T$ varies from 200 to 800, and $m= 5,000$ is kept fixed. The results are reported in Table~\ref{table:logit_correct_small}, with default probabilities kept at 1\% and 3\% per year. As mentioned earlier, the RMSE  of both the estimators is quite small. For instance, when the number of firms is 7,000 and the data is generated for 200 months, when the annual default probability is about 1\%, the RMSE of the proposed estimator is about 11\% of the absolute value of the underlying $\beta$, while that of the MLE is about 8.5\%.  When the annual default probability is about 3\%, all else being the same, the RMSE of the proposed estimator is about 8.9\% of the absolute value of the underlying $\beta$, while that of the MLE is about 6.5\%. The corresponding figures for $\alpha$ are 1.6\% and 1.2\% when the default probability is 1\%, and 1.3\% and 0.8\% when the default probability is 3\%.

We observe that under Test 2, both the proposed estimator and the MLE give a large and approximately equal percentage overlap with the true model in higher deciles. 
For example, the proposed estimator has an average overlap of 95.39\% in the first decile the true model, while the MLE has an average overlap of 96.18\% with the true model (see Table~\ref{table:Gaussian}).

\noindent\textbf{2) Proposed Estimator as initial seed:} When using intensity based models, computational effort to estimate the MLE can be large. In order to reduce this, we use the proposed estimator as an initial seed for solving the MLE. Considering a representative example where the number of firms $N=10,000$, and the number of time periods $T=200$, we solve the MLE using the proposed estimator as an initial seed, and by selecting the initial seed using a Gaussian random vector, centred at the true parameters. Using the proposed estimator gives a 5 times reduction in the number of iterations required to convergence to the MLE over the case where the variance of the Gaussian is identity, and a 7 times reduction over the case where its variance is three times the identity. Further details are given in \ref{EC:Supnum}.\\
\noindent\textbf{3) Approximate MLE with regularisation:} We test the accuracy of the approximate MLE with ridge regularisation. Recall that here, one can derive a closed form approximation to the regularised MLE. The covariates evolve according to \eqref{eqn:Size-large}, and we set the covariance of the ridge regulariser, $\mathbf{Z}=\mathbf{I}$. In this case it can be seen that the solution of \eqref{eqn:MAP-FOC-1} and \eqref{eqn:MAP-FOC-2} converges in probability to $\frac{\beta}{2}$ (see \ref{sec:E.C.RegProof} for details). We hence compare both the proposed approximation and the regularised MLE to $\frac{\beta}{2}$ and observe that the both the regularised MLE and the proposed approximation give similar errors. Table~\ref{Reg-full} reports the results.\\
\noindent\textbf{4) Approximate MLE with corrupted data:}
To test robustness of the proposed estimator to missing data, we consider a toy example, where we simulate default data using the intensity model, where covariates follow an evolution given by~\eqref{eqn:Size-large}, for $N=10,000$ firms observed over $T=200$ months. Recall that in this case, the RMSEs in absence of missing data were 0.1192 and 0.1743 for the MLE and the proposed estimator, respectively.  To test the effect of missing data, we delete 1000 firm-period of entries (roughly 0.05\% of total available data) where a default is not observed. We then compute the MLE and the proposed estimator, and find that they give RMSEs of 0.1218 and 0.1743, respectively. Conversely, if the same number of entries are deleted just before the time of default of firms (i.e., at $(i,\tau_i)$), the corresponding RMSEs are 0.3752 and 0.4083 for the MLE and the proposed estimator, respectively.  This suggests that while estimating $\beta$, it is more important to acquire accurate data just before the time of default of a firm.

In order to capture small corruptions in the data collection, we generate defaults using the above set-up and add i.i.d. $N(0,0.25)$ noise to the covariates. Recall that $\Sigma$ denotes the covariance matrix of the uncorrupted data. We conduct simulation experiments covering the following cases:
\begin{enumerate}
    \item We first assume that all the data is corrupted by noise (that is, the $N(0,0.25)$ noise is added to all the data), but that $\Sigma$ is known. We then estimate parameters using both the proposed estimator and the MLE. In this case, we find that the RMSEs for the MLE and the proposed estimator are $0.2893$ and $ 0.1863$, respectively. 
    \item Next, we assume that the data just before the default of firms is accurate (that is, no noise is added to these data points), but the rest of the data is corrupted by additive $N(0,0.25)$ noise, and also that $\Sigma$ is known. Here, we find that the RMSEs for the MLE and the proposed estimator are $0.2657$ and $ 0.1723$, respectively. This suggests that in presence of noisy data, if $\Sigma$ is known, the proposed estimator outperforms the MLE.
    \item As mentioned in Remark~\ref{rem:Cov-Noise}, in practice, $\Sigma$ is often estimated empirically. To compare the proposed estimator to the MLE under this condition, we repeat step 1, but this time, estimate $\Sigma$ from the data.  We find that the RMSEs for the MLE and the proposed estimator are $0.2893$ and $0.2677$, respectively. 
        \item Lastly, we consider a case where 80\% of the data is corrupted by noise, while the remaining 20\% is accurate, and estimate parameters using the proposed estimator and the MLE. For the proposed estimator, the covariance matrix, $\Sigma$ is estimated using the accurate 20\% data. We find that the RMSEs for the MLE and the proposed estimator are $0.2435$ and $0.1830$, respectively. On the other hand, computing the MLE using only the accurate data gives an RMSE of $0.2687$.
    \item When the corruption in data is $N(0.1,0.25)$, that is, the noise has a non-zero mean, RMSEs of 0.4523 and 0.4362 are obtained in the MLE and the proposed estimator, respectively.
\end{enumerate}
Hence, our results suggests that in presence of noisy data, both estimators perform equally well even if the number of firms of data observed is of order tens of thousands. Further, so long as a fraction of data can be accurately obtained, the proposed estimator out-performs the MLE.\\
\noindent \textbf{5) Approximate MLE with non-Gaussian covariates:} In practice, covariates are often non-Gaussian. Observe that while the MLE is unaffected by the distribution of the covariates, the proposed estimator requires transformation to Gaussian marginals. Since transformed data is used as a covariate, computing the RMSE between the estimated and true parameters may no longer be accurate. Hence, in order to test the predictive power of the proposed estimator under such transformations, we use a modification of Test 2. We assume that the covariates follow the evolution $Y_{t,k} = W_{t,k}$,  and $X_{i,t,k}= Z_{i,t,k}$. Here, $W_{t,k}$ and $Z_{i,t,k}$ independent log normal. We generate defaults according to the intensity model using these covariates, and estimate the true parameters using the MLE.  Then, the covariates $(Y_{t,k},X_{i,t,k})$ are transformed to have an approximately Gaussian distribution, and the parameters are estimated using the proposed method. To perform this transformation, we match the quantile plot of the transformed covariate with that of a standard Gaussian. While comparing the proposed estimator to the MLE using Test 2, when computing default probabilities in Step 2 there for the proposed estimator, we transform the data to have a marginally Gaussian distribution. We observe that the proposed estimator gives 86.12\% cumulative overlap in the first decile, while the MLE gives 88.38\% overlap (we obtain similar conclusions even if the covariates were originally Pareto distributed). Thus, while ranking firms in terms of default probabilities, both estimators are almost equally accurate. More details are given in Table \ref{table:Non_Gaussian}.\\
\textbf{6) Effect of missing covariates:} To test for model mis-specification, we consider a simple, three factor model, with $d_1=2$ and $d_2=1$ in \eqref{eqn:Size-large}. Here, the calibrator assumes that only factors $(Y_{1,t}, X_{i,t,1})$ determine the default likelihood for firm $i$ at time $t+1$, and uses this data to estimate parameters $(\beta_1, \beta_2, \alpha)$.  The RMSE is estimated  under the two approaches for $(\beta_1, \beta_2,\alpha)$ in Table~\ref{table:logit_incorrect_small}. These experiments are carried out for default probabilities of 1\% and 3\%. As mentioned earlier, the estimation bias dominates in this case, and both the proposed estimator and the MLE have similar RMSE's in these experiments. 
\section{Performance on C-DATA} \label{sec:corp_data}
\subsection{Description of Data}
A data set provided by  \href{http://www.rmi.nus.edu.sg/duanjc/}{Risk Management Institute, National University of Singapore}, is used for estimation and testing of corporate defaults in the US.  The data consists of standard market, accounting and macroeconomic variables used for identification in other default studies. The specific variables considered are the following - trailing 1 year return on the S\&P 500 index, 3-month U.S treasury rates, Distance to default (DTD), cash to assets ratio (CASH/TA), that is the ratio sum of total cash and short term investments to assets,  the ratio of net income to total assets (NI/TA), the logarithm of the ratio of the firms equity value to the average equity value of the S\&P 500 firms (Size), market to book asset ratio (M/B), and the 1 year idiosyncratic firm volatility (Sigma). Of these DTD, CASH/TA, NI/TA and Size are assigned two separate covariates, level and trend, as in Duan et al. (2012). The level covariate is the average value over the past 12 months, while trend represents  the current value minus 12-month moving average. A summary statistics of the data are as follows: number of companies: 15,644 ,time duration: 306 months (1992-2017), number of effective company-month observations: 1,658,617, total number of defaults observed: 1123 (for year-wise split, refer to Table~\ref{1}). In Table~\ref{1}, the number of active firms in a year are calculated by averaging the number of active firms in each month of the year. The default percentage is then the ratio of the number of defaults recorded in a particular year to the number of firms active in that year. Since data is only available for 6 time periods in 2017, the default percentage reported is six monthly, for that year.
\subsection{Pre-Calibration Processing}
Covariates are transformed using log and power law transformations so that their empirical distribution is approximately Gaussian. First, the covariates are all made positive (so that logarithms exist) by subtracting the least value of a particular covariate available to us from the rest. A suitable transformation is then applied (see Table~\ref{transforms}) to convert the covariates to approximately Gaussian. This is done by matching the quantile plot of the covariate in question to that of a standard Gaussian. The transformed variables are then standardized by subtracting their empirical mean and dividing the result by variables standard deviation.

\subsection{Comparing different estimators}
In order to compare the proposed estimator with those obtained by using the MLE associated with default intensity model and the logit model, comparison tests (identical to the ones used by Duffie et al. 2007, Duan et al. 2012 and Duan et al. 2013) are conducted. 
The testing procedure is iterative:
\begin{enumerate}
	\item   For each $t$, starting at $t_0=150$, the data-set is separated into 1 to t (fitting data-set) and t to t+12 (testing data-set).
	
\item The parameters are estimated from data in the training set using the proposed estimator and the MLE. Firms in the testing data-set are ranked in the decreasing order of  their conditional default probabilities and bucketed into ten deciles. 
\item Defaults in each decile bucket were noted and a cumulative coverage of the defaults is reported next to the decile. As an example, if for $t=186$, 16 defaults occur in the next year (that is,  period 186 to 198), 10 from the firms listed in the top decile of risk, 5 from the next decile and 1 from the fifth decile, then the first decile is allocated number 10, second decile number 15, and the fifth decile number 16, the rest are allocated 16 for this iteration.
\item This process is continued for each year beginning at $t=150$ to $t=306$, and the numbers in each decile are added.
\item Finally, these numbers are averaged and  reported for each method, and the cumulative percentage coverage is reported with each decile.
\item The procedure is then repeated by adding a covariate for contagion, whose value is the number of defaults observed in the previous month divided by the total number of active firms in that month. As before, this is normalized and transformed so that the empirical distribution is approximately Gaussian.
\end{enumerate}

As is apparent, a better predictive method is likely to have higher percentages of defaults allocated to higher deciles. Note that this test is similar in spirit to Test 2 from Section 4.1. However, here, since the true model is latent to the calibrator, we test how the three estimations methods fare against each other while ranking defaulting firms (since our objective to start with was to predict defaulting firms accurately). Due to the small number of defaults in the data, we use 12 months of test data instead of the 1 month test data used in Section 4.1 (see also for e.g, Duan et al. (2012, 2013)).




\subsection{Results} \label{sec:result}
We denote the forward discrete intensity method from Duan et al. (2012, 2013) by DI, and logistic regression method by logit. Raw and transformed data, respectively refer to the cases when we use the data as it is and where we transform it to have an approximately Gaussian distribution. Tables~\ref{2} through \ref{2-Guass+cont} gives the results of estimation. The key points to note are -
\begin{enumerate}
    \item After applying simple transformations and converting the marginals of the data to approximately standard Gaussian (see Table \ref{transforms}), our method ranks about 3\% firms more than the logit method, and only 0.9\% firms less DI method in the first decile (see Table~\ref{2-Gauss} for the full details). Further, in the higher deciles, the proposed estimator gives a coverage almost identical to the DI method.
    \item Transforming the data to have approximately Gaussian marginals appears to significantly improve the accuracy of predictions using the proposed approximations . Interestingly, we observe that in the case of C-DATA, transforming the covariates to have Gaussian marginals also slightly improves the accuracy of predictions of the MLE, suggesting that such transformations may be useful in practice.
   \item Inclusion of contagion leads to a slight improvement in the predictive power of the estimator in all three methods (see Table \ref{2-Guass+cont}).
\end{enumerate}



\section{Conclusion}

We considered the popular default intensity based as well as logit models that have been used in the past to model corporate defaults.  We developed an approximate closed form estimator for parameters - we showed that each parameter maybe approximated by  a weighted average of the corresponding covariate observed just before default occurrences. We further  evaluated the performance of this estimator and developed an asymptotic expansion for the MSE. We showed both theoretically and numerically that the proposed estimator performs about as well as those obtained by using far more computationally intensive maximum likelihood methods
when the underlying default generating model is correctly specified.   Often in practice, regularisation is used to provide robustness to parameter over-fitting. We adapted our asymptotic regime to accommodate  regularisation, and showed that the when ridge regularisation is used, the MLE admits a closed form approximation. In general, we showed that the regularised MLE may be approximated by the solution to a simple convex program.  Realistically, the default generating mechanism is unknown, and any proposed model is misspecified. In this case we argued that the proposed estimators are as effective as those obtained using the maximum likelihood method. We further argued that in presence of missing or corrupted data, the proposed estimator outperforms the MLE.  Further, on C-DATA, we observed that the proposed estimator performs as well as MLE under logit and default intensity models. We also observed that using the proposed estimator as a starting point for MLE algorithms substantially speeds up their performance.  \\
\textbf{{Acknowledgements:}} The authors would like to thank the associate editor and two referees for their comments, which have helped significantly improve the quality of this paper. They would also like to thank Kay Giesecke, Michael Gordy and N.R.Prabhala for their useful suggestions to the draft, Aakash Kalyani for his help with the initial numerical experiments, and Jin-Chuan Duan and Risk Management Institute, National University of Singapore, for providing the data on US firms. This work was initiated while the second author was an adjunct at CAFRAL, Reserve Bank of India.

\bibliographystyle{nonumber}


%


\clearpage
\appendix


\section{MLE in presence of other exits, Multiple class extension}\label{EC:Extensions}
In Sections~\ref{EC:other1} and \ref{EC:other2}, we consider extensions of the proposed estimator to the case where firms may exit due to censoring. We outline the parameter estimation method in Section~\ref{EC:other3}. In Sections~\ref{EC:Multi} and \ref{EC:Multi2}, we extend the proposed estimator to the case where heterogeneity is allowed among firms.
{\subsection{Discrete default intensity model with censoring exits}}\label{EC:other1}
Recall that $({v_{i,t}:i \leq m, s_i \leq t \leq \tau_i})$  are sub-realisations of a stationary process $(V_{i,t}, i \leq m, t=0, \ldots, T-1)$, which is assumed to be multivariate Gaussian,  normalised to have stationary mean zero and variance one. In discrete default intensity model setting (see, e.g.,  Lando 2009, Duffie and Singleton 2012 for continuous default intensity models), we assume that given $v_{i,t}$ at time $t$, firm $i$ either defaults or has a censoring exit within time $[t,t+1)$ with the default intensity and the censoring exit intensity given, respectively, by $\psi(v_{i,t},b_1,a_1) \triangleq \exp(b_1^{\intercal}v_{i,t} -a_1)$ and $\phi(v_{i,t},b_2,a_2) \triangleq \exp({b}_2^{\intercal}v_{i,t} -a_2)$, where $b_1,b_2 \in \Re^{d_1+d_2}$ and  $a_1,a_2 \in \Re$ are the parameters to be estimated from data.
Then, the conditional probability that a firm $i$ that has survived till time $t$, survives till time $t+1$ is given by $\exp(- \xi(v_{i,t}) )$, where $ \xi(v_{i,t}) =\psi(v_{i,t}) +\phi(v_{i,t})$. Since both defaults and censoring exit cannot simultaneously occur at an interval $[t, t+1)$, as in Duan et al (2012), we deviate mildly from assuming that the default event and censoring exit event in an interval $[t, t+1)$ are independent (conditionally, given $v_{i,t}$) and instead assign respective probabilities $1- \exp(- \psi(v_{i,t}))$ and $\exp(- \psi(v_{i,t})) (1- \exp( -\phi(v_{i,t})))$ to these events. Alternate adjustments, e.g.,  where $1- \exp(- \psi(v_{i,t}))$  is instead set to  $\exp(- \phi(v_{i,t})) (1- \exp(- \psi(v_{i,t}))$ do not affect our proposed estimator below. Let $\tau_i$ be the exit time of firm $i$. Observe that the log likelihood function  of default observations becomes
 \begin{equation}\label{eqn:quasilikelihood}
    \log{ \mathcal{L}(b_1,b_2,a_1,a_2)} = \sum_{i=1}^{m}\sum_{t=s_{i}}^{\tau_i}\mathcal{L}_{i,t}(b_1,b_2,a_1,a_2)
 \end{equation}
  where again, the contribution to it by firm $i$ surviving at time $t <T$, is denoted by
 $\mathcal{L}_{i,t}(b_1,b_2,a_1,a_2)$, and equals
\begin{align*}
&  \mathbb{I}(\tau_{i} \geq t+1)\exp(- \xi(v_{i,t},b_1,b_2,a_1,a_2)) +
 \mathbb{I}(\tau_{i}=\tau_{D_i} =t)(1-\exp(-\psi(v_{i,t}))) + \mathbb{I}(\tau_{i} = t<\tau_{D_i}) \times  \\
 & \exp(-(\psi(v_{i,t},b_1,a_1))(1-\exp(-\phi(v_{i,t},b_2,a_2))).
  \end{align*}
  Then, $\log \mathcal{L}_{i,t}(b_1,b_2,a_1,a_2)$ simplifies to
  \begin{align*}
&       -(1-d_{i,t+1}-m_{i,t+1}){\xi(v_{i,t},b_1,b_2,a_1,a_2)}
+{d_{i,t+1}}\log\left(1-\exp(-\psi(v_{i,t}b_1,a_1))\right)  \\
&
 +{m_{i,t+1}}\log\left(\exp(-\psi(v_{i,t})) - \exp(-\xi(v_{i,t},b_1,b_2,a_1,a_2))\right).
\end{align*}

 Component-wise setting the derivatives with respect to $b_1$ and $a_1$, and then with respect to  $b_2$ and $a_2$, to zero, the solutions to the resulting equations, $\hat{\beta}_{M}$, $\hat{\vartheta}_M$, $\hat{\alpha}_{1,M}$ and $\hat{\alpha}_{2,M}$ satisfy:
\begin{align}
 \sum_{i\leq m,t=s_{i}}^{\tau_{i}} v_{i,t}d_{i,t+1}\left(\frac{\mathrm{e}^{(\hat{\beta}_{M}^{\intercal}v_{i,t} -\hat{\alpha}_{1,M})} \exp(-\mathrm{e}^{(\hat{\beta}^{\intercal}_{M}v_{i,t} -\hat{\alpha}_{1,M})})}{1- \exp(-\mathrm{e}^{(\hat{\beta}^{\intercal}_{M}v_{i,t} -\hat{\alpha}_{1,M})})}+\mathrm{e}^{(\hat{\beta}^{\intercal}_{M}v_{i,t}-\hat{\alpha}_{1,M})}\right)  &= \sum_{i\leq m,t=s_{i}}^{\tau_{i}} v_{i,t}\mathrm{e}^{(\hat{\beta}_{M}^{\intercal}v_{i,t}-\hat{\alpha}_{1,M})}\label{eqn:duanapprox1},\\
 \sum_{i\leq m,t=s_{i}}^{\tau_{i}} d_{i,t+1}\left(\frac{\mathrm{e}^{(\hat{\beta}_{M}^{\intercal}v_{i,t} -\hat{\alpha}_{1,M})} \exp(-\mathrm{e}^{(\hat{\beta}_{M}^{\intercal}v_{i,t} -\hat{\alpha}_{1,M})})}{1- \exp(-\mathrm{e}^{(\hat{\beta}_{M}^{\intercal}v_{i,t} -\hat{\alpha}_{1,M})})}+\mathrm{e}^{(\hat{\beta}_{M}^{\intercal}v_{i,t}-\hat{\alpha}_{1,M})}\right)  &= \sum_{i\leq m,t=s_{i}}^{\tau_{i}} \mathrm{e}^{(\hat{\beta}_{M}^{\intercal}v_{i,t}-\hat{\alpha}_{1,M})},
 \label{eqn:duanapprox2}
 \end{align}

 \begin{align}
 & \sum_{i\leq m,t=s_{i}}^{\tau_{i}} v_{i,t}m_{i,t+1}\left(\frac{\mathrm{e}^{(\hat{\vartheta}_{M}^{\intercal}v_{i,t} -\hat{\alpha}_{2,M})} \exp(-\mathrm{e}^{(\hat{\vartheta}_{M}^{\intercal}v_{i,t} -\hat{\alpha}_{2,M})})}{1- \exp(-\mathrm{e}^{(\hat{\vartheta}_{M}^{\intercal}v_{i,t} -\hat{\alpha}_{2,M})})}+\mathrm{e}^{(\hat{\vartheta}_{M}^{\intercal}v_{i,t}-\hat{\alpha}_{2,M})}\right)+ v_{i,t}d_{i,t+1}\mathrm{e}^{(\hat{\vartheta}_{M}^{\intercal}v_{i,t}-\hat{\alpha}_{2,M})}\nonumber  \\
 &= \sum_{i\leq m,t=s_{i}}^{\tau_{i}} v_{i,t}\mathrm{e}^{(\hat{\vartheta}_{M}^{\intercal}v_{i,t}-\hat{\alpha}_{2,M})}  \label{eqn:duanapprox3},\\
 & \sum_{i\leq m,t=s_{i}}^{\tau_{i}} m_{i,t+1}\left(\frac{\mathrm{e}^{(\hat{\vartheta}_{M}^{\intercal}v_{i,t} -\hat{\alpha}_{2,M})} \exp(-\mathrm{e}^{(\hat{\vartheta}^{\intercal}_{M}v_{i,t} -\hat{\alpha}_{2,M})})}{1- \exp(-\mathrm{e}^{(\hat{\vartheta}^{\intercal}_{M}v_{i,t} -\hat{\alpha}_{2,M})})}+\mathrm{e}^{(\hat{\vartheta}_{M}^{\intercal}v_{i,t}-\hat{\alpha}_{2,M})}\right) +d_{i,t+1}\mathrm{e}^{(\hat{\vartheta}_{M}^{\intercal}v_{i,t}-\hat{\alpha}_{2,M})}\nonumber  \\
  &= \sum_{i\leq m,t=s_{i}}^{\tau_{i}} \mathrm{e}^{(\hat{\vartheta}_{M}^{\intercal}v_{i,t}-\hat{\alpha}_{2,M})}.
 \label{eqn:duanapprox4}
 \end{align}

As noted in  Duffie et al. (2007) and Duan et al. (2012), the problem factors into two independent sub-problems, the first dependent on $(\hat{\beta}, \hat{\alpha}_{1,M})$ and the second on $(\hat{\vartheta},\hat{\alpha}_{2,M})$.\\
\subsection{Logit model with censoring exits}\label{EC:other2}
We now consider the case where  the default and censoring  exit probabilities have a logit structure. Let $p(v_{i,t})$ denote the conditional probability that  firm  $i$,
surviving at time $t$, defaults  between  time $t$ and $t+1$. Similarly, let $q(v_{i,t}) $ denote
the analogous conditional probability of censoring  exit, where again these are assumed to depend upon
parameters $\beta,{\vartheta} \in \Re^{d}$ and  $\alpha_1,\alpha_2 \in \Re$ and only on $v_{i,t}$ given the history till time $t$. Specifically,
\begin{equation}\label{eqn:M-def}
p(v_{i,t}) = \frac{\exp(\beta^{\intercal} v_{i,t}-\alpha_1) }
{1+\exp(\beta^{\intercal} v_{i,t}-\alpha_1) + \exp({\vartheta}^{\intercal} v_{i,t}-\alpha_2)}
\end{equation}
and
\begin{equation}\label{eqn:M-other}
    q(v_{i,t}) = \frac{\exp({\vartheta}^{\intercal} v_{i,t}-\alpha_2) }
{1+\exp(\beta^{\intercal} v_{i,t}-\alpha_1) +  \exp({\vartheta}^{\intercal} v_{i,t}-\alpha_2)}.
\end{equation}
The likelihood, again call it ${\cal L}$, of seeing the exit data $(d_{i,t},m_{i,t}: i \leq m, s_{i} < t \leq \tau_{i})$, is given by
\begin{equation} \label{eqn:LR}
{\cal L} = \prod_{i \leq m} \prod_{t=s_{i}}^{ \tau_{i}-1}
\left ( p(v_{i,t})^{d_{i,t+1}}(q(v_{i,t}))^{m_{i,t+1}}(1- p(v_{i,t}) - q(v_{i,t}))^{1- m_{i,t+1} - d_{i,t+1}} \right ).
\end{equation}
Again ML estimation  corresponds to finding parameters that maximize ${\cal L}$, or equivalently, $\log {\cal L}$.  As is well known, in this case  the function  $\log {\cal L}$  is a concave function of underlying parameters $(\beta, \alpha_1, {\theta},\alpha_2)$.
Component-wise, setting the partial derivatives with these parameters to zero, the solutions, $\hat{\beta}_M$, $\hat{\vartheta}_M$, $\hat{\alpha}_{1,M}$ and $\hat{\alpha}_{2,M}$, satisfy:
\begin{align}
\sum_{i\leq m, t=s_{i}}^{ \tau_{i}-1}v_{i,t} d_{i,t+1}&=
\sum_{i \leq m, t=s_{i}}^{ \tau_{i}-1}v_{i,t}
\frac{\exp(\hat{\beta}^{\intercal}_{M} v_{i,t}-\hat{\alpha}_{1,M}) }
{1+\exp(\hat{\beta}^{\intercal}_{M} v_{i,t}-\hat{\alpha}_{1,M}) +  \exp({\vartheta}_{M}^{\intercal} v_{i,t}-\hat{\alpha}_{2,M})}\label{logit:001i},\\
\sum_{i \leq m, t=s_{i}}^{ \tau_{i}-1}d_{i,t+1}&=
\sum_{i \leq m, t=s_{i}}^{ \tau_{i}-1}
\frac{\exp(\hat{\beta}^{\intercal}_{M} v_{i,t}-\hat{\alpha}_{1,M}) }
{{1+\exp(\hat{\beta}^{\intercal}_{M} v_{i,t}-\hat{\alpha}_{1,M}) +  \exp(\hat{\vartheta}^{\intercal}_{M} v_{i,t}-\hat{\alpha}_{2,M})}}\label{logit:003i},\\
\sum_{i\leq m, t=s_{i}}^{ \tau_{i}-1}v_{i,t} m_{i,t+1}&=
\sum_{i \leq m, t=s_{i}}^{ \tau_{i}-1}v_{i,t}
\frac{\exp(\hat{\vartheta}^{\intercal}_{M} v_{i,t}-\hat{\alpha}_{2,M}) }
{1+\exp(\hat{\beta}^{\intercal}_{M} v_{i,t}-\hat{\alpha}_{1,M}) +  \exp(\hat{\vartheta}^{\intercal}_{M} v_{i,t}-\hat{\alpha}_{2,M})}\label{logit:002i},\\
\sum_{i \leq m, t=s_{i}}^{ \tau_{i}-1}m_{i,t+1} &=
\sum_{i \leq m, t=s_{i}}^{ \tau_{i}-1}
\frac{\exp(\hat{\vartheta}^{\intercal}_{M} v_{i,t}-\hat{\alpha}_{2,M}) }
{{1+\exp(\hat{\beta}^{\intercal}_{M} v_{i,t}-\alpha_{1,M}) +  \exp(\hat{\vartheta}^{\intercal}_{M} v_{i,t}-\alpha_{2,M})}}, \label{logit:004i}
\end{align}
Let $\tau= \sum_{i \leq m} (\tau_{i}-s_{i})$ denote the firm periods of
data available. \\
\subsection{Parameter estimation}\label{EC:other3}
When the defaults and other exits occur with small probabilities, (so that $\exp(\hat{\beta}_M^{\intercal} v_{i,t}-\hat{\alpha}_{1,M})$
and $\exp(\hat{\vartheta}_M^{\intercal} v_{i,t}-\hat{\alpha}_{2,M})$ are typically small)
the RHS of (\ref{eqn:duanapprox1}) and
(\ref{logit:001i}) divided by $\tau$ may be approximated  by
\[
E(V_{i,t} \exp(\beta^{\intercal} V_{i,t}-\alpha_1))=\Sigma \beta   \exp\left(\frac{1}{2} \beta^{\intercal} \Sigma \beta - \alpha_1\right),
\]
where  $\Sigma$ denotes the correlation matrix of $(V_{i,t})$ and is assumed to be independent of $i$. Similarly,  the RHS of (\ref{eqn:duanapprox2}) and (\ref{logit:002i}) divided by $\tau$ may be approximated  by $\exp(\frac{1}{2}\beta^{\intercal}\Sigma\beta-\alpha_1)$. Further, observe that when $\exp(\hat{\beta}_M^{\intercal} v_{i,t}-\hat{\alpha}_{1,M})$ is small, the LHS of
(\ref{eqn:duanapprox1}) and (\ref{eqn:duanapprox2})
may be approximated by $\sum_{i\leq m, t=s_{i}}^{ \tau_{i}-1}v_{i,t} d_{i,t+1}$
and $\sum_{i\leq m, t=s_{i}}^{ \tau_{i}-1} d_{i,t+1}$, respectively.
Assuming that $\Sigma$ is known and invertible, the above discussion suggests that
$\beta$ may be approximated by
\begin{equation}  \label{eqn:interesting}
 \Sigma^{-1} \frac{\sum_{i \leq m, t=s_{i}}^{ \tau_i}v_{i,t} d_{i,t+1}}
 {\sum_{i \leq m, t=s_{i}}^{ \tau_i}d_{i,t+1}}.
\end{equation}
Similarly,    ${\vartheta}$ may be approximated by
\begin{equation}\label{eqn:thetaapprox}
\Sigma^{-1} \frac{\sum_{i \leq m, t=s_{i}}^{ \tau_i}v_{i,t} m_{i,t+1}}
 {\sum_{i \leq m, t=s_{i}}^{ \tau_i}m_{i,t+1}}.
\end{equation}
Then, the RHS above, call them $\hat{\beta}$ and $\hat{\vartheta}$, respectively, are our proposed estimators for $\beta$ and ${\vartheta}$.  Further,
\begin{align}
    \hat{\alpha}_1 & = \log\left({\frac{\sum_{i\leq m, t=s_{i}}^{\tau_{i}}\exp(\hat{\beta}^{\intercal}v_{i,t})}{\sum_{i\leq m, t=s_{i}}^{\tau_{i}} d_{i,t+1} }}\right) \textrm{, and}\\
    \hat{\alpha}_2 & = \log\left({\frac{\sum_{i\leq m, t=s_{i}}^{\tau_{i}}\exp(\hat{\vartheta}^{\intercal}v_{i,t})}{\sum_{i\leq m, t=s_{i}}^{\tau_{i}} m_{i,t+1} }}\right),
\end{align}
are our estimators for $\alpha_1$ and  $\alpha_2$, respectively. Suppose we model the true intercepts as $\alpha_1(\gamma) = \log\frac{1}{\gamma}-\log c_1$ and $\alpha_2(\gamma) = \log\frac{1}{\gamma} - \log c_{2}$ (that is defaults and exits both become rare), then in the limit, we obtain the following accuracy bounds on errors in parameter estimation:
\begin{proposition}\label{prop:Multitypeerror}
Let Assumptions~\ref{Assumption:Geometric} and \ref{assumption:TSremainder} hold. Then, 
\begin{align*}
    \|\beta - \hat{\beta}(\gamma)\|_{2}^{2} &=K_{1,1}\gamma^{\delta+\zeta-1} + K_{1,2}\gamma^{\zeta} \\
    \|\vartheta - \hat{\vartheta}(\gamma)\|_{2}^{2} &=K_{2,1}\gamma^{\delta+\zeta-1} + K_{2,2}\gamma^{\zeta} \\
    \|\alpha_1(\gamma) - \hat{\alpha}_1(\gamma)\|_{2}^{2} &=O_{\mathcal{P}}(\gamma^{\delta+\zeta-1}) +O_{\mathcal{P}}(\gamma^{\zeta}) \\
    \|\alpha_2(\gamma) - \hat{\alpha}_2(\gamma)\|_{2}^{2} &=O_{\mathcal{P}}(\gamma^{\delta+\zeta-1}) +O_{\mathcal{P}}(\gamma^{\zeta}) 
\end{align*}
for some computable positive constants, $K_{i,j}$.
\end{proposition}
The proof of Proposition~\ref{prop:Multitypeerror} is essentially identical to that of Theorem~\ref{theorem:1}, and thus is omitted.


\subsection{Parameter estimation with multiple classes of firms}\label{EC:Multi}
We consider $m$ firms, divided into $K\geq 1$ classes, such that $m_k$ fraction of total firms are in class $k$, $\sum_k m_k=1$. Firms in each class are assumed to have a similar risk profile and  dependence structure. Recall that  $Y_t \in \Re^{d_1}$ denotes the vector of common market information at time $t$, and  for firm $i \in {\cal C}_k$, let $X_{i,k,t} \in \Re^{d_2}$ denote a vector of company specific information at time $t$. Again, let $(Y_t, X_{i,k,t}: i \leq m_k, k \leq K)_{ t \leq T}$ denote a stationary Gaussian process. We let $(Y, X_{i,k}: i \leq m_k, k \leq K)$ denote  random variables with the associated stationary distribution. Let $\Sigma_{YY} \in \Re^{d_1 \times d_1}$ denote the correlation matrix corresponding to $Y$,  $\Sigma_{YX_k} \in \Re^{d_1 \times d_2}$ denote the correlation matrix between $Y$ and $X_{i,k}$ which we assume to be same for all $i \in {\cal C}_k$ (also denoted by $\Sigma_{X_k Y}^T$). Similarly, let $\Sigma_{X_k X_k} \in \Re^{d_2 \times d_2}$ denote the correlation matrix between the components of $X_{i,k}$ again assumed to be same for all $i$. Further, for each $k$ let
\begin{equation} \label{eqn:matrix}
\Sigma_k \triangleq
\begin{pmatrix}
\Sigma_{YY}  &  \Sigma_{YX_k} \\
\Sigma_{X_k Y} & \Sigma_{X_k X_k}\\
\end{pmatrix}.
\end{equation}

Consider a case where the only form of exits are defaults. In this framework, $p_k(y_t, x_{i,k,t})$ denotes the conditional probability that a firm  $i \in {\cal C}_k$,
surviving at time $t$, defaults  at time $t+1$, and is assumed to be a function
of $(y_t, x_{i,k,t})$ given $(y_s, x_{j,k,s}: s \leq t, j \in {\cal C}_k, k \leq K)$.
Then, the likelihood, call it ${\cal L}$ of seeing the 
default data $(d_{i,k,t}: s_{i,k} < t \leq \tau_{i,k})$ for each $i \in {\cal C}_k$ for $k \leq K$,
is given by
\[
{\cal L}
= \prod_{i \in {\cal C}_k, k \leq K} \prod_{t=s_{i,k}}^{ \tau_{i,k}-1}
\left ( p_k(y_t, x_{i,k,t})^{d_{i,k,t+1}}(1- p_k(y_t, x_{i,k,t}))^{1-d_{i,k,t+1}} \right )
\]
Suppose that the conditional default probabilities have a logit form, $p_k(y_t, x_{i,k,t}) = \frac{\exp(\theta_k^{\intercal} y_t + \eta_k^{\intercal}x_{i,k,t}-\alpha_k) }
{1+\exp(\theta_k^{\intercal} y_t + \eta_k^{\intercal}x_{i,k,t}-\alpha_k)}$ (a similar discussion holds for the default intensity case as well), then the first order conditions for a given $k$, are
\begin{equation}  \label{logit:001A}
\sum_{i\in {\cal C}_k, t=s_{i,k}}^{ \tau_{i,k}}y_t d_{i,k,t+1}
= 
\sum_{i \in {\cal C}_k, t=s_{i,k}}^{ \tau_{i,k}-1}y_t 
\frac{\exp(\theta_k^{\intercal} y_t + \eta_k^{\intercal}x_{i,k,t}-\alpha_k) }
{1+\exp(\theta_k^{\intercal} y_t + \eta_k^{\intercal}x_{i,k,t}-\alpha_k)}
\end{equation}
\begin{equation}  \label{logit:002A}
\sum_{i\in {\cal C}_k, t=s_{i,k}}^{ \tau_{i,k}}x_{i,k,t} d_{i,k,t+1}
= 
\sum_{i \in {\cal C}_k, t=s_{i,k}}^{ \tau_{i,k}} x_{i,k,t} 
\frac{\exp(\theta_k^{\intercal} y_t + \eta_k^{\intercal}x_{i,k,t}-\alpha_k) }
{1+\exp(\theta_k^{\intercal} y_t + \eta_k^{\intercal}x_{i,k,t}-\alpha_k)},
\end{equation}
and
\begin{equation}  \label{logit:003A}
\sum_{i\in {\cal C}_k, t=s_{i,k}}^{ \tau_{i,k}}d_{i,k,t+1}
= 
\sum_{i \in {\cal C}_k, t=s_{i,k}}^{ \tau_{i,k}} 
\frac{\exp(\theta_k^{\intercal} y_t + \eta_k^{\intercal}x_{i,k,t}-\alpha_k) }
{1+\exp(\theta_k^{\intercal} y_t + \eta_k^{\intercal}x_{i,k,t}-\alpha_k)},
\end{equation}
Note that if all the ${\beta_{k}}=(\theta_k,\eta_k)$ are different, then the problem decouples, and we may apply the previously developed estimators class-wise. \label{ss:multi1}
\subsection{Parameter estimation when all $\theta_k$ are equal}\label{EC:Multi2}
In practice, class specific data may be limited, and hence, it is reasonable to work in a framework where $\theta$ is common across classes. A similar procedure can be developed when some parameters are assumed to be common  across some of the classes. We outline an estimation methodology for this case. A few definitions are needed first. Let
\begin{align}
    e & = \frac{E(\sum_{i,t,k}Y_{t}D_{i,t+1,k})}{E(\sum_{i,t,k}D_{i,t+1,k})}\label{eqn:errorglobbal}\\
    g_{k}& = \frac{E(\sum_{i,t} X_{i,t,k}D_{i,t+1,k})}{E(\sum_{i,t}D_{i,t+1,k})}.\label{eqn:errorclass}
\end{align}
Observe that from (\ref{logit:001A}), (\ref{logit:002A}) and (\ref{logit:003A}),
\begin{equation}\label{eqn:approx1}
    e \approx \frac{\sum_{i=1}^{K}m_{k} E  Y\exp(\theta^{\intercal}{Y}+\eta_{k}^{\intercal}X_{k} -\alpha_k)}{\sum_{i=1}^{K} m_{k}E\exp(\theta^{\intercal}{Y}+\eta_{k}^{\intercal}X_{k} -\alpha_k) }.
\end{equation}
The RHS above equals
\[
\frac{\sum_{i=1}^{K}m_k  (\Sigma_{YX_k}\eta_k + \Sigma_{YY}\theta)\exp(\frac{1}{2}\beta_k^{\intercal} \Sigma_k \beta_k-\alpha_k)} {\sum_{i=1}^{K}m_{k}\exp(\frac{1}{2}\beta_k^{\intercal} \Sigma_k \beta_k -\alpha_k)}.
\]
We define
\begin{equation}\label{eqn:classdef}
    f_{k} = \frac{m_k\exp(\frac{1}{2}\beta_k^{\intercal} \Sigma_k \beta_k-\alpha_k)} {\sum_{i=1}^{K}m_{k}\exp(\frac{1}{2}\beta_k^{\intercal} \Sigma_k \beta_k-\alpha_k)}.
\end{equation}
Then, the LHS of (\ref{eqn:approx1}) is approximately
\[
\sum_{k=1}^{K} f_k (\Sigma_{YX_k}\eta_k + \Sigma_{YY}\theta).
\]
Now, consider (\ref{eqn:errorclass}). It can be similarly shown that
\[
\eta_{k} \approx \Sigma^{-1}_{X_kX_k} (g_k - \Sigma_{XY_k}\theta).
\]
Plugging this into (\ref{eqn:approx1}), it is easily seen that
\begin{equation}
    \theta \approx \left(\Sigma_{YY} - \sum_{k=1}^{K} f_k \Sigma_{XY_K}\Sigma^{-1}_{X_kX_k}\Sigma_{YX_k}\right)^{-1} \left(e- \sum_{k} f_k\Sigma_{YX_k}\Sigma_{X_kX_k} g_{k}\right),
\end{equation}
where the inverse is assumed to exist. This motivates the proposed estimator. Define 
\begin{align}
    \Hat{e} &= \frac{\sum_{i,k,t} Y_tD_{i,t+1,k}}{\sum_{i,j,k} D_{i,k,t+1}},\nonumber\\
    \Hat{g}_k &= \frac{\sum_{i,t} X_{i,k,t}D_{i,t+1,k}}{\sum_{i,j,k}D_{i,k,t+1}},\nonumber\\ 
    \hat{f}_{k} &= \frac{\sum_{i,t} D_{i,k,t+1}}{\sum_{i,k,t}D_{i,k,t+1}}.\label{eqn:empall}
\end{align}
Using the notation of (\ref{eqn:empall}), the proposed estimator for $\theta$ becomes
\begin{equation}\label{eqn:hatbeta}
\Hat{\theta}(\gamma)= \left(\Sigma_{YY} - \sum_{k=1}^{K} \hat{f}_k \Sigma_{XY_K}\Sigma^{-1}_{X_kX_k}\Sigma_{YX_k}\right)^{-1} \left(\hat{e}- \sum_{k} \hat{f}_k\Sigma_{YX_k}\Sigma_{X_kX_k} \hat{g}_{k}\right).
\end{equation}
Then, we calculate $\hat{\eta}_k$ as
\begin{equation}\label{eqn:hateta}
\hat{\eta}_{k} = \Sigma^{-1}_{X_kX_k} (\hat{g}_k - \Sigma_{XY_k}\hat{\theta}).
\end{equation}
To analyse the performance of the above estimator, we again embed the system into an asymptotic regime, indexed by $\gamma\to 0$, such that for a class $k$, $\alpha_{k}(\gamma) = \log\frac{c_{k}}{\gamma}$, for constants $c_{k}>0$, and append $\gamma$ to the notation used above, to indicate that $\Hat{\theta}$,$\Hat{\eta}_{k}$,$\Hat{g}_k$, $\Hat{e}$ and $\Hat{f}_{k}$ are all functions of $\gamma$.
\begin{proposition}\label{prop:multiclass}
Let $\hat{\theta}(\gamma)$ and $\hat{\eta}_{k}(\gamma)$ be defined by (\ref{eqn:hatbeta}) and (\ref{eqn:hateta}). Then, under Assumption~\ref{Assumption:Geometric}, 
\begin{align}
    \|\theta-\hat{\theta}(\gamma)\|^{2}_{2} &= O_{\mathcal{P}}(\gamma^{\delta+\zeta-1}) + O_{\mathcal{P}}(\gamma^{\zeta}), \nonumber\\
    \|\eta-\hat{\eta}_{k}(\gamma)\|^{2}_{2} &= O_{\mathcal{P}}(\gamma^{\delta+\zeta-1}) + O_{\mathcal{P}}(\gamma^{\zeta}).\label{eqn:multclassthm1}
\end{align}
\end{proposition}
Having estimated $\theta$ and $\eta_k$, we calculate the approximation to $\alpha_k$, that is 
\begin{equation}\label{eqn:alphamulti}
    \Hat{\alpha}_{k}(\gamma) = \log \left( \frac{\sum_{i=1}^{m_{k}} \sum_{t=1}^{\tau_{i,k}  } \exp(\Hat{\theta}^{\intercal}y_{t} + \Hat{\eta}_{k}^{\intercal}x_{i,t,k})}{\sum_{i=1}^{m_{k}} \sum_{t=1}^{\tau_{i,k}} d_{i,t+1,k}}\right).
\end{equation}
\section{Causal ARMA Processes, Discrete Intensity models}\label{sec:ARMA-Int}
\subsection{ARMA Processes}
In the following, without loss of generality, assume $i=j$.  As a concrete example, consider the ARMA(p,q) model, which is the stationary solution of the system of linear difference equations
\begin{equation}\label{eqn:ARMA}
    V_{i,t} - \Phi_{1} V_{i,t-1} - \cdots - \Phi_{p} V_{i,t-p} = Z_{i,t} +\Theta_{1} Z_{i,t-1} + \cdots +\Theta_{q} Z_{i,t-q},  
\end{equation}
for some $\Phi_{i},\Theta_{j} \in \Re^{d\times d}$ and $Z_{i,t} = (\zeta_{i,t},\phi_{t})^{\intercal}$ an  $N(0,\mathbf{I})$ independent identically distributed (iid) noise. Define the matrix valued polynomial $\mathrm{P}(z) = \mathbf{I}- \Phi_1 z- \cdots -\Phi_p z^p$, where  $\mathbf{I} \in \Re^{d\times d}$ is an identity matrix. To ensure the stationary solution is causal, we impose the condition that $\det(\mathrm{P}(z)) \neq0$ for all $ \{z\in \mathbb{C}:|z|\leq1$\}. Note that this condition is a general version of the causality condition for a stationary one dimensional AR(1) process. Define $\Psi_{j} = \sum_{i=1}^{j} \Phi_{i}\Psi_{j-i}+\Theta_{j}$. Then, the unique stationary solution of (\ref{eqn:ARMA}) has a representation $V_{i,t} = \sum_{j=0}^{\infty} \Psi_{j}Z_{t-j} $ (see Brockwell and Davis 2013, Theorem 11.3.1). The covariance matrix of $V_{i,t}$ becomes
\[
\Gamma(h) = E(V_{i,t+h}V_{i,t}^{\intercal}) = \sum_{k=0}^{\infty} \Psi_{k+h}\Psi_{k}^{\intercal}.
\]
 Further, it can be shown that for causal ARMA(p,q) processes, there exist constants $B$ and $\rho\in(0,1)$ which depend on $\Phi_{i}$ and $\Theta_{j}$, such that components of $\Gamma(h)$, $\gamma_{i,j}(h)$ satisfy $|\gamma_{i,j}(h)| < B \rho^{|h|}$  for all $i,j$ and $h$ (see Brockwell and Davis 2013). Then, $\|\Gamma(h)\| < B d\rho^{|h|}$. Thus causal ARMA(p,q) processes satisfy Assumption~\ref{Assumption:Geometric}.
 \subsection{Discrete Intensity based models}
In this section we provide a detailed discussion on the discrete intensity type models used for the conditional default probabilities. Consider a system where there are $m$ firms observed for $T$ time periods. We consider the doubly stochastic framework of Duffie et al. (2007). For simplicity, assume that the only source of exits are defaults. Here, default of a bank is modelled as the first arrival time of a doubly  stochastic  Poisson process, which we will refer to as the default process. More specifically, we assume that there exists a probability space $(\Omega,\mathcal{F},\mathbb{P})$ which supports a $d_1$ dimensional Brownian motion, ${Y}_t$ and $d_2$ dimensional Brownian motions $\{{X}_{i,t}\}_{i=1}^{m}$. The covariate $V_{i,t}\in \Re^d$ follow the stochastic differential equation,
\begin{equation}\label{eqn:ARMA-Disc}
    \mathrm{d}V_{i,t}=\mu(V_{i,t})\mathrm{d}t + b(V_{i,t})\mathrm{d}Z_{i,t} \textrm{ where $Z_{i,t}=(Y_t,X_{i,t})$}.
\end{equation}
Here $\mu(\cdot)$ and $b(\cdot)$ are an $\Re^d$ valued map, and a $\Re^{d\times d}$ matrix respectively, and satisfy standard conditions guaranteeing the existence of a unique strong solution to the SDE \eqref{eqn:ARMA-Disc}. We assume that the intensity of the default process is given by 
\begin{equation}\label{eqn:Stoch-Int}
    \lambda_{i,t} = \exp(\beta^\intercal V_{i,t}-\alpha), 
\end{equation}
Recall that $\mathcal{F}_t=\sigma(V_{i,s}:s\leq t)$ denotes the filtration of covariate information available till time $t$ Now, the conditional  probability of survival till time $(t+1)$ given covariate information up to time $t$ is given by 
\begin{equation}\label{eqn:ARMA-disc-2}
    \bar p(t,t+1) = \mathbf{E}_{t} \left( \int_{t}^{t+1} \exp\left(-\mathrm{e}^{\beta^\intercal V_{i,s}-\alpha}\right)ds\right),
\end{equation}
where $\mathbf{E}_t$ denotes the conditional expectation given $\mathcal{F}_t$.  We now proceed with the discretisation. We assume that time total time period of observation is divided into $T$ unit time intervals, and that the value of the covariate remains constant between time intervals. Then \eqref{eqn:ARMA-Disc} becomes
\begin{equation*}
    V_{i,t+1} = \mu(V_{i,t}) + \sigma(V_{i,t})Z_{i,t+1}.
\end{equation*}
Now, the conditional probability of default between time $t$ and $t+1$ becomes
\begin{equation}\label{eqn:Pdef-CDI}
 p(V_{i,t}) = 1-\exp(-\mathrm{e}^{\beta^\intercal V_{i,t}-\alpha}),
\end{equation}
where the integral in \eqref{eqn:ARMA-disc-2} simplifies as $\exp(-\mathrm{e}^{\beta^\intercal V_{i,t}-\alpha})$, since we assume covariates are held constant throughout an interval. This gives the discrete intensity model Section \ref{sec:MLE}. The distribution of covariates now depends on the choice of $\mu(\cdot)$ and $\sigma(\cdot)$. For example, setting $\mu(t,V_{i,t})=\mathbf{A}V_{i,t}$, where $\mathbf{A}\in\Re^{d\times d}$ is invertible and $\sigma(V_{i,t})=\mathbf{I}$ gives $V_{i,t+1}=\mathbf{A}V_{i,t}+Z_{i,t+1}$, which recovers the AR(1) model for covariate evolution used in Duffie et.al (2007). This model can incorporate exits due to censoring by assuming that censoring exits are the first arrival time of another doubly stochastic Poisson process. We refer the reader to Duffie et al. (2007) for more details.\\
\clearpage

\Large{\textbf{Proofs of main results}}\\

\noindent\normalsize\noindent \textbf{Outline:} Sections E.C.3 to E.C.8 contain the proofs of the main results of this paper. In E.C.3, we prove Theorem~\ref{theorem:1} and Proposition \ref{theorem:2}. E.C.4 contains the preliminaries and the proof of Theorem~\ref{thm:CLT} and Proposition~\ref{prop:varianceratio}. \ref{EC:Misp} contains the proof of Theorem~\ref{thm:Mis-spec} and Lemma~\ref{lemma:Bias-MLE}, while the proof outline of Proposition~\ref{lemma:MAP-Reg} is given in E.C.7. We collect the proofs of all intermediate lemmas in E.C.8. Finally, the discussion on the high dimensional approximate MLE is given E.C.6.\\

\section{Proofs of Theorems~\ref{theorem:1} and \ref{thm:CLT}}\label{EC:Proofs}
We first prove Theorem~\ref{theorem:1}.
We first need the following- 
\begin{lemma}\label{lemma:expVCV}
{
Suppose that $\{Y_{t}\}_{t\geq0}$ is Gaussian stationary process, in $\Re$ with $Y_{0} \sim N\left(0, \sigma^{2}\right)$. Let $j_{1} < \ldots <j_{k}$, and $\mathcal{C}_{2}^{k}$ be the set of all their pair-wise combinations. Then,
}
\begin{equation}\label{eqn:lemmaexp}
E\left(\exp\left(\beta\sum_{r=1}^{k}{Y_{j_{r}} }\right)\right) = \exp\left(\beta^{2}\left(\frac{1}{2}k\sigma^{2}+\sum_{(i_m,i_n)\in \mathcal{C}_{2}^{k}: i_m>i_n} \sigma_{i_{m},i_{n}}\right)\right),
\end{equation}
{and}
\begin{equation}
E\left(Y_{j_{k}}\exp\left(\beta\sum_{r=1}^{k}{Y_{j_{r}}}\right)\right) ={\beta} \left(\sigma^{2} + \sum_{r=1}^{k-1}\sigma_{j_k,j_r}\right) \exp\left({\beta^{2}}\left(\frac{1}{2}k\sigma^{2}+\sum_{(i_{m},i_n)\in \mathcal{C}_{2}^{k} : (i_m>i_n)  } \sigma_{i_{m},i_{n}} \right)\right),
\end{equation}
where $\sigma_{i_m,i_n}=EY_{i_m}Y_{i_n}$.
\end{lemma}
\textbf{Proof of Lemma~\ref{1Dbound} :}
To keep the notation simple,  we prove the lemma for the one dimensional case. The proof is  extended to dimension $d \geq 2$ by essentially following the same steps.
We let $V_{i,t}$ be a Gaussian stationary process with $V_{i,0}\sim N(0,\sigma^{2})$. Then, $(\Sigma \beta)_{1} ={\beta}\sigma^{2}$.\\
Observe that
\begin{align*}
    E D_{i,t+1}& =E\left(p(\gamma, V_{i,t}) \prod_{j=0}^{t-1}(1- p(\gamma, V_{i,j}))\right)\\
    E V_{i,t}D_{i,t+1}& =E \left(V_{i,t}p(\gamma, V_{i,t})\prod_{j=0}^{t-1}(1- p(\gamma, V_{i,j}))\right),
\end{align*}
where recall that $p(\gamma, V_{i,j}) = c \gamma \exp( \beta V_{i,j})(1+ H(\gamma,V_{i,j}))$.
Hence,
\begin{align*}
    E\hat{D}_{\gamma} & = \frac{1}{\gamma T(\gamma)m(\gamma)}\sum_{i=1}^{m(\gamma)}\sum_{t=0}^{T(\gamma)-1} E \prod_{j=0}^{t-1}(1- p(\gamma, V_{i,j}))p(\gamma, V_{i,t}) \textrm{ and }\\
    E\hat{V}_{\gamma} & = \frac{1}{\gamma T(\gamma)m(\gamma)}\sum_{i=1}^{m(\gamma)}\sum_{t=0}^{T(\gamma)-1} E \prod_{j=0}^{t-1}(1- p(\gamma, V_{i,j}))V_{i,t}p(\gamma, V_{i,t}).
\end{align*}
It can be seen that through expansion,
\[
\sum_{t=0}^{T(\gamma)-1}\prod_{j=0}^{t-1}(1-x_{j})x_{t} = \sum_{k=1}^{T(\gamma)}(-1)^{k+1} \sum_{i_{k}=k-1}^{T(\gamma)-1} \cdots \sum_{i_{1}=0}^{i_2-1} x_{i_{1}} \cdots x_{i_{k}}.
\]
Using the above, and removing $m(\gamma)$ as all firms are homogeneous:
\begin{align}
    E\hat{D}_{\gamma} &= \frac{1}{\gamma T(\gamma)}E\left(\sum_{k=1}^{T(\gamma)} (-1)^{k+1}\sum_{j_{k}=k-1}^{T(\gamma)-1} \cdots \sum_{j_{1}=0}^{i_{2}-1} p(\gamma,V_{i,j_{1}})\cdots p(\gamma,V_{i,j_{k}})\right)\label{bigeqn1} \textrm{, and,} \\
    E\hat{V}_{\gamma} &= \frac{1}{\gamma T(\gamma)} E\left(\sum_{k=1}^{T(\gamma)} (-1)^{k+1}\sum_{j_{k}=k-1}^{T(\gamma)-1} \cdots \sum_{j_{1}=0}^{i_{2}-1} V_{i,j_{k}}p(\gamma,V_{i,j_{1}})\cdots  p(\gamma,V_{i,j_{k}})\right).\label{bigeqn2}
\end{align}

The proof follows once we  show that $E\hat{V}_{\gamma} -  \sigma^{2}{\beta} E\hat{D}_{\gamma} = O(\gamma)$
while $E\hat{D}_{\gamma}$ is greater than or equal to a positive constant as $\gamma \rightarrow 0$.  Using Lemma~\ref{lemma:expVCV} and the definition of $p(\gamma,V_{i,t})$, it can be seen that ${E\hat{V}_{\gamma}} - \beta\sigma^{2}{E\hat{D}_{\gamma}}$
equals $\beta$ times
\begin{equation}\label{eqn:remainder}
 \frac{1}{\gamma T(\gamma)}\sum_{k=2}^{T(\gamma)}  (-1)^{k+1}c^k \gamma^{k}  \sum_{j_{k}=k-1}^{T(\gamma)-1} \cdots \sum_{j_{1}=0}^{j_{2}-1} \left(\sum_{r=1}^{k-1}\sigma_{j_{k},j_{r}} \right) \exp\left({\beta^{2}}\left(\frac{1}{2}k\sigma^{2}+\sum_{(i_{m},i_n)\in \mathcal{C}_{2}^{k}:i_m>i_n} \sigma_{i_{m},i_{n}}\right) \right) + O(\gamma).
\end{equation}
 Note that the sums in \eqref{bigeqn1} and \eqref{bigeqn2} also contain several terms involving $H(\gamma, V_{i,t})$.  However, recall that $|H(\gamma,V_{i,t})| \leq c \gamma \exp(\beta^\intercal V_{i,t})$. Hence, these do not contribute asymptotically and can be absorbed in the remainder $O(\gamma)$. Under Assumption~\ref{Assumption:Geometric}, the absolute value of (\ref{eqn:remainder}) can be upper bounded by 
 \[
 \frac{1}{\gamma T(\gamma)}\sum_{k=2}^{T(\gamma)} K c^k \gamma^{k}  \sum_{j_{k}=k-1}^{T(\gamma)-1} \cdots \sum_{j_{1}=0}^{j_{2}-1} \left(\sum_{r=1}^{k-1}\rho^{j_{k}-j_{r}} \right) \exp\left({\beta^{2}}\left(\frac{1}{2}k\sigma^{2}+\sum_{(i_{m},i_n)\in \mathcal{C}_{2}^{k}:i_m>i_n} K \rho^{(i_{m}-i_{n})}\right) \right) + O(\gamma)
 \]
 Let
\[
S_{k}\triangleq \sum_{j_{k}=k-1}^{T(\gamma)-1} \cdots \sum_{j_{1}=0}^{j_{2}-1} \left(\sum_{r=1}^{k-1}\rho^{(j_{k}-j_{r})} \right)\exp\left({\beta^{2}}\left(\frac{1}{2}k\sigma^{2}+\sum_{(i_{m},i_n)\in \mathcal{C}_{2}^{k}:i_m>i_n} K \rho^{(i_{m}-i_{n})}\right) \right).
\]
Then,
\begin{equation}\label{eqnL5.2}
   |{E\hat{V}_{\gamma}} - {\beta}\sigma^{2}{E\hat{D}_{\gamma}}| \leq  \frac{\beta}{\gamma T(\gamma)}\sum_{k=2}^{T(\gamma)} K c^{k} \gamma^{k}  S_{k} + O(\gamma).
\end{equation}

To proceed further we need an upper bound on the exponential component of $S_{k}$ which is independent of the choice of $j_{1},\ldots j_{k}$, but dependent on $k$. The following is useful to this end.\\
Let $j_{1}<\ldots<j_{k}$ and let $\mathcal{C}_{2}^{k}$ be the set of all their pair-wise combinations. Then,
\begin{equation}\label{eqn:sum1}
\sum_{i_{m},i_{n} \in \mathcal{C}_{2}^{k} : i_m>i_n} \rho^{(i_{m}-i_{n})}     \leq \frac{k}{1-\rho}.
\end{equation}
Further,
\begin{equation}\label{upper5.2}
S_{k} \leq \exp\left({\beta^{2}}\left(\frac{1}{2}\sigma^{2}+\frac{K}{1-\rho}\right)k\right)
 \sum_{j_{k}=k-1}^{T(\gamma)-1} \cdots \sum_{j_{1}=0}^{j_{2}-1} \left(\sum_{r=1}^{k-1} \rho ^{(j_{k}-j_{r})} \right).
\end{equation}
Now,
\begin{align*}
    \sum_{j_{k}=k-1}^{T(\gamma)-1} \cdots \sum_{j_{1}=0}^{j_{2}-1} \left(\sum_{r=1}^{k-1}\rho^{(j_{k}-j_{r})} \right) & =\sum_{j_{k}=k-1}^{T(\gamma)-1} \sum_{j_{k-1}=k-2}^{j_k-1}   \rho^{(j_{k}-j_{k-1})}   \sum_{j_{k-2}=k-3}^{j_{k-1}-1} \cdots \sum_{j_{1}=0}^{j_{2}-1} \left(1 + \sum_{r=1}^{k-2}\rho^{j_{k-1} - j_{r}} \right)\\
    &\leq k \sum_{j_{k}=k-1}^{T(\gamma)-1} \sum_{j_{k-1}=k-2}^{j_k-1} \rho^{(j_{k}-j_{k-1})}   (j_{k-1})^{k-2}\\
    &\leq k T(\gamma)^{k-1}.
\end{align*}
Thus,
\begin{equation}\label{eqnL5.2221}
   |{E\hat{V}_{\gamma}} - {\beta}\sigma^{2}{E\hat{D}_{\gamma}}| \leq
    \frac{1}{\gamma T(\gamma)}\sum_{k=2}^{T(\gamma)}  \tilde{c}^k k \gamma^{k}  T(\gamma)^{k-1} + O(\gamma),
\end{equation}
for an appropriate constant $\tilde{c}$. Further, it is easy to see that since $T(\gamma)
= \gamma^{-\zeta}$, the RHS above is $O(\gamma)$.\\
To see (\ref{eqn:sum1}), let $m,n$ be such that $k > m> n >0$. Since $i_{1}<i_{2}\cdots < i_{k}$, $i_{m} -i_{n} > m-n.$ We now note that a term of the form $i_{m} -i_{m-r}$ occurs exactly $k-r$ times in the summation and further, that $\rho^{(i_{m} -i_{m-r})}$ is upper bounded by $\rho^{r}$. Then the summation in (\ref{eqn:sum1}) is upper bounded by $\sum_{r=1}^{k} (k-r)\rho^{r}$,
which gives (\ref{eqn:sum1}). Similarly, $E\hat{D}_{\gamma}$ equals
\begin{equation}
 \frac{1}{\gamma T(\gamma)}\sum_{k=1}^{T(\gamma)}(-1)^{k+1}\gamma^{k}  \sum_{j_{k}=k-1}^{T(\gamma)-1} \cdots \sum_{j_{1}=0}^{j_{2}-1} \exp\left({\beta^{2}}\left(\frac{1}{2}k\sigma^{2}+\sum_{(i_{m},i_n)\in \mathcal{C}_{2}^{k}:i_m>i_n} \sigma_{i_{m},i_{n}}\right) \right) +O(\gamma).
\end{equation}

This equals  $c \exp(\frac{\beta^{2}\sigma^{2}}{2})$ plus
\begin{equation} \label{eq:resid}
 \frac{1}{\gamma T(\gamma)}\sum_{k=2}^{T(\gamma)}(-1)^{k+1}\gamma^{k}  \sum_{j_{k}=k-1}^{T(\gamma)-1} \cdots \sum_{j_{1}=0}^{j_{2}-1} \exp\left({\beta^{2}}\left(\frac{1}{2}k\sigma^{2}+\sum_{(i_{m},i_n)\in \mathcal{C}_{2}^{k}:i_m>i_n} \sigma_{i_{m},i_{n}}\right) \right) +O(\gamma).
\end{equation}
As before, (\ref{eq:resid}) is $O(\gamma)$, and the result follows.
\qed

We use $\xrightarrow[]{\mathcal{P}}$ and $\xrightarrow[]{\mathcal{D}}$ respectively, to denote convergence in probability and distribution. We further need the following definiton:
 \begin{definition}
A collection of random variables, $X_{\gamma}$, indexed by a continuous parameter $\gamma>0$ is said to be $o_{p}(a_{\gamma})$ if as $\gamma\to 0$,
\[
\frac{X_{\gamma}}{a_{\gamma}} \xrightarrow{\mathcal{P}} 0.
\]
 \end{definition}
\textbf{Proof of Lemma~\ref{lemma:keylem} :}\\
Since we can write
\[
\left(\beta^{*}(\gamma)-\hat{\beta}(\gamma)\right) = \Sigma^{-1}\left(\frac{E\hat{V}_{\gamma}}{E\hat{D}_{\gamma}} - \frac{\hat{V}_{\gamma}}{\hat{D}_{\gamma}}\right),
\]
establishing Lemma~\ref{lemma:keylem} reduces to estimating the expectation of the square of the error
\begin{equation}\label{eqn:basic}
    \frac{\hat{V}_{\gamma}^{(1)}}{\hat{D}_{\gamma}}
- \frac{E\hat{V}_{\gamma}^{(1)}}{E \hat{D}_{\gamma}},
\end{equation}
where $\hat{V}_{\gamma}^{(1)}$ denotes the first component of vector
$\hat{V}_{\gamma}$. Similar proofs will follow for other terms and the overall result can be obtained by linearity of expectation.
To evaluate (\ref{eqn:basic}), we use the Taylor series (\ref{eqn:TSEbeta}) and then show that $E(\hat{D}_{\gamma} - E\hat{D}_{\gamma})^{2} = O(\gamma^{\zeta})+O(\gamma^{\delta+\zeta-1})$. The result will then follow from Chebeshyev's inequality. Note that since both $E\hat{D}_{\gamma}$ and $E\hat{V}_{\gamma}^{(1)}$ are bounded away from $0$, $f(x,y)=\frac{x}{y}$ is differentiable at   $\frac{E\hat{V}_{\gamma}^{(1)}}{E\hat{D}_{\gamma}}$. Then, using the Taylor series expansion in probability, (see Section 6.5 in Brockwell and Davis 2013), the remainder term in (\ref{eqn:TSEbeta}) is
\begin{equation}\label{eqn:TSrem}
R((\hat{D}_{\gamma},\hat{V}_{\gamma}^{(1)}) , (E\hat{D}_{\gamma},   E\hat{V}_{\gamma}^{(1)})) =  o_{p}(\gamma^{\frac{1}{2}\zeta})+ o_{p}(\gamma^{\frac{1}{2}(\zeta+\delta-1)})
\end{equation}
In the Taylor series (\ref{eqn:TSEbeta}), we first consider $(\hat{D}_{\gamma}- E\hat{D}_{\gamma})$ and express it as
\[
\frac{1}{\gamma T(\gamma)m(\gamma)}\sum_{t=0}^{T(\gamma)-1}
\sum_{i=1}^{{m}(\gamma)}(D_{i,t+1} - E (D_{i,t+1}|{\cal F}_t))
+
\frac{1}{\gamma T(\gamma)m(\gamma)}\sum_{t=0}^{T(\gamma)-1}
\sum_{i=1}^{{m}(\gamma)}(E (D_{i,t+1}|{\cal F}_t)- ED_{i,t+1}),
\]
then the error $E(\hat{D}_{\gamma}- E\hat{D}_{\gamma})^2$ is the expectation of
\begin{equation}\label{var}
\left (\frac{1}{\gamma T(\gamma)m(\gamma)}\sum_{t=0}^{T(\gamma)-1}
\sum_{i=1}^{{m}(\gamma)}(D_{i,t+1} - E (D_{i,t+1}|{\cal F}_t))
+
\frac{1}{\gamma T(\gamma)m(\gamma)}\sum_{t=0}^{T(\gamma)-1}
\sum_{i=1}^{{m}(\gamma)}(E (D_{i,t+1}|{\cal F}_t)- ED_{i,t+1})
 \right )^2.
\end{equation}
Note that the cross terms where one term has the form
 $D_{i,t+1} - E (D_{i,t+1}|{\cal F}_t)$,
 have an expectation zero. Thus, the expectation of (\ref{var}) equals
\begin{equation}  \label{eqnn:1001}
\frac{1}{\gamma^2 T^2(\gamma) m^2(\gamma)}\sum_{t=0}^{T(\gamma)-1}
\sum_{i=1}^{{m}(\gamma)}E(D_{i,t+1} - E (D_{i,t+1}|{\cal F}_t))^2 \,\,\,\,\,\,\,  +
\end{equation}

\begin{equation}  \label{eqnn:1002}
\frac{1}{\gamma^2 T^2(\gamma) m^2(\gamma)}\sum_{t_1=0}^{T(\gamma)-1}
\sum_{t_2=0}^{T(\gamma)-1}
\sum_{i_1=1}^{{m}(\gamma)} \sum_{i_2=1}^{{m}(\gamma)}
E\left((E (D_{i_1,t_1+1}|{\cal F}_{t_1})- ED_{i_1,t_1+1}) (E (D_{i_2,t_2+1}|{\cal F}_{t_2})- ED_{i_2,t_2+1})\right).
\end{equation}
Notice that $E(D_{i,t+1} - E (D_{i,t+1}|{\cal F}_t))^2 = E(D_{i,t+1}) - E((ED_{i,t+1}\vert \mathcal{F}_{t})^{2})$. Hence,
\begin{equation}\label{1001:bound}
\frac{1}{\gamma^2 T^2(\gamma) m^2(\gamma)}\sum_{t=0}^{T(\gamma)-1}
\sum_{i=1}^{{m}(\gamma)}E(D_{i,t+1} - E (D_{i,t+1}|{\cal F}_t))^2 =\gamma^{\zeta +\delta - 1} (C+o(1)),
\end{equation}
for an appropriate constant $C$.
\vspace{0.2in}

We now evaluate (\ref{eqnn:1002}). First fix an $i$ and $t$ and consider
\begin{align}\label{exp:1002}
    & E\left(E\left(D_{i_{1},t_{1}+1}\vert \mathcal{F}_{t_1}) - ED_{i_{1},t_{1}+1} \right)(E\left(D_{i_{2},t_{2}+1}\vert \mathcal{F}_{t_2}) - ED_{i_{2},t_{2}+1} \right)\right)\nonumber\\
    & = E\left(E\left(D_{i_1 , t_{1}+1} \vert \mathcal{F}_{t_{1}}\right) E\left(D_{i_2 , t_{2}+1} \vert \mathcal{F}_{t_{2}}\right)\right) - E\left(D_{i_1 , t_{1}+1}\right)E\left(D_{i_2 , t_{2}+1}\right) .
\end{align}
Using arguments similar to \eqref{eq:resid}, for all $(i,t)$ $E(D_{i,t+1})=\exp(\frac{1}{2}\beta^\intercal\Sigma\beta -\alpha(\gamma))(1+o(\gamma))$, which equals  $E\exp(\beta^\intercal V_{i}-\alpha(\gamma))(1+o(\gamma))$. Further, note that since $\zeta<1$, for all $t<T(\gamma)$, $\mathbb{I}(\tau_{i} < t) = o(1)$. Then,
\[
E(D_{i,t+1}\vert \mathcal{F}_{t})=p(\gamma, V_{i,t})\mathbb{I}(\tau_{i} \geq t) = c\gamma \exp(\beta^{\intercal} V_{i})
(1+H(\gamma,V_{i,t}))(1+o(1)) \textrm{ a.s.} \textrm{ } \forall i,t
\]
Evaluating (\ref{exp:1002}) hence boils down to evaluating
\begin{equation}\label{neterrorterm}
      \gamma^{2} c^{2} \left (E(\exp({\beta}^{{\intercal}} V_{i,t_{1}})\exp({\beta}^{{\intercal}}  V_{j,{t_{2}}})) - (E\exp({\beta}^{{\intercal}} V_{i,t}))^2 \right),
\end{equation}
By stationarity,
\begin{equation}\label{exp2}
  (E\exp({\beta}^{{\intercal}} V_{i,t}))^2 =  \exp \left(\beta^{{\intercal}} \Sigma \beta \right) \forall i,t.
\end{equation}
To analyze the first term of (\ref{neterrorterm}), we need to get a handle on $E((V_{i, t_{1}} + V_{j,t_{2}})(V_{i, t_{1}} + V_{j,t_{2}})^{\intercal})$. Note that $V_{i,t_1}+V_{j,t_2}$ is Gaussian, with mean $0$. The covariance can be calculated to be $E(V_{i,t_1}V_{i,t_1}^{\intercal}) + E(V_{j,t_2}V_{j,t_2}^{\intercal})+ E(V_{i,t_1}V_{j,t_2}^{\intercal})+ E(V_{j,t_2}V_{i,t_1}^{\intercal})$. By stationarity, this is $2\Sigma$ plus $E(V_{i,t_1}V_{j,t_2}^{\intercal}) + E(V_{j,t_2}V_{i,t_1}^{\intercal})$. Then, 
\begin{equation}\label{mgf1}
    E(\exp({\beta}^{{\intercal}} V_{i,t_{1}})\exp({\beta}^{{\intercal}}  V_{j,{t_{2}}})) = \exp \left(\beta^{\intercal} \Sigma \beta + \beta^{{\intercal}} \cdot E(V_{i,t_1}V_{j,t_2}^{\intercal}) \beta \right).
\end{equation}
Now note that we can use  (\ref{exp2}) and (\ref{mgf1}) to rewrite (\ref{eqnn:1002}) as
\begin{equation}\label{totalsum}
\frac{\exp \left(\beta^T \cdot \Sigma \beta\right)}{T^2(\gamma) m^2(\gamma) }\sum_{i_1=0}^{m(\gamma)}\sum_{i_2=0}^{m(\gamma)}\sum_{t_1=0}^{T(\gamma)}\sum_{t_2=0}^{T(\gamma)} \left( \exp \left( \beta^{\intercal} \cdot E(V_{i,t_1}V_{j,t_2}^{\intercal}) \beta\right) -1 \right)
\end{equation}
plus lower order remainder terms. We divide the summation in three parts and analyze each part subsequently.\\
\textbf{Part 1:}
First consider the case where $i_1 \neq i_2$ and $t_1 \neq t_2$.
We first evaluate the outer sum over all $m(\gamma)(m(\gamma)-1)$ cases where $i_1 \neq i_2$. By using the fact that all firms are statistically identical, the contribution to (\ref{totalsum}) is
\begin{equation}\label{eqnt1neqt2}
    \frac{\exp \left(\beta^{\intercal} \cdot  \Sigma \beta\right)}{T(\gamma) }\left(\sum_{t_1=0}^{T(\gamma)}\sum_{t_2=0}^{T(\gamma)} \frac{1}{T(\gamma)}\left( \exp \left( \beta^{{\intercal}} E(V_{1,t_1}V_{2,t_2}^{\intercal}) \beta\right) -1 \right)(1 + o(1))\right),
\end{equation}
 By Assumption~\ref{Assumption:Geometric}, each term of the summation in (\ref{eqnt1neqt2}) is bounded between $\exp(-K\|\beta\|^{2}_{2}\rho^{|t_2-t_1|}) -1$ and $\exp(K\|\beta\|^{2}_{2}\rho^{|t_2-t_1|})-1$. Then, the entire sum lies between
 \begin{equation}\label{eqn:vardecayupper}
     \frac{\exp \left(\beta^{\intercal} \cdot  \Sigma \beta\right)}{T(\gamma) }\left(\sum_{t_1=0}^{T(\gamma)}\sum_{t_2=0}^{T(\gamma)} \frac{1}{T(\gamma)}\left( \exp \left(K\|\beta\|^{2}_{2}\rho^{|t_2-t_1|} )\right) -1 \right)(1 + o(1))\right)
 \end{equation}
 and 
 \begin{equation}\label{eqn:vardecaylower}
     \frac{\exp \left(\beta^{\intercal} \cdot  \Sigma \beta\right)}{T(\gamma) }\left(\sum_{t_1=0}^{T(\gamma)}\sum_{t_2=0}^{T(\gamma)} \frac{1}{T(\gamma)}\left( \exp \left(-K\|\beta\|^{2}_{2}\rho^{|t_2-t_1|} )\right) -1 \right)(1 + o(1))\right)
 \end{equation}
 which depend only on $|t_{2}-t_{1}|$. First consider (\ref{eqn:vardecayupper}). Note that the number of times where $|t_2 - t_1| =k$ is exactly $(T(\gamma) -k)$. Then, (\ref{eqn:vardecayupper}) becomes (ignoring the o(1) term),
\[
 \frac{{\exp \left(\beta^T \Sigma \beta\right)}}{T(\gamma)}\left(\sum_{k=1}^{T(\gamma)-1} \frac{T(\gamma) -k}{{T(\gamma) }}\left( \exp \left( K\|\beta\|^{2}_{2}\rho^{k}  \right) -1 \right)\right).
\]
To complete the argument, consider the following:\\
Let $\theta \in \Re$, $\rho \in (0, 1)$.  Then for all sufficiently large $k$,
\begin{equation}\label{aux1}
    |{\exp\left(\theta\cdot\rho^{k}\right)-1}| < 2|\theta| \rho^{k}.
\end{equation}
With $\theta = K\|\beta\|^{2}_{2}$ and $\theta = -K\|\beta\|^{2}_{2}$  the summations in (\ref{eqn:vardecayupper}) and (\ref{eqn:vardecaylower}) are convergent series. In particular, it means that their tails converge to 0. By the bounds developed above, the tail of the summation (\ref{eqnt1neqt2}) is sandwiched between two vanishing quantities, and thus itself is vanishing. But then the summation converges to  finite value. Hence, (\ref{eqnt1neqt2}) can be written as $(B_1+o(1))\gamma^{\zeta}$, for some constant $B_{1}$.\\
To see (\ref{eqn:vardecayupper}), note that
\begin{align*}
   \beta^{\intercal} \cdot E(V_{i,t_1}V_{j,t_2}^{\intercal}) \beta &\leq \|\beta\|_{2} \|E(V_{i,t_1}V_{j,t_2}^{\intercal}) \beta\|_{2}\\
   &\leq \|\beta\|^{2}_{2}\|E(V_{i,t_1}V_{j,t_2}^{\intercal})\| \leq K\|\beta\|^{2}_{2} \rho^{|t_2-t_1|},
\end{align*}
where the first inequality follows from the Cauchy-Schwarz inequality, the second is a result of the definition of operator norm, and the third follows from Assumption~\ref{Assumption:Geometric}. (\ref{eqn:vardecaylower}) can be easily seen by reversing the direction of the Cauchy-Schwarz and following the same steps. To see (\ref{aux1}), it is sufficient to show that if $\theta >0$,
\[
{\exp\left(\theta\cdot\rho^{k}\right)-1} < 2\theta \rho^{k}.
\]
for each large enough $k$. To this end, note that if $x>0$,  $f(x) = \frac{\exp(x)-1}{x}$ is an increasing function of $x$ and hence $u_{k} \triangleq \frac{\exp\left(\theta\cdot\rho^{k}\right)-1}{\theta\cdot\rho^{k}}$ decreases in $k$, such that $\lim_{k \to\infty}u_{k} =1$. Hence, $\exists k_1 : \forall k \geq k_1$, $u_{k} < 2$. Then, we have $\forall k \geq k_{1}$, ${\exp\left(\theta\cdot\rho^{k}\right)-1} < 2\theta\rho^{k}$. The other direction can be reasoned similarly.\\

\textbf{Part 2:}
Suppose $t_2 =t_1$, $i_1\neq i_2 $. The outer sum over $i_1$ and $i_2$ can be handled as in Part 1. Note that for this case, for all $i$ and $j$, $\Vert\mathbf{E}(V_{i,t_1}V_{j,t_1}^{\intercal})\Vert \leq K$. Further, the number of instances of the inner sum in which this occurs is exactly equal to $T(\gamma)$, and the expression inside the summation, $ \exp(\beta^{\intercal} E(V_{1,1}V_{2,1}^{\intercal}) \beta)-1$ is bounded. To see this, consider the following
\begin{align*}
    0\leq \langle \beta ,E(V_{1,1}V_{2,1}^{\intercal}) \beta \rangle  &\leq \|\beta\|_{2}^{2} \| E(V_{1,1}V_{2,1}^{\intercal})\|_{2}\\
    & \leq \|\beta\|_{2}^{2}K < \infty  
\end{align*}
Proceeding as in Part 1, the contribution to the total sum is
\[
\left(\exp \left( \frac{1}{2}\beta^{{\intercal}} \Sigma\beta\right)\right)\left(\exp \left(\beta^{\intercal} \cdot E(V_{1,1}V_{2,1}^{\intercal}) \beta\right)-1\right)\gamma^{\zeta} = B_2\gamma^{\zeta},
\]
for some constant $B_{2}$.\\

\textbf{Part 3:}
Consider the remaining cases, where $i_1 = i_2$. There are $m(\gamma)$ identical cases of this. The analysis proceeds exactly as in the previous cases, and regardless of whether $t_1=t_2$, an additional $\gamma^{\delta}$ factor is introduced due to only $m(\gamma)$ terms being present in the sum, rather than $m^{2}(\gamma)$ terms. Hence, this case gives a rate of the form $B_{3}\gamma^{\delta+\zeta}$, for some constant $B_{3}$, and does not contribute to the main sum asymptotically.\\
From the above three cases, there is some constant, call it $C^{'}$, such that (\ref{totalsum}) is $(C^{'}+o(1))\gamma^{\zeta}$.\\
The other two terms in the Taylor series, $E(\hat{V}_{\gamma}^{(1)} - E\hat{V}_{\gamma}^{(1)})^2$ and $E(( \hat{V}_{\gamma}^{(1)} - E\hat{V}_{\gamma}^{(1)})( \hat{D}_{\gamma} - E\hat{D}_{\gamma}))$ can be analyzed similarly.


Then, we have that 
\begin{equation}\label{eqn:smallprob}
    \left(\frac{\hat{V}_{\gamma}^{(1)}}{\hat{D}_{\gamma}} - \frac{E\hat{V}_{\gamma}^{(1)}}{E\hat{D}_{\gamma} }\right)^{2} = O_{\mathcal{P}}(\gamma^{\delta+\zeta-1}) + O_{\mathcal{P}}(\gamma^{\zeta}).
\end{equation}

Since 
\[
\|\hat{\beta}(\gamma) -{\beta^{*}}(\gamma)\|_{2}^{2} \leq \|\Sigma^{-1}\|\Big\|\left(\frac{\hat{V_{\gamma}}}{\hat{D_{\gamma}}} -  \frac{\hat{EV_{\gamma}}}{\hat{ED_{\gamma}}}\right)\Big\|_{2}^{2} ,
\]
we have the overall result. \qed

\textbf{Proof of Corollary~\ref{Cor:thm1}:}\\
To prove Corollary~\ref{Cor:thm1}, recall that $\mu=(\delta+\zeta-1)\land \zeta$. We need the following:\\
 Let $X_{\gamma}$ be random vectors which converge to $X$ in distribution. Then, it is known that if $X_{\gamma}$ are uniformly integrable, $EX_{\gamma} \to EX$ (Theorem 3.5 in Billingsley 1999).\\ 
Now, (\ref{eqn:TSrem}) together with the mapping theorem implies that
$\gamma^{-\mu} R^{2}((\hat{D}_{\gamma}, \hat{V}_{\gamma}^{(1)}),(E\hat{D}_{\gamma},   E\hat{V}_{\gamma}^{(1)})) \xrightarrow{D} 0$. Assumption~\ref{assumption:TSremainder} and Theorem 3.5 from Billingsley (1999) hence imply that 
\[
E(\gamma^{-\mu} R^{2}((\hat{D}_{\gamma}, \hat{V}_{\gamma}^{(1)}),(E\hat{D}_{\gamma},   E\hat{V}_{\gamma}^{(1)}))) \to 0,
\]
which means that the remainder is of lesser order in variance. Hence, we now write $E\left(\frac{\hat{V}_{\gamma}^{(1)}}{\hat{D}_{\gamma}}
- \frac{E\hat{V}_{\gamma}^{(1)}}{E \hat{D}_{\gamma}}\right)^2
$ as
\[
\frac{1}{(E \hat{D}_{\gamma})^2}E( \hat{V}_{\gamma}^{(1)} - E\hat{V}_{\gamma}^{(1)})^2
+\frac{(E\hat{V}_{\gamma}^{(1)})^2}{(E \hat{D}_{\gamma})^4}E(\hat{D}_{\gamma}- E\hat{D}_{\gamma})^2
- 2 \frac{E\hat{V}_{\gamma}^{(1)}}{(E \hat{D}_{\gamma})^3}E \left (( \hat{V}_{\gamma}^{(1)} - E\hat{V}_{\gamma}^{(1)})
(\hat{D}_{\gamma}- E\hat{D}_{\gamma})\right )
+ o(\gamma^{\mu}).
\]
Then, (\ref{eqn:smallprob}) can be strengthened to 
\begin{equation}\label{eqn:smallvariance}
    E\left(\frac{\hat{V}_{\gamma}^{(1)}}{\hat{D}_{\gamma}} - \frac{E\hat{V}_{\gamma}^{(1)}}{E\hat{D}_{\gamma} }\right)^{2} = C_{1,1}\gamma^{\delta+\zeta-1} + C^{\prime}_{1,1}\gamma^{\zeta},
\end{equation}
for some constants $C_{1,1}$ and $C_{1,1}^{\prime}$. The overall expansion of $\|\hat{\beta}(\gamma) -{\beta}^{*}(\gamma)\|_{2}^{2}$ contains cross terms of the form $\left(\frac{\hat{V}_{\gamma}^{i}}{\hat{D}_{\gamma}}-\frac{E\hat{V}_{\gamma}^{i}}{E\hat{D}_{\gamma}}\right)\left(\frac{\hat{V}_{\gamma}^{j}}{\hat{D}_{\gamma}}-\frac{E\hat{V}_{\gamma}^{j}}{E\hat{D}_{\gamma}}\right) $, $(i,j) \in \{1,2,\ldots,d\}$. For these cases, constants $C_{i,j}$ and $C_{i,j}^{'}$ can be calculated as above. With $\mathrm{A}=\Sigma^{-1}(\Sigma^{-1})^{\intercal}$ and $\mathbf{v}= \frac{\hat{V_{\gamma}}}{\hat{D_{\gamma}}}- \frac{\hat{EV_{\gamma}}}{\hat{ED_{\gamma}}}$, $E(\|\hat{\beta}(\gamma) -\beta\|_{2}^{2})=   E\left(\sum_{i,j}\mathrm{A}_{i,j}v_i v_j\right)$. Then, $K_{1} = \sum_{i,j}C_{i,j}\mathrm{A}_{i,j}$ and $K_{2}=\sum_{i,j}C_{i,j}^{'}\mathrm{A}_{i,j}$, which gives Corollary~\ref{Cor:thm1}.
\qed

\textbf{Cross terms in \eqref{eqnn:1001} have expectation zero:} \\
Consider a typical cross-term of \eqref{eqnn:1001}: 
\begin{equation}\label{eqn:cross-1}
    E((D_{i_1,t_1+1}-ED_{i_1,t_1+1}\vert \mathcal{F}_{t_1})(D_{i_2,t_2+1}-ED_{i_2,t_2+1}\vert \mathcal{F}_{t_2})).
\end{equation}

Consider two cases:\\
\textbf{Case 1:} Here, $t_1=t_2$. Then, \eqref{eqn:cross-1} can be re-written using iterated conditioning as
\[
E(E((D_{i_1,t_1+1}-ED_{i_1,t_1+1}\vert \mathcal{F}_{t_1})(D_{i_2,t_2+1}-ED_{i_2,t_1+1}\vert \mathcal{F}_{t_1})\vert \mathcal{F}_{t_1}))
\]
Recall that for $i\neq j$, conditioned on $\mathcal{F}_t$, $D_{i,t+1}$ and $D_{j,t+1}$ are independent (since they only depend on the i.i.d. uniform random variable). Then, one may write the above as 
\[E(E((D_{i_1,t_1+1}-ED_{i_1,t_1+1}\vert \mathcal{F}_{t_1}\vert \mathcal{F}_{t_1})E(D_{i_2,t_1+1}-ED_{i_2,t_1+1}\vert \mathcal{F}_{t_1}\vert \mathcal{F}_{t_1}))) =0.\]
\textbf{Case 2:} Here $t_2> t_1$. Then, We have that $t_2\geq t_1+1$. Now, write \eqref{eqn:cross-1} as
\begin{equation}\label{eqn:cross-2}
    E(E((D_{i_1,t_1+1}-ED_{i_1,t_1+1}\vert \mathcal{F}_{t_1})(D_{i_2,t_2+1}-ED_{i_2,t_2+1}\vert \mathcal{F}_{t_2})\vert \mathcal{F}_{t_1+1})).
\end{equation}
Observing that $(D_{i_1,t_1+1}-ED_{i_1,t_1+1}\vert \mathcal{F}_{t_1})$ is $\mathcal{F}_{t_1+1}$ measurable, \eqref{eqn:cross-2} becomes 
\[
  E((D_{i_1,t_1+1}-ED_{i_1,t_1+1}\vert \mathcal{F}_{t_1})E(D_{i_2,t_2+1}-ED_{i_2,t_2+1}\vert \mathcal{F}_{t_2})\vert \mathcal{F}_{t_1+1})
\]
Finally, noting that $\mathcal{F}_{t_1+1}\subseteq \mathcal{F}_{t_2}$, $E(D_{i_2,t_2+1}-E(D_{i_2,t_2+1}\vert \mathcal{F}_{t_2})\vert \mathcal{F}_{t_1+1})=0$ by the tower property of conditional expectation. Now, summing over all $i,j,t_1,t_2$ gives the result.
\ \\

\noindent {\large\textbf{Proof of \eqref{eqnlemma2}:}}\\
Recall that 
\begin{align}
\hat{\alpha}(\gamma) - \alpha(\gamma) = & \log\left(\frac{\sum_{i=1}^{m(\gamma)}\sum_{t=1}^{T(\gamma)} \exp({\beta}^{\intercal}V_{i,t})}{\sum_{i=1}^{m(\gamma)}\sum_{t=1}^{T(\gamma)} E\exp(\beta^{\intercal}V_{i,t})}\right) \label{eqn:alpha1}\\
& +\log\left(\frac{\sum_{i=1}^{m(\gamma)}\sum_{t=1}^{T(\gamma)} ED_{i,t+1}}{\sum_{i=1}^{m(\gamma)}\sum_{t=1}^{T(\gamma)} D_{i,t+1}}\right) \label{eqn:alpha2}\\
 &+\log\left(\frac{\sum_{i=1}^{m(\gamma)}\sum_{t=1}^{T(\gamma)} \exp(\hat{\beta}^{\intercal}(\gamma)V_{i,t})}{\sum_{i=1}^{m(\gamma)}\sum_{t=1}^{T(\gamma)} \exp(\beta^{\intercal}V_{i,t})}\right) \label{eqn:alpha3}
\end{align}
plus smaller order terms. We evaluate each of the above separately. The key idea is to notice the following:\\
Let $X_{\gamma}$ be an indexed set of random variables, such that $X_{\gamma} = O_{\mathcal{P}}(f(\gamma))$, for some $f(\gamma)\to 0$. Then, using a Taylor series yields
\begin{equation}\label{eqn:TSalpha}
    \log(1+X_{\gamma}) = X_{\gamma} + o_{\mathcal{P}}(f(\gamma)) = O_{\mathcal{P}}(f(\gamma)).
\end{equation}
We divide the rest of the proof into three parts:\\

\noindent\textbf{Part 1:}\\
Consider (\ref{eqn:alpha1}). Rewrite this as $\log\left(1+ U_{\gamma}\right)$, where
\[
U_{\gamma}=\frac{\sum_{i=1}^{m(\gamma)}\sum_{t=1}^{T(\gamma)} \exp({\beta}^{\intercal}V_{i,t}) -E\exp(\beta^{\intercal}V_{i,t}) }{\sum_{i=1}^{m(\gamma)}\sum_{t=1}^{T(\gamma)} E\exp(\beta^{\intercal}V_{i,t})}.
\]
We show that $U_{\gamma}= 
O_{\mathcal{P}}(\gamma^{\frac{1}{2}(\delta+\zeta-1)}) + O_{\mathcal{P}}(\gamma^{\frac{1}{2}\zeta})$, and then use (\ref{eqn:TSalpha}). Note that $\sum_{i=1}^{m(\gamma)}\sum_{t=1}^{T(\gamma)} E\exp(\beta^{\intercal}V_{i,t})$ divided by $T(\gamma)m(\gamma)$ is  $\Theta(1)
$. Now, consider
\begin{equation}\label{eqn:break}
    \frac{1}{T(\gamma)m(\gamma)}\sum_{i=1}^{m(\gamma)}\sum_{t=1}^{T(\gamma)} \left(\exp({\beta}^{\intercal}V_{i,t}) -E\exp(\beta^{\intercal}V_{i,t})\right).
\end{equation}
To bound \eqref{eqn:break}, we evaluate its second moment. Squaring and taking the expected value, we get
\[
\frac{1}{T^{2}(\gamma)m^{2}(\gamma)}\sum_{i_1,i_2=1}^{m(\gamma)}\sum_{t_1,t_2=1}^{T(\gamma)}E\exp(\beta^{\intercal}(V_{i_1,t_1} + V_{i_2,t_2})) - (E(\exp(\beta^{\intercal}V_{1,1})))^{2}.
\]
Following the proof of Lemma~\ref{lemma:keylem} (see \ref{neterrorterm} onward), this is $O(\gamma^{\zeta})$, which together with (\ref{eqn:TSalpha}) gives us the desired bound. \\

\noindent \textbf{Part 2:}

\noindent Next, we must evaluate (\ref{eqn:alpha2}). Rewrite it as
\[
\log\left(1+ \frac{\sum_{i=1}^{m(\gamma)}\sum_{t=1}^{T(\gamma)} (D_{i,t+1} -ED_{i,t+1})}{\sum_{i=1}^{m(\gamma)}\sum_{t=1}^{T(\gamma)} ED_{i,t+1}}\right)
\]
Now, $\frac{1}{\gamma T(\gamma)m(\gamma)}$ times $\sum_{i=1}^{m(\gamma)}\sum_{t=1}^{T(\gamma)} ED_{i,t+1}$ is $\Theta(1)$, and it follows from Lemma~\ref{lemma:keylem}, that $$\frac{1}{\gamma T(\gamma)m(\gamma)}\sum_{i=1}^{m(\gamma)}\sum_{t=1}^{T(\gamma)} (D_{i,t+1} -ED_{i,t+1}) = O_{\mathcal{P}}(\gamma^{\frac{1}{2}(\delta+\zeta-1)}) + O_{\mathcal{P}}(\gamma^{\frac{1}{2}\zeta}).$$
Hence,
\[
 \frac{\sum_{i=1}^{m(\gamma)}\sum_{t=1}^{T(\gamma)} (D_{i,t+1} -ED_{i,t+1})}{\sum_{i=1}^{m(\gamma)}\sum_{t=1}^{T(\gamma)} ED_{i,t+1}} = O_{\mathcal{P}}(\gamma^{\frac{1}{2}(\delta+\zeta-1)}) + O_{\mathcal{P}}(\gamma^{\frac{1}{2}\zeta})
\]
Then, we use (\ref{eqn:TSalpha}) to get that (\ref{eqn:alpha2}) is  $O_{\mathcal{P}}(\gamma^{\frac{1}{2} (\delta+\zeta-1)}) + O_{\mathcal{P}}( \gamma^{\frac{1}{2}\zeta})$.\\

\noindent \textbf{Part 3:}\\
 We now evaluate (\ref{eqn:alpha3}).Note that this is 
\[
\log\left(1+ \frac{\sum_{i=1}^{m(\gamma)}\sum_{t=1}^{T(\gamma)} \exp({\hat{\beta}}^{\intercal}(\gamma)V_{i,t}) - \exp(\beta^{\intercal}V_{i,t}) }{\sum_{i=1}^{m(\gamma)}\sum_{t=1}^{T(\gamma)} \exp(\beta^{\intercal}V_{i,t})}\right),
\]
 Again, $\sum_{i=1}^{m(\gamma)}\sum_{t=1}^{\tau_{i}}  \exp(\beta^{\intercal}V_{i,t})$, normalised by $\frac{1}{T(\gamma)m(\gamma)}$ is $O_{\mathcal{P}}(1)$. To proceed, observe that the since $\|\beta-\hat\beta(\gamma)\|$ is small, one can linearise $\exp({\hat{\beta}}^{\intercal}(\gamma)V_{i,t}) - \exp(\beta^{\intercal}V_{i,t})$.  Specifically, by the mean value  form of the Taylor remainder theorem, \eqref{eqn:alpha3} is
\begin{equation}\label{eqn:alphacrit}
    \frac{1}{T(\gamma)m(\gamma)}\sum_{i,t}\exp(\beta^{\intercal}V_{i,t}) (\hat{\beta}(\gamma)-\beta)^{\intercal}V_{i,t} 
\end{equation}
plus 
\begin{equation}\label{eqn:alpharem}
    \frac{1}{2T(\gamma)m(\gamma)}\sum_{i,t}\exp(\bar{\beta}_{i,t}^{\intercal}(\gamma)V_{i,t}) ((\hat{\beta}(\gamma)-\beta)^{\intercal}V_{i,t})^{2} 
\end{equation}
for some $\bar{\beta}_{i,t}(\gamma)$ on the line joining $\beta$ and $\hat{\beta}(\gamma)$. By Cauchy-Schwarz, (\ref{eqn:alphacrit}) is bounded above by
\[
\|\hat{\beta}(\gamma)-\beta\|_2\frac{1}{T(\gamma)m(\gamma)} \sum_{i,t}\|V_{i,t}\|_2\exp(\beta^{\intercal}V_{i,t}) . 
\]
By the Weak law of Large Numbers (henceforth referred to as WLLN), $\frac{1}{T(\gamma)m(\gamma)} \sum_{i,t}\|V_{i,t}\|_2\exp(\beta^{\intercal}V_{i,t})$ is $O_{\mathcal{P}}(1)$, and from Theorem~\ref{theorem:1}, $\|\hat{\beta}(\gamma)-\beta\|_2=O_{\mathcal{P}}(\gamma^{\frac{1}{2}(\delta+\zeta-1)}) + O_{\mathcal{P}}(\gamma^{\frac{1}{2}\zeta})$.\\  
We now bound (\ref{eqn:alpharem}). Notice that this may be bounded above by 
\[
 \frac{1}{2T(\gamma)m(\gamma)}\sum_{i,t}\max(\exp(\beta^{\intercal}V_{i,t}) , \exp(\Hat{\beta}^{\intercal}(\gamma) V_{i,t})) ((\hat{\beta}(\gamma)-\beta)^{\intercal}V_{i,t})^{2} 
\]
Using Cauchy-Schwarz, this is bounded by
\begin{equation}\label{eqn:alpha1.0}
    \|\hat{\beta}(\gamma)-\beta\|_{2}^{2}\frac{1}{2T(\gamma)m(\gamma)}\sum_{i,t}\max(\exp(\|\beta\|_{2}\|V_{i,t}\|_{2}) , \exp(\|\Hat{\beta}(\gamma)\|_{2}\|V_{i,t}\|_{2})) \|V_{i,t}\|^{2}_{2}. 
\end{equation}
 Now,  consider
\begin{equation}\label{eqn:alpha1.1}
    \frac{1}{2T(\gamma)m(\gamma)}\sum_{i,t}\max(\exp(\|\beta\|_{2}\|V_{i,t}\|_{2}) , \exp(\|\Hat{\beta}(\gamma)\|_{2}\|V_{i,t}\|_{2})) \|V_{i,t}\|^{2}_{2}. 
\end{equation}
First note that $\|\hat{\beta}(\gamma)-\beta\|_{2}^{2}$ is $O_{\mathcal{P}}(\gamma^{\delta+\zeta-1}) + O_{\mathcal{P}}(\gamma^{\zeta})$, that is, for any $\epsilon>0$, we can find a set, call it $\mathrm{A}$, of probability larger than $1-\epsilon$, such that for some sufficiently large $C$, $\|\Hat{\beta}(\gamma)\| \leq C$ on $\mathrm{A}$. Then, on this set, (\ref{eqn:alpha1.1}) is bounded by 
\begin{equation}\label{eqn:alpha1.2}
    \frac{1}{2T(\gamma)m(\gamma)}\sum_{i,t}\exp((\|\beta\|_{2} + C)\|V_{i,t}\|_{2})\|V_{i,t}\|_2^2,
\end{equation}
which is $O_{\mathcal{P}}(1)$. Then,
\[
\frac{1}{T(\gamma)m(\gamma)}\sum_{i=1}^{m(\gamma)}\sum_{t=1}^{T(\gamma)} \exp({\hat{\beta}}^{\intercal}(\gamma)V_{i,t}) - \exp(\beta^{\intercal}V_{i,t}) = O_{\mathcal{P}}(\gamma^{\frac{1}{2}(\delta+\zeta-1)}) + O_{\mathcal{P}}(\gamma^{\frac{1}{2}\zeta}).
\]
This completes the proof.\qed
\ \\
\large{\textbf{Proof of Proposition~\ref{theorem:2}:}}\\
\normalsize
We first note that 
\[
\frac{p(\gamma, V_i) -\hat{p}(\gamma,V_i)}{p(\gamma, V_i)}
\]
can be rewritten as 
\begin{equation}\label{eqn:thm3int}
     1 - \exp((\beta - \hat{\beta}(\gamma))^{\intercal}V_{i} +(\alpha(\gamma)-\hat{\alpha}(\gamma))) + G(\gamma,V_{i}),
\end{equation}
where $|G(\gamma,V_{i})| =O_{\mathcal{P}}( \gamma \exp(\beta^{\intercal}V_{i}))$. 
 By Theorem \ref{theorem:1}, $\| (\beta - \hat{\beta}(\gamma))^{\intercal}V_{i}\|_{2}^{2} $ and $\|\alpha(\gamma)-\hat{\alpha}(\gamma)\|_{2}^{2}$ are $O_{\mathcal{P}}(\gamma^{\delta+\zeta-1}) + O_{\mathcal{P}}(\gamma^{\zeta})$. The Taylor series for $1 - \exp((\beta - \hat{\beta}(\gamma))^{\intercal}V_{i} +(\alpha(\gamma)-\hat{\alpha}(\gamma)))$ is hence
\[
1 - \exp((\beta - \hat{\beta}(\gamma))^{\intercal}V_{i} +(\alpha(\gamma)-\hat{\alpha}(\gamma)) = (\beta - \hat{\beta}(\gamma))^{\intercal}V_{i} +(\alpha(\gamma)-\hat{\alpha}(\gamma)) + o_{p}(\frac{1}{2}\gamma^{\zeta}) + o_{p}(\frac{1}{2}\gamma^{\zeta+\delta-1})
\]
and (\ref{eqn:thm3int}) becomes  
\[
1 - \exp((\beta - \hat{\beta}(\gamma))^{\intercal}V_{i} +(\alpha(\gamma)-\hat{\alpha}(\gamma)) = O_{p}(\gamma^{\frac{1}{2}(\delta+\zeta-1)}) + O_{p}(\gamma^{\frac{1}{2}\zeta}).
\]
Squaring both sides and using the mapping theorem gives the final result.
\qed
\section{Proof of Theorem~\ref{thm:CLT} and preliminaries}
\subsection{Preliminaries:}
We first give a few mathematical preliminaries necessary for the proof:\\
\textbf{A central limit theorem for martingale differences:}\\
 A few definitions are needed first. An array of random variables, $\{X_{n,k}\}$, $k \in \{1,2,\cdots k_n\}$, $n\geq1$ along with an array of filtrations $\{\mathcal{F}_{n,k}\}$, such that for each $n$, $\mathcal{F}_{n,k}$ is non-decreasing in $k$, is called a martingale difference array if
\begin{itemize}
    \item $\forall k,n$, $X_{n,k}$ is $\mathcal{F}_{n,k}$ measurable and
    \item For each $n$, $\{X_{n,k}\}$, $k\in\{1,2,\ldots k_n\}$ is a martingale difference with respect to the filtration $\{\mathcal{F}_{n,k}$\}, that is, $E(X_{n,k}\vert\mathcal{F}_{n,k-1})=0$ a.s.
\end{itemize}
Define its conditional variance sum as
\[
V_{n,k} = \sum_{i=1}^{k_n}E(X_{n,i}^{2}\vert \mathcal{F}_{n,i-1}).
\]
We now state the theorem we need to prove our result -
\begin{theorem}\label{thm:MgleCLT}
Let $k_{n}\to \infty $ as $n\to\infty$. For each $n$, let $\{X_{n,k}\}$, $k\in\{1,2,\ldots k_n\}$ be a zero mean, square integrable martingale difference with respect to the filtration $\{\mathcal{F}_{n,k}$\}. Further, as $n \to\infty$ -
\begin{equation}\label{eqn:condvarsum}
    V_{n,k_{n}} \xrightarrow[]{\mathcal{P}} \sigma^{2},
\end{equation}
where $\sigma^{2}$ is an almost surely finite random variable, and for every $\kappa >0$
\begin{equation}\label{Lindeberg}
    \sum_{i=1}^{k_{n}}E(X_{n,i}^{2}\mathbb{I}(\vert X_{n,i}\vert > \kappa)\vert \mathcal{F}_{n,i-1}) \xrightarrow[]{\mathcal{P}} 0.
\end{equation}
Then, $S_{n,k_n} \triangleq \sum_{i=1}^{k_n}X_{n,i}$ converges in distribution as $n\to\infty$ to a random variable with the characteristic function $E\exp(-\frac{1}{2}\sigma^{2}t^{2})$ (recall that $\sigma^2$ here is a random variable).
\end{theorem}
\begin{remark}{
In general, a stronger condition is required that along with (\ref{eqn:condvarsum}) and (\ref{Lindeberg}), that $\forall k$,
\[
\mathcal{F}_{n,k} \subseteq \mathcal{F}_{n+1,k} \ \forall n.
\]
However, we will not need this in our proof as $\sigma$ in (\ref{eqn:condvarsum}) is constant and thus measurable with respect to each $\mathcal{F}_{n,k}$. This ensures that $\sigma$ is in the completion of each $\mathcal{F}_{n,k}$, and hence, that the above condition is not required for convergence in distribution - see Theorem 3.2 and the remark following it in Hall and Heyde (1980) for more details.}
\end{remark}
Lemma~\ref{rem:Vectors} extends Theorem~\ref{thm:MgleCLT} to random vectors:
\begin{lemma}\label{rem:Vectors}
\em{ Let $\{X_{n,k}\}\in\Re^{d}$ be a square integrable martingale difference array corresponding to the filtration $\{\mathcal{F}_{n,k}\}$. Suppose
\begin{align}
    M_{n,k_n} = \sum_{i=1}^{k_n} E(X_{n,i}X_{n,i}^{\intercal}\vert \mathcal{F}_{n,i-1})  \xrightarrow[]{\mathcal{P}} \Lambda \label{eq:Vectorcv}
\end{align}
and $\forall \kappa >0$,
\begin{align}
    \sum_{i=1}^{k_{n}}E(\|X_{n,i}\|^{2}_{2}\mathbf{I}(\| X_{n,i}\|_{2} > \kappa)\vert \mathcal{F}_{n,i-1}) \xrightarrow[]{\mathcal{P}} 0\label{eqn:vectorCLT},
\end{align}
then, as $n\to\infty$
\[
\sum_{i=1}^{k_{n}} X_{n,k_n} \xrightarrow[]{\mathcal{D}} \mathrm{N}(0,\Lambda).
\]
}
\end{lemma}
\textbf{Z-estimators:}
Define a random function $\mathbf{F}(\cdot)$ as a map from a parameter space $\mathcal{X}$ which assigns a random element $\mathbf{F}(\theta)$ to each $\theta\in\mathcal{X}$, that is $\mathbf{F}: \mathcal{X} \rightarrow \Omega\times \mathcal{Y}$, for some target space $\mathcal{Y}$. Consider a sequence of random functions $\Psi_{n}(\cdot)$. A random element $\theta_n$ which satisfies
\[
\Psi_{n}(\theta_n) =0
\]
is called a Z-estimator (zero- estimator) for $\Psi_{n}$. Recall that a sequence of random variables $\hat{\theta}_n$ used to estimate a parameter $\theta$ is called asymptotically consistent to $\theta$, if $\hat{\theta}_n \xrightarrow[]{\mathcal{P}} \theta$ as $n\to\infty$. For the present case, note that $\mathcal{X},\mathcal{Y}=\Re^{d}$.\\
Let $\Psi(\cdot): \Re^d \to \Re^d$ be a deterministic function, such that 
\begin{equation}\label{eqn:Z-est-Conv}
    \Psi_{n}(\theta) \xrightarrow[]{\mathcal{P}} \Psi(\theta)
\end{equation}
point-wise on a compact $\Theta \subset \Re^d $. Let $\theta_{0} \in \Theta$ be a zero of $\Psi(\cdot)$. The following theorem gives conditions under which $\hat{\theta}_n - \theta_0$, appropriately scaled is asymptotically normal (see Van der Vaart 1998, Theorem 5.41 for a related discussion and the proof).\\
\begin{theorem}\label{thm:Z-est}
\em{
Let $\Psi_{n}(\cdot)$ be a sequence of random vector valued functions, and let $\Psi(\cdot)$ be a deterministic function, such that $\Psi_{n}(\cdot) \xrightarrow[]{\mathcal{P}}\Psi(\cdot)$ point-wise on a compact parameter set $\Theta$. Let the following hold:
\begin{enumerate}
    \item  $\theta_{o}$ is a zero of $\Psi(\cdot)$, and $\Psi(\cdot)$ has an 
    invertible derivative at $\theta_{o}$, call it $\dot{\Psi}^{-1}_{\theta_{o}}$. \label{C1}
    \item The first derivative with respect to $\theta$ of $\Psi_{n}(\cdot)$ converges at $\theta_{o}$, and that the second derivative is bounded at least in some neighbourhood $\mathcal{N}$ of $\theta_{o}$.\label{C2}
    \item Further, for some $r_n\to\infty$, $\sqrt{r_{n}}\Psi_{n}(\theta_{o}) \xrightarrow[]{\mathcal{D}} \mathbf{Z}$, where $\mathbf{Z}$ is a Gaussian vector with covariance matrix $\mathrm{V}$\label{C3}.
\end{enumerate}
}
   Then, for any asymptotically consistent sequence $\hat{\theta}_{n}$, such that $\Psi_{n}(\hat{\theta}_{n})=0$, $\sqrt{r_n}(\hat{\theta}_{n}-\theta_{o}) $ is asymptotically normal, with covariance matrix $(\dot{\Psi}^{-1}_{\theta_{0}})^{\intercal} \mathrm{V} \dot{\Psi}^{-1}_{\theta_{0}}$.
\end{theorem}
\begin{remark}{
The above theorem does not guarantee a sequence of asymptotically consistent Z-estimators, $\hat{\theta}_{n}$. However, in our set-up, existence can be verified using stocahstic equicontinuity (see Van der Vaart (1998))}. We will verify existence of an aysmptotically consistent sequence in Section~\ref{sec:Consistent}.
\end{remark}
\subsection{Proof of Theorem~\ref{thm:CLT}:}
For this proof, assume $\frac{1}{\gamma}$ is an integer to avoid routine technicalities. Some notation is needed to proceed. Recall that under the assumptions of the theorem, the conditional default probability is given by
\[
p(\gamma,V_{i,t}) = q(\beta^\intercal V_{i,t}-\alpha(\gamma)),
\]
where $q(x) =  e^x(1+ h(x))$ and write $q^{\prime}(x)$ for its derivative. Let $\hat{\beta}_M(\gamma)$ solve the MLE. Then define
\begin{align}
    q^{\prime}(V_{i,t}) &=  q^{\prime}(\hat\beta^{\intercal}_M(\gamma) V_{i,t}-\alpha(\gamma))\label{eqn:Prob-der}\\
    q(V_{i,t})&=q(\hat\beta^{\intercal}_M(\gamma) V_{i,t}-\alpha(\gamma))\nonumber
\end{align}
The proof proceeds in two steps.\\
\textbf{Step 1: Establish point-wise convergence to a limit}\\ 
Recall the definition of the likelihood function given in \eqref{eqn:LR1}. Differentiating this with respect to $\beta$ and setting the derivative to zero gives  the following first order conditions: 
\[
\frac{1}{\gamma^{-(\zeta+\delta-1)}}\sum_{i=1}^{m(\gamma)}\sum_{t=0}^{T(\gamma)-1}  V_{i,t} D_{i,t+1}\frac{q^{\prime}(V_{i,t}) }{q(V_{i,t})(1-q(V_{i,t})) }  = \frac{1}{\gamma^{-(\zeta+\delta-1)}}\sum_{i=1}^{m(\gamma)}\sum_{t=0}^{T(\gamma)-1} V_{i,t}\frac{q^{\prime}(V_{i,t})}{1-q(V_{i,t})}\mathbb{I}(\tau_{i}\geq t)
\]
Define
\begin{equation}\label{psi_n2}
\Psi_{\gamma}(\mathbf{u})\triangleq \frac{1}{\gamma^{-(\zeta+\delta-1)}}\sum_{i=1}^{m(\gamma)}\sum_{t=0}^{T(\gamma)-1}  \left(\frac{V_{i,t} D_{i,t+1}q^{\prime}(\mathbf{u}^\intercal V_{i,t}-\alpha(\gamma)) }{q(\mathbf{u}^\intercal V_{i,t}-\alpha(\gamma))(1-q(\mathbf{u}^\intercal V_{i,t}-\alpha(\gamma))) }-V_{i,t}\frac{q^{\prime}(\mathbf{u}^\intercal V_{i,t}-\alpha(\gamma))}{1-q(\mathbf{u}^\intercal V_{i,t}-\alpha(\gamma))}
\right)\mathbb{I}(\tau_{i}\geq t).
\end{equation}
Note that the maximum likelihood estimate $\hat{\beta}_{M}(\gamma)$ is a zero of $\Psi_{\gamma}(\cdot)$. From \eqref{eqn:Prob-der} and the assumption of the theorem, for any $\mathbf{u}\in\Re^2$,
\begin{align}
    \Psi_{\gamma}(\mathbf{u})&= \frac{1}{\gamma^{-(\zeta+\delta-1)}}\sum_{i=1}^{m(\gamma)}\sum_{t=0}^{T(\gamma)-1}  \left(V_{i,t} D_{i,t+1}-V_{i,t}\mathrm{e}^{\mathbf{u}^{\intercal} V_{i,t}-{\alpha}(\gamma) }\right)\nonumber\\ &-\frac{1}{\gamma^{-(\zeta+\delta-1)}}\sum_{i=1}^{m(\gamma)}\sum_{t=0}^{T(\gamma)-1}  \left(V_{i,t} D_{i,t+1}- V_{i,t}\mathrm{e}^{\mathbf{u}^{\intercal} V_{i,t}-{\alpha}(\gamma) } \right)\mathbb{I}(\tau_{i}<t) + o_{\mathcal{P}}(1) \label{eqn:CLT-int}\\
&=  \frac{1}{\gamma^{-(\zeta+\delta-1)}}\sum_{i=1}^{m(\gamma)}\sum_{t=0}^{T(\gamma)-1}  \left(V_{i,t} D_{i,t+1}- V_{i,t}\mathrm{e}^{\mathbf{u}^{\intercal} V_{i,t}-{\alpha}(\gamma) }\right) + o_{\mathcal{P}}(1)\label{eqn:CLTkey}.
\end{align}
Equation \eqref{eqn:CLTkey} follows from the fact that  since $\zeta<1$, for all $t\in\{1,\ldots T(\gamma)\}$, $\mathbb{I}(\tau_i<t)=o_{\mathcal{P}}(1)$. Since $\alpha(\gamma)=\log\frac{1}{\gamma}$, by the WLLN, $\Psi_{\gamma} (\cdot) \xrightarrow[]{\mathcal{P}} \Psi(\cdot)$ point-wise on $\Theta$, where
\begin{equation}\label{psi2}
\Psi(\mathbf{u}) = \Sigma \beta \exp(\frac{1}{2}\beta^{\intercal}\Sigma\beta)-E(V_{i}\exp (\mathbf{u}^{\intercal}V_{i})).
\end{equation}
\textbf{Step 2: Check the conditions of Theorem~\ref{thm:Z-est}}\\
\textbf{Condition~\ref{C1}}  For this, we establish a convexity condition on the anti-derivative of $\Psi(\cdot)$. Observe that
\[
\Psi(\mathbf{u}) = \Sigma\left(\beta\exp(\frac{1}{2}\beta^{\intercal}\Sigma\beta) - \mathbf{u}\exp(\frac{1}{2}\mathbf{u}^{\intercal}\Sigma\mathbf{u})\right)
\]
is the derivative of the real valued function  
\[
\Phi(\mathbf{u}) = \mathbf{u}^{\intercal} \left(\Sigma\beta\exp(\frac{1}{2}\beta^{\intercal}\Sigma\beta)\right) - \exp(\frac{1}{2}\mathbf{u}^{\intercal}\Sigma\mathbf{u}).
\]
By the positive definiteness of the matrix $\Sigma$, $ \frac{1}{2}\mathbf{u}^{\intercal}\Sigma\mathbf{u}$ is strictly convex, and hence, $\exp(\frac{1}{2}\mathbf{u}^{\intercal}\Sigma\mathbf{u})$ is also convex. $\Phi(\cdot)$ is the sum of a linear function and a strictly concave function, and thus is itself strictly concave. Further, $\Phi(\mathbf{u}) \to -\infty$ as $\|\mathbf{u}\|_{2} \to \infty$. This means that its derivative has a unique 0. Additionally, this means that the Hessian of $\Phi(\cdot)$, a $\Re^{(d_1+d_2)\times (d_1+d_2)}$ matrix, is negative definite, and thus invertible. Letting $\mathbb{D}^{k}$ denote the differential operator of order $k$,
\[
\mathbb{D}^{2}\Phi(\mathbf{u}) = \mathbb{D} \Psi(\mathbf{u}) \prec 0,
\]
which implies that the derivative of $\Psi(\cdot)$ is invertible everywhere. 

\textbf{Condition~\ref{C2}} Observe from \eqref{eqn:Prob-der} that,
\begin{align*}
\mathbb{D}{\Psi}_{\gamma}(\mathbf{u}) &= \frac{1}{\gamma^{-(\delta+\zeta)}} \sum_{i=1}^{m(\gamma)}\sum_{t=0}^{T(\gamma)-1} \mathbb{D}\left(V_{i,t} D_{i,t+1}- V_{i,t}\mathrm{e}^{\mathbf{u}^{\intercal} V_{i,t}-{\alpha}(\gamma) } \right)+o_{\mathcal{P}}(1)\\ 
 &= -\frac{1}{\gamma^{-(\delta+\zeta)}} \sum_{i=1}^{m(\gamma)}\sum_{t=0}^{T(\gamma)-1} \mathbb{D}\left(V_{i,t}\exp(\mathbf{u}^{\intercal} V_{i,t})\right) + o_{\mathcal{P}}(1).
\end{align*}
The RHS, however, by the WLLN, converges in probability to 
\[
E(\mathbb{D}(V_{i}\exp(\mathbf{u}^{\intercal}V_i))) = \mathbb{D}{\Psi}(\mathbf{u}).
\]
Similarly,
\[
\mathbb{D}^{2}{\Psi}_{\gamma}(\mathbf{u}) = \frac{1}{\gamma^{-(\delta+\zeta)}} \sum_{i=1}^{m(\gamma)}\sum_{t=0}^{T(\gamma)-1} \mathbb{D}^{2}\left(V_{i,t}\exp(\mathbf{u}^{\intercal} V_{i,t})\right) + o_{\mathcal{P}}(1).
\]
 The sum on the RHS converges by the WLLN to the integrable random  variable $\mathbb{D}^{2}{\Psi}(\cdot)$ for each $\mathbf{u}$ at least in some neighbourhood of $\beta$. The RHS is $\left(\mathbb{D}^{2}{\Psi}(\mathbf{u}) +o_{\mathcal{P}}(1)\right)$, taking values in $\Re^{(d_1+d_2)\times (d_1+d_2)\times (d_1+d_2)}$. In particular, it is stochastically bounded in that neighbourhood.\\

\textbf{Condition~\ref{C3}} For each $\gamma$, let $ k_{i,t}= (i-1)T(\gamma) +t$, $t\in\{0,1,\cdots T(\gamma)\}$, let $D_{k_{i,t}}(\gamma)\triangleq D_{i,t+1}$ and $\mathcal{D}_{k_{i,t}}(\gamma)$ be its generated $\sigma$-algebra.  Similarly, let $V_{k_{i,t}}(\gamma) \triangleq V_{i,t}$ and let $\mathcal{G}_{ k_{i,t}}(\gamma)$ be the $\sigma$-algebra generated by it. Recall that the true conditional default probability, $p(\gamma, V_{i,t}) = q(\beta^\intercal V_{i,t}-\alpha(\gamma))$. Write $r(V_{i,t})$ and $r^{\prime}(V_{i,t})$ for $q(\beta^\intercal V_{i,t}-\alpha(\gamma))$ and $q^{\prime}(\beta^\intercal V_{i,t}-\alpha(\gamma))$, respectively.  Define an array of random variables -
\begin{equation}\label{eqn:MartingaleCLT}
    \Gamma_{k_{i,t}+1}(\gamma) = \frac{1}{\gamma^{-\frac{\zeta+\delta-1}{2}}}\left( V_{k_{i,t}}(\gamma)D_{ k_{i,t}}(\gamma) \frac{r^{\prime}(V_{k_{i,t}}(\gamma))}{r(V_{k_{i,t}}(\gamma))(1-r(V_{k_{i,t}}(\gamma)))} -V_{k_{i,t}}(\gamma)\frac{r^{\prime}(V_{k_{i,t}}(\gamma))}{1-r(V_{k_{i,t}}(\gamma))} \right)\mathbb{I}(\tau_{i}\geq t).
\end{equation}
For notational convenience, we will drop the subscript from $k_{i,t}$, that is, we will write $\Gamma_{k}(\gamma)$, $\mathcal{D}_{k}(\gamma) $ and $\mathcal{G}_{k}(\gamma) $ instead of $\Gamma_{k_{i,t}}(\gamma)$, $\mathcal{D}_{k_{i,t}}(\gamma)$ and $\mathcal{G}_{k_{i,t}}(\gamma)$, and $V_{k}(\gamma)$, $D_{k}(\gamma)$ instead of $V_{k_{i,t}}(\gamma)$ and $D_{k_{i,t}}(\gamma)$, respectively. Then,
\begin{equation}
    S_{\gamma,k(\gamma)} = \sum_{k=0}^{k(\gamma)-1}\Gamma_{k+1}(\gamma) = \gamma^{-\frac{\zeta+\delta-1}{2}}\Psi_{\gamma}(\beta).
\end{equation}
 Define the following filtrations recursively. Let
\begin{equation}\label{eqn:filterbase}
    \mathcal{F}_{0}(\gamma) = \sigma(V_{1,0})
\end{equation}
and 
\begin{equation}\label{eqn:filterrecursion}
    \mathcal{F}_{k+1}(\gamma) = \sigma(\mathcal{F}_{k}(\gamma) \cup \mathcal{D}_{k}(\gamma) \cup \mathcal{G}_{k+1}(\gamma)).
\end{equation}
Since $E(D_k(\gamma)\vert\mathcal{F}_{k}) = p(
V_k(\gamma))\mathbb{I}(\tau_i\geq t)$, and $p(V_k(\gamma))$ is $\mathcal{F}_{k}(\gamma)$ measurable,
\begin{equation}\label{eqn:martingaleCLT}
    E(\Gamma_{k+1}(\gamma)\vert \mathcal{F}_{k}(\gamma) ) = 0\textrm{ a.s.}, 
\end{equation}
and $\Gamma_{k}(\gamma)$ is $\mathcal{F}_{k}(\gamma)$ measurable. Then, $\{\Gamma_{k}(\gamma)\}_{k}$ is a martingale difference array with respect to the filtration $\mathcal{F}_{k}(\gamma)$. Next, note that from \eqref{eqn:MartingaleCLT}, $\Gamma_{k+1}(\gamma)\Gamma_{k+1}^{\intercal}(\gamma)$ can be written as
\begin{equation}\label{eqn:MLE-critSq}
\Gamma_{k+1}(\gamma)\Gamma_{k+1}^{\intercal}(\gamma)= \frac{1}{\gamma^{-(\delta + \zeta -1)}} V_{k}(\gamma)V_{k}^{\intercal}(\gamma)\left(D_{k}(\gamma)\frac{r^{\prime}(V_{k}(\gamma))}{r(V_{k}(\gamma))(1-r(V_k(\gamma)))} -\frac{r^{\prime}(V_{k}(\gamma))}{1-r(V_{k}(\gamma))}\right)^{2}\mathbb{I}(\tau_i\geq t).    
\end{equation}
From \eqref{eqn:Prob-der},
\[
\frac{r^{\prime}(V_{k}(\gamma))}{1-r(V_k(\gamma))} = \gamma\exp(\beta^{\intercal}V_{k})(1+o_{\mathcal{P}}(1)).
\]
Then, the scalar square term in \eqref{eqn:MLE-critSq} may be further expanded as
\[
D_{k}(\gamma) \left(\frac{r^{\prime}(V_{k}(\gamma)))}{r(V_{k}(\gamma))(1-r(V_k(\gamma)))}\right)^{2} +o_{\mathcal{P}}(\gamma).
\]
From \eqref{eqn:Prob-der},
\[
\left(\frac{r^{\prime}(V_{k}(\gamma)))}{r(V_{k}(\gamma))(1-r(V_k(\gamma)))}\right)^{2} = 1+o_{\mathcal{P}}(1).
\]
Simplifying further, the conditional expectation of \eqref{eqn:MLE-critSq} with respect to $\mathcal{F}_{k}(\gamma)$ equals
\[
\frac{1}{\gamma^{-(\zeta+\delta-1)}}V_{k}(\gamma)V_{k}^{\intercal}(\gamma)\left( r(V_k(\gamma)) + o_{\mathcal{P}}(\gamma)\right)\mathbb{I}(\tau_i\geq t).
\]
But then, 
\begin{equation}\label{eqn:CLT-asymptotic}
    \sum_{i=1}^{m(\gamma)}\sum_{t=0}^{T(\gamma)-1} E\left(\Gamma_{k}(\gamma)\Gamma_{k}^{\intercal}(\gamma)\vert\mathcal{F}_{k-1}(\gamma)\right) \xrightarrow[]{\mathcal{P}} \Lambda,
\end{equation}
where $\Lambda = E\left(V_{i,t}V_{i,t}^{\intercal}{\exp(\beta^{\intercal}V_{i,t})}\right)$. Now, by Lemma~\ref{rem:Vectors} and Lemma \ref{lemma:LindebergCLT} stated below, 
\begin{equation}\label{eqn:AsymCLT}
 \gamma^{-\frac{1}{2}(\zeta+\delta-1)}\left(\beta-\hat{\beta}_{M}(\gamma)\right) \xrightarrow[]{\mathcal{D}} N(0,(\dot{\Psi}^{-1}(\beta))^{\intercal}\Lambda\dot{\Psi}^{-1}(\beta)).
\end{equation}
\begin{lemma}\label{lemma:LindebergCLT}
For every $\kappa >0$
\begin{equation}\label{eqn:Lindeberg}
    \sum_{k=1}^{k_{\gamma}} E\left(\|\Gamma_{k}(\gamma)\|^{2}_{2}\mathbb{I}(\|\Gamma_{k}(\gamma)\|_{2}  > \kappa)\big\vert\mathcal{F}_{k-1}(\gamma)\right) \xrightarrow[\gamma\to 0]{\mathcal{P}} 0.
\end{equation}
\end{lemma}
The covariance matrix of the RHS of (\ref{eqn:AsymCLT}) can be calculated by noting that both $\Lambda$ as well as $\dot{\Psi}(\beta)$ are equal to $E(V_{i}V_{i}^{\intercal}\exp(\beta^{\intercal}V_{i}))$, and that $\dot{\Psi}(\beta)$ is symmetric. Then, the covariance becomes $(E(V_{i}V_{i}^{\intercal}\exp(\beta^{\intercal}V_{i})))^{-1}= (\Sigma + \Sigma\beta\beta^{\intercal}\Sigma)^{-1}\exp(-\frac{1}{2}\beta^{\intercal}\Sigma\beta)$, while $(\Sigma+\Sigma\beta\beta^{\intercal}\Sigma)^{-1}$ exists as $\Sigma$ is assumed to be positive definite. \qed
To prove the second part of the theorem, recall that $X_{\gamma} \triangleq \gamma^{-\frac{1}{2}(\zeta+\delta-1)}\left(\beta-\hat{\beta}_{M}(\gamma)\right)$. By (\ref{eqn:AsymCLT}), $X_{\gamma}$ is relatively compact. Thus, by Prokhorov's Theorem (see Billingsley (1999), Theorem 5.2), it is tight. Hence for every $\epsilon>0$, there exists an $M$ large enough such that $P(\|X_{\gamma}\|_{2}^{2} > M)<\epsilon$. Hence $\|X_{\gamma}\|^{2}_{2}=O_{\mathcal{P}}(1)$ or $\|\beta-\hat{\beta}_{M}(\gamma)\|_{2}^{2}=O_{\mathcal{P}}(\gamma^{\delta+\zeta-1})$.
\subsection{Verifying consistency of the estimator}\label{sec:Consistent}
In order to verify consistency of the estimators $\hat\beta_M(\gamma)$, we use the following characterisation from van der Vaart and Wellner (1996):\\
\textbf{Existence of consistent estimators:}
Let $\Psi_{\gamma}(\cdot)$ be a sequence of random functions which converge point-wise in probability to a deterministic function, $\Psi(\cdot)$. Let the following hold:
\begin{enumerate}
    \item The zero of $\Psi(\cdot)$, $\theta_{0}$ is well separated in the sense that 
    \[
    \inf_{\theta\not\in G}\|\Psi(\theta_0)\| >0,
    \]
    for every open $G$ that contains $\theta_0$.
    \item Uniform convergence in probability holds, that is, $\sup_{\theta\in\Theta}\|\Psi(\theta) - \Psi_{\gamma}(\theta)\|=o_{\mathcal{P}}(1)$.
\end{enumerate}
Then, any sequence $\hat\theta_\gamma$, such that $\Psi_{\gamma}(\hat\theta_\gamma) = o_{\mathcal{P}}(1)$ satisfies $\theta_{\gamma}\xrightarrow{\mathcal{P}}\theta_0$.\\

In our set-up, point-wise convergence follows from the WLLN. To see the separation of the zero  of $\Psi(\cdot)$ at $\theta_0$, recall that $\Psi(\cdot)$
was the derivative of a strictly convex function, and hence has a well separated zero. However, verifying uniform convergence is not straightforward. To this end we use the following characterisation from Newey (1991):\\
\textbf{Characterisation of uniform convergence in probability (Newey, 1991 - Corollary 2.2)}
Let $M_{\gamma}\to\ M$ point-wise in probability and $\Theta$ be compact. Further for some $B_{\gamma} = O_{\mathcal{P}}(1)$, 
\[
\|M_{\gamma}(\theta) -M_{\gamma}(\theta^{'})\|\leq B_{\gamma} d(\theta,\theta^{'}).
\]
Then, $\sup_{\theta\in\Theta}\|M(\theta) - M_{\gamma}(\theta)\|=o_{\mathcal{P}}(1)$.\\
From \eqref{eqn:Prob-der}, for any $\theta$,
\[
r^{\prime}(V_{i,t}) = \gamma\exp(\theta^{\intercal}V_{i,t})(1+\gamma h(\theta^\intercal V_{i,t}-\alpha(\gamma)) + \gamma h^{'}(\theta^\intercal V_{i,t}-\alpha(\gamma)))
\] 
We now re-write $\Psi_{\gamma}(\theta) = M_{\gamma}(\theta) + (\Psi_{\gamma}-M_{\gamma})(\theta)$, where
\[
M_{\gamma}(\theta) = \frac{1}{\gamma T(\gamma)m(\gamma)} \sum_{i,t} \left(V_{i,t}D_{i,t+1} - \gamma V_{i,t}\exp(\theta^{\intercal}V_{i,t})\right).
\]
Notice that we have the point-wise convergence in probability, $M_{\gamma}(\theta) \to \Psi(\theta)$. From the Mean Value Theorem,
\[
\|M_{\gamma}(\theta) - M_{\gamma}(\theta^{'})\| \leq \left(\frac{1}{T(\gamma)m(\gamma)}\sum_{i,t} \exp(M\|V_{i,t}\|)\right) \|\theta-\theta^{'}\| = B_{\gamma} \|\theta-\theta^{'}\|,
\]
where $M=\sup_{\theta\in\Theta}\|\theta\|<\infty$. Now, from the WLLN, $B_{\gamma}=O_{\mathcal{P}}(1)$, and thus, we have 
\[
\sup_{\theta}\|M_{\gamma}(\theta) - \Psi(\theta)\| = o_{\mathcal{P}}(1).
\]
Finally, observe that for all $\theta\in\Theta$, $(\Psi_{\gamma}-M_{\gamma})(\theta)$ consists of all the residual terms in the expression \eqref{psi_n2}, which are $O_{\mathcal{P}}(\gamma)$. Combining these, we have
\[
\sup_{\theta\in\Theta}\|\Psi(\theta) -\Psi_\gamma(\theta)\| = o_{\mathcal{P}}(1) + O_{\mathcal{P}}(\gamma),
\]
which gives the uniform convergence in probability, and hence the consistency of $\hat\theta_\gamma$.

\subsection{{Proof of Proposition~\ref{prop:varianceratio}}}

In the one dimensional setting, with $V_{i} \sim \mathrm{N}(0,1)$, from Theorem~\ref{thm:CLT}, for the MLE,
\begin{equation}\label{eqn:MLEerror}
    \gamma^{-(\delta +\zeta -1)}E(\|\beta - \hat{\beta}_{M}(\gamma)\|_{2}^{2}) = \frac{1}{(1+\beta^{2})}\exp\left(-\frac{\beta^{2}}{2}\right) + o(1).
\end{equation}
While $\delta<1$, the dominant rate term in $E(\|\hat{\beta}(\gamma)-\beta\|_{2}^{2})$ is $K_{1}\gamma^{\zeta+\delta-1}$. From~\ref{EC:Proofs}, $K_{1}$ is the sum
\begin{align}
\frac{1}{\gamma^{2}T^{2}(\gamma)m^{2}(\gamma)}E\sum_{i=1}^{m(\gamma)}\sum_{t=0}^{T(\gamma)-1}&\frac{(E\hat{V}_{\gamma})^{2}}{(E\hat{D}_{\gamma})^{4}}\left( D_{i,t+1}-ED_{i,t+1}\vert\mathcal{F}_{t}\right)^{2} +\frac{1}{(E\hat{D}_{\gamma})^{2}} \left( V_{i,t}D_{i,t+1}-EV_{i,t}D_{i,t+1}\vert\mathcal{F}_{t}\right)^{2}\nonumber\\
&-2\frac{(E\hat{V}_{\gamma})}{(E\hat{D}_{\gamma})^{3}} \left( D_{i,t+1}-ED_{i,t+1}\vert\mathcal{F}_{t}\right)\left( V_{i,t}D_{i,t+1}-EV_{i,t}D_{i,t+1}\vert\mathcal{F}_{t}\right).\label{eqn:sumA4}
\end{align}
We explicitly evaluate the first sum in (\ref{eqn:sumA4}).
\begin{align*}
E\left( D_{i,t+1}-ED_{i,t+1}\vert\mathcal{F}_{t}\right)^{2}  &= E\left(E\left( D_{i,t+1}-ED_{i,t+1}\vert\mathcal{F}_{t}\right)^{2}\vert \mathcal{F}_{t}\right)\\
& =E\left(E \left(D_{i,t+1}^{2} -2D_{i,t+1}ED_{i,t+1}\vert\mathcal{F}_{t} + (ED_{i,t+1}\vert\mathcal{F}_{t})^{2} \right)\vert\mathcal{F}_{t}\right)\\
& =ED_{i,t+1} + O(\gamma^{2}) = \gamma \exp\left(\frac{\beta^{2}}{2}\right) + O(\gamma^{2}).
\end{align*}
Further, we have seen that (see the proof of Lemma~\ref{1Dbound})
\begin{align*}
E\hat{V}_{\gamma} = \beta E\hat{D}_{\gamma} + O(\gamma).
\end{align*}
Then, summing  over all $i,t$, the first sum of (\ref{eqn:sumA4}) becomes $\gamma^{\delta + \zeta-1} \beta^{2} \exp\left(-\frac{\beta^{2}}{2}\right)  + o(\gamma^{\delta+\zeta-1})$. Similarly, the second and third sums can be seen to equal
\[
\gamma^{\delta + \zeta-1} (1+\beta^{2}) \exp\left(-\frac{\beta^{2}}{2}\right)  + o(\gamma^{\delta+\zeta-1})
\]
and
\[
-2\gamma^{\delta + \zeta-1} \beta^{2} \exp\left(-\frac{\beta^{2}}{2}\right)  + o(\gamma^{\delta+\zeta-1}).
\]
respectively. Then (\ref{eqn:sumA4}) is equal to
\begin{equation}\label{eqn:laststep}
 \gamma^{\delta + \zeta-1} \exp\left(-\frac{\beta^{2}}{2}\right) + o(\gamma^{\delta+\zeta-1}).   
\end{equation}
Now, taking the ratio of (\ref{eqn:MLEerror}) and (\ref{eqn:laststep}) and then the limit completes the proof.
\qed
\section{Proof of Theorem~\ref{thm:Mis-spec} and Lemma~\ref{lemma:Bias-MLE}}\label{EC:Misp}
\subsection{Proof of Lemma~\ref{lemma:Bias-MLE}}

To see this, observe that here, finding the MLE amounts to solving
\begin{equation}\label{eqn:MLE-Corrupt}
    \max_{b,a} \left( \sum_{i\leq m(\gamma),t\leq \tau_i(\gamma)} D_{i,t+1} \log{p(V_{i,t}^c,b,a)} + \sum_{i\leq m(\gamma),t \leq \tau_i(\gamma)}(1-D_{i,t+1})\log{(1-p(V_{i,t}^c),b,a))} \right).
\end{equation}

Recall that here $p(V_{i,t}^c,b,a) = q(b^\intercal V_{i,t}-a)$,
where $q(\cdot)$ is defined in Section~\ref{sec:MAP}. Applying the first order conditions to \eqref{eqn:MLE-Corrupt} yields
\begin{align}
     \frac{1}{\tilde{p}(\gamma) T(\gamma)m(\gamma)}\sum_{i \leq m(\gamma), t=1}^{ \tau_i(\gamma)}  \frac{V_{i,t}^c q^{\prime}(V_{i,t}^c)}
{q(V_{i,t}^c)(1-q(V_{i,t}^c))}
   D_{i,t+1}
&=  \frac{1}{\tilde{p}(\gamma) T(\gamma)m(\gamma)} \sum_{i \leq m(\gamma), t=1}^{ \tau_{i}(\gamma)}V_{i,t}\frac{q^{\prime}(V_{i,t}^c)}{1-q(V_{i,t}^c)}\label{eqn:MLE-Noise-1}\\
\frac{1}{\tilde{p}(\gamma) T(\gamma)m(\gamma)}\sum_{i \leq m(\gamma), t=1}^{ \tau_i(\gamma)}  \frac{ q^{\prime}(V_{i,t}^c)}
{q(V_{i,t}^c)(1-q(V_{i,t}^c))}
   D_{i,t+1}
&=  \frac{1}{\tilde{p}(\gamma) T(\gamma)m(\gamma)} \sum_{i \leq m(\gamma), t=1}^{ \tau_{i}(\gamma)}\frac{q^{\prime}(V_{i,t}^c)}{1-q(V_{i,t}^c)}\label{eqn:MLE-Noise-2},
\end{align}
where, $q(V_{i,t})$ and $q^{\prime}(V_{i,t})$ respectively denote $q(\hat\beta_{1,M}^\intercal(\gamma)V_{i,t}^c-\hat\alpha_{c,M}(\gamma))$ and  $q^{\prime}(\hat\beta_{1,M}^\intercal(\gamma)V_{i,t}^c-\hat\alpha_{c,M}(\gamma))$. Since defaults are rare,
\[
\frac{1}{\tilde{p}(\gamma)T(\gamma)m(\gamma)} \sum_{i,t} V_{i,t}^c D_{i,t+1} =\frac{1}{\tilde{p}(\gamma)T(\gamma)m(\gamma)}\sum_{i,t} V_{i,t}^c \exp(\hat{\beta}_{c,M}^\intercal(
\gamma) V_{i,t}^c -\hat{\alpha}_{c,M}(
\gamma)))+o_{\mathcal{P}}(1)
\]
and 
\[
\frac{1}{\tilde{p}(\gamma)T(\gamma)m(\gamma)} \sum_{i,t} D_{i,t+1}  =\frac{1}{\tilde{p}(\gamma)T(\gamma)m(\gamma)} \sum_{i,t} \exp(\hat{\beta}_{c,M}^\intercal(
\gamma) V_{i,t}^c -\hat{\alpha}_{c,M}(
\gamma))+o_{\mathcal{P}}(1),
\]
where $\hat{\beta}_{c,M}(\gamma)$ and $\hat{\alpha}_{c,M}(\gamma)$ solve \eqref{eqn:MLE-Noise-1} and \eqref{eqn:MLE-Noise-2}. Suppose $V_{i,t}\sim N(0,\mathbf{I})$ and $A_{i,t}\sim N(0,c\mathbf{I})$ for some $c>0$. Hence, $V_{i,t}^c\sim N(0,(1+c^2)\mathbf{I})$. Observe that by the Law of Large Numbers,
\begin{equation}\label{eqn:Noise-imp}
    \beta = \hat\beta_{c,M}(\gamma)(1+c^2)\exp\left(-\frac{1}{2}\|\beta\|^2\right)\cdot\exp\left(\frac{\|\hat{\beta}_{c,M}(\gamma)\|^2(1+c^2)}{2} - (\hat\alpha_{c,M}(\gamma) -\alpha(\gamma)) \right) + o_{\mathcal{P}}(1).
\end{equation}
and
\begin{equation}\label{eqn:noise-imp-0}
\exp\left( -\frac{1}{2}\|\beta\|^2\right)\cdot\exp\left(\frac{\|\hat{\beta}_{c,M}(\gamma)\|^2(1+c^2)}{2} - (\hat\alpha_{c,M}(\gamma) -\alpha(\gamma)) \right)= 1+o_{\mathcal{P}}(1).    
\end{equation}
Here, the LHS of \eqref{eqn:Noise-imp} follows from the fact that $A_{i,t}$ is independent of $V_{i,t}$ and $D_{i,t+1}$, and hence, $E(V_{i,t}^cD_{i,t+1}) = E(V_{i,t}D_{i,t+1})$. From \eqref{eqn:noise-imp-0}, 
$\beta = \hat{\beta}_{c,M}(\gamma)(1+c^2) +o_{\mathcal{P}}(1)$, that is $\|\beta-\hat\beta_{c,M}(\gamma)\| \xrightarrow{P} \frac{c^2\|\beta\|}{1+c^2} \neq 0$.

\subsection{Proof of Theorem~\ref{thm:Mis-spec}}

Let $q(Y_{1,t})$ be as defined in Section 3.7, and let $q^{\prime}(Y_{1,t})$ be its derivative.
For $(\theta,\xi)\in\Re\times\Re$, define the sequence of random functions, $M^1_{\gamma}(\cdot):\Re^2\to\Re^2$:
\begin{equation}\label{eqn:Random-mis-spec2}
   M^{1}_\gamma(\theta,\xi) = \left( \begin{array}{c}
     \frac{1}{\tilde{p}(\gamma) T(\gamma)m(\gamma)}\sum_{i \leq m(\gamma), t=1}^{ \tau_i(\gamma)} \left( Y_{1,t} D_{i,t+1} -Y_{1,t} {q^{\prime}(\vartheta Y_{1,t} -\xi)} \right), \\
     \frac{1}{\tilde{p}(\gamma) T(\gamma)m(\gamma)}\sum_{i \leq m(\gamma), t=1}^{ \tau_i(\gamma)} \left( D_{i,t+1} - {q^{\prime}(\vartheta Y_{1,t}-\xi)} \right)
\end{array}\right)
\end{equation}
and define  the deterministic function, $M(\cdot):\Re^2\to\Re^2$:
\begin{equation}\label{eqn:Limit-mis-spec}
    M(\vartheta,\xi) = \left( \begin{array}{c}
    (\beta_1+ \rho \beta_2)  - \vartheta \exp \left ( \frac{1}{2}\vartheta^2  - \xi\right)\cdot\exp \left ( -\frac{1}{2}(\beta_1^2 + 2 \rho \beta_1 \beta_2 + \beta_2^2) \right ),   \\
     1-\exp \left ( \frac{1}{2}\vartheta^2  - \xi\right) \cdot    \exp \left ( -\frac{1}{2}(\beta_1^2 + 2 \rho \beta_1 \beta_2 + \beta_2^2) \right  ).
\end{array}\right)
\end{equation}
Observe that the difference between the MLE Equations \eqref{logit:0011aa} and \eqref{logit:0011ab}, and $M_\gamma^{1}$ is due to the asymptotically smaller terms in default probability and is hence $o_{\mathcal{P}}(1)$. Let $(\hat\beta_{1,M}(\gamma),\hat\alpha_M(\gamma))$ be a zero to the MLE equations. Then, $M^{1}_\gamma( \hat\beta_{1,M}(\gamma),\hat\alpha_M(\gamma)) =o_{\mathcal{P}}(1)$. Define
\begin{equation}\label{eqn:Random-mis-spec}
   M_\gamma(\vartheta,\xi) = \left( \begin{array}{c}
       \frac{1}{\tilde{p}(\gamma) T(\gamma)m(\gamma)}\sum_{i \leq m(\gamma), t=1}^{ \tau_i(\gamma)}  Y_{1,t} D_{i,t+1}- \frac{1}{ c_{\beta}T(\gamma)m(\gamma)} \sum_{i \leq m(\gamma), t=1}^{ \tau_{i}(\gamma)}Y_{1,t}\exp(\vartheta Y_{1,t} - \xi), \\
       \frac{1}{\tilde{p}(\gamma) T(\gamma)m(\gamma)}\sum_{i \leq m(\gamma), t=1}^{ \tau_i(\gamma)}  
D_{i,t+1}-\frac{1}{c_{\beta} T(\gamma)m(\gamma)} \sum_{i \leq m(\gamma), t=1}^{ \tau_{i}(\gamma)} \exp(\vartheta Y_{1,t} - \xi)
 \end{array}\right)
\end{equation}
where $c_{\beta} = \exp \left ( \frac{1}{2}(\beta_1^2 + 2 \rho \beta_1 \beta_2 + \beta_2^2) \right  )$ and let $\hat{\xi}_M(\gamma) =\hat\alpha_M(\gamma)-\alpha(\gamma)$. Since $\tilde{p}(\gamma) = c_{\beta} \exp(-\alpha(\gamma))(1+o(\gamma))$, if $(\hat\beta_{1,M}(\gamma),\hat\alpha_M(\gamma))$ is an approximate zero to $M_\gamma^{1}(\cdot)$, $(\hat\beta_{1,M}(\gamma),\hat\xi_M)$ is an approximate zero of $M_{\gamma}(\cdot)$. Now, $\frac{\tilde{p}(\gamma)}{\gamma} \to \exp \left ( \frac{1}{2}(\beta_1^2 + 2 \rho \beta_1 \beta_2 + \beta_2^2) \right  )$. Then, by the WLLN,
\[
\frac{1}{\tilde{p}(\gamma) T(\gamma)m(\gamma)}\sum_{i \leq m(\gamma), t=1}^{ \tau_i(\gamma)}  Y_{1,t} D_{i,t+1} \xrightarrow{\mathcal{P}} (\beta_1+\rho\beta_2) \textrm{ \ \ and \ \ } \frac{1}{\tilde{p}(\gamma) T(\gamma)m(\gamma)}\sum_{i \leq m(\gamma), t=1}^{ \tau_i(\gamma)}  D_{i,t+1} \xrightarrow{\mathcal{P}} 1
\]
The point-wise convergence, $M_\gamma(\cdot)  \xrightarrow{\mathcal{P}} M(\cdot)$  on $\Re^2$ follows. Notice that, $M(\mathbf{u})=0$ has a solution given by 
\begin{align*}
\theta^* & =\beta_1+\rho\beta_2\\
\xi^*&=\frac{\beta_2^2(1-\rho^2)}{2}, 
\end{align*}
Further, outside any open set $G$ containing $(\theta^*,\xi^*)$, it is easy to see that $M(\cdot)$ is not zero. Hence, with $d(\cdot,\cdot)$ as the Euclidean distance on $\Re^2$, the identifiablity condition, 
\begin{equation}\label{eqn:identifiable}
\inf_{(\theta,\xi):d((\theta,\xi),(\theta^*,\xi^*)) \geq \delta} \|M(\theta,\xi)\|_2 >0
\end{equation}
is satisfied for all $\delta>0$. Finally, uniform convergence in probability of $M_\gamma(\cdot)$ to $M(\cdot)$ follows using the arguments from Section~\ref{sec:Consistent}.

Since $(\hat\beta_{1,M},\hat{\xi}_M)$ is an approximate zero to $M_\gamma(\cdot)$, and $(\theta^*,\xi^*)$ is the zero to $M(\cdot)$,
\[
(\hat\beta_{1,M}, \hat\xi_{M})\xrightarrow{\mathcal{P}} (\theta^*,\xi^*).
\]
or $\hat{\beta}_{1,M} \to \beta_1+\rho\beta_2$ and $\alpha(\gamma)-\hat{\alpha}_M(\gamma) \xrightarrow{\mathcal{P}} \frac{\beta_2^2(1-\rho^2)}{2}$.\\

\noindent To see that the proposed estimator also converges to the same values, recall that by the law of large numbers,
\begin{equation}\label{eqn:Misp-1}
    \frac{1}{\tilde{p}(\gamma) T(\gamma)m(\gamma)} \sum_{i\leq m(\gamma),t\leq T(\gamma)} D_{i,t+1}Y_{1,t} \xrightarrow{\mathcal{P}}(\beta_1+\rho\beta_2)
\end{equation}
and 
\begin{equation}\label{eqn:Misp-2}
    \frac{1}{\tilde{p}(\gamma) T(\gamma)m(\gamma)} \sum_{i\leq m(\gamma),t\leq T(\gamma)}D_{i,t+1} \xrightarrow{\mathcal{P}}1.
\end{equation}
Taking the ratio of \eqref{eqn:Misp-1} and\eqref{eqn:Misp-2} completes the proof.
\qed
\section{High-Dimensional simple approximate MLE}\label{EC:HD}
\color{blue}
\color{black}
To keep the discussion simple, we assume that the only source of exits are defaults. Further, the number of covariates, $d(\gamma)\to\infty$ as $\gamma\to 0$. The conditional probability of default given covariate information and survival till time $t$ is 
\begin{equation}\label{eqn:High-dimensional-defprob}
   E(D_{i,t+1} \ \vert \ \mathcal{F}_{t})= p(\gamma,V_{i,t}) = e^{\frac{\beta^{\intercal}V_{i,t}}{\sqrt{d(\gamma)}} - \alpha(\gamma)}(1+H(\gamma,V_{i,t})),
\end{equation}
where $\vert H(\gamma,V_{i,t})\vert \leq c\gamma e^{\frac{\beta^{\intercal}V_{i,t}}{\sqrt{d(\gamma)}}}$.  For simplicity, $V_{i,t}$ is a $d(\gamma)$ dimensional i.i.d standard Gaussian process. As before, $\alpha(\gamma)=-c\log\gamma$ and $\beta\in\Re^{d(\gamma)}$ are parameters to be estimated. The normalisation in \eqref{eqn:High-dimensional-defprob} is selected so that the variance of $\frac{\beta^{\intercal}V_{i,t}}{\sqrt{d(\gamma)}}$ is finite, and the average default probability, given by $\exp\left(\frac{\|\beta\|_2^2}{d(\gamma)}-\alpha(\gamma)\right) + O(\gamma^2)$ remains small.

\subsection{Parameter estimation}
Recall the ML equations for the intensity model, which in this setting become (where $d$ is the number of dimensions)
\begin{equation}  \label{logit:001-HD}
\sum_{i \leq m, t=s_{i}}^{ \tau_{i}-1}  \frac{v_{i,t} e^{\frac{\hat{\beta}^{\intercal} v_{i,t}}{\sqrt{d}}  -\hat{\alpha}}}
{1-\exp(-e^{\frac{\hat{\beta}^{\intercal} v_{i,t}}{\sqrt{d}}-\hat{\alpha}})}
   d_{i,t+1}
=  \sum_{i \leq m, t=s_{i}}^{ \tau_i-1}v_{i,t} e^{\frac{\hat{\beta}^{\intercal} v_{i,t}}{\sqrt{d}}-\hat{\alpha}},
\end{equation}
and
\begin{equation}  \label{logit:003-HD}
\sum_{i\leq m, t=s_{i}}^{ \tau_{i}-1}  \frac{e^{\frac{\hat{\beta}^{\intercal} v_{i,t}}{\sqrt{d}}-\hat{\alpha}}}
{1-\exp(-e^{\frac{\hat{\beta}^{\intercal} v_{i,t}}{\sqrt{d}}-\hat{\alpha}})}
   d_{i,t+1}
=  \sum_{i \leq m, t=s_{i}}^{ \tau_{i}-1}e^{\frac{\hat{\beta}^{\intercal} v_{i,t}}{\sqrt{d}}-\hat{\alpha}}.
\end{equation}
Now, repeating the argument in Section~\ref{sec:MLE}, 
\begin{align*}
    {\sum_{i,t}}{v_{i,t}d_{i,t+1}} \approx \frac{\beta}{\sqrt{d}} \exp\left(\frac{\|\beta\|_2^2}{2d} -\alpha\right)\\
    \sum_{i,t} d_{i,t+1} \approx \exp\left(\frac{\|\beta\|_2^2}{2d} -\alpha\right).
\end{align*}
This gives the estimator for $\beta$ in the high dimensional setting:
\begin{equation}\label{eqn:High-dim-beta}
    \hat{\beta}(\gamma) = \sqrt{d(\gamma)} \frac{\hat{V}_{\gamma}}{\hat{D}_\gamma},
\end{equation}
where $\hat{V}_{\gamma}$ and $\hat{D}_{\gamma}$ are as defined before. To derive the accuracy of the estimator, we follow the template of the proof of Theorem~\ref{theorem:1}. Accordingly, write 
\begin{equation}\label{eqn:betastarHD}
    \beta^{*}_H(\gamma) = \sqrt{d(\gamma)}\frac{E\hat V_\gamma}{E\hat D_\gamma},
\end{equation}
and consider $\|\beta-\beta^{*}_H(\gamma)\|_2$. We outline the crucial steps of the analysis. First, recall that
\begin{align}
    E\hat{D}_{\gamma} &= \frac{1}{\gamma T(\gamma)}E\left(\sum_{k=1}^{T(\gamma)} (-1)^{k+1}\sum_{j_{k}=k-1}^{T(\gamma)-1} \cdots \sum_{j_{1}=0}^{i_{2}-1} p(\gamma,V_{i,j_{1}})\cdots p(\gamma,V_{i,j_{k}})\right)\label{eqn:bigeqn1-HD} \textrm{, and,} \\
    E\hat{V}_{\gamma} &= \frac{1}{\gamma T(\gamma)} E\left(\sum_{k=1}^{T(\gamma)} (-1)^{k+1}\sum_{j_{k}=k-1}^{T(\gamma)-1} \cdots \sum_{j_{1}=0}^{i_{2}-1} V_{i,j_{k}}p(\gamma,V_{i,j_{1}})\cdots  p(\gamma,V_{i,j_{k}})\right).\label{eqn:bigeqn2-HD}
\end{align}
Since we have assumed the covariates to be independent across time,  from \eqref{eqn:bigeqn1-HD} and \eqref{eqn:bigeqn2-HD} 
\begin{align}
     E\hat{D}_{\gamma} &= \frac{1}{\gamma T(\gamma)}\left(\sum_{k=1}^{T(\gamma)} (-1)^{k+1}c^k\gamma^k\sum_{j_{k}=k-1}^{T(\gamma)-1} \cdots \sum_{j_{1}=0}^{i_{2}-1} \exp(\frac{k\|\beta\|^2_2}{2d(\gamma)})  \right)+ O(\gamma)\nonumber \textrm{, and,}\\
     E\hat{V}_{\gamma} &= \frac{1}{\gamma T(\gamma)} \left(\sum_{k=1}^{T(\gamma)} (-1)^{k+1}c^k\gamma^k\sum_{j_{k}=k-1}^{T(\gamma)-1} \cdots \sum_{j_{1}=0}^{i_{2}-1} \frac{\beta}{\sqrt{d(\gamma)}}\exp(\frac{k\|\beta\|^2_2}{2d(\gamma)})\right)+ O(\gamma)\label{eqn:HD-Vgamma}.
\end{align}
Observe that $\frac{\|\beta\|_2^2}{d(\gamma)}=O(1)$. Then, similar to before, $E\hat D_\gamma = O(1)$, and
$\Big\Vert\left(\beta- \sqrt{d(\gamma)}\frac{E\hat V_\gamma}{E\hat D_\gamma}\right)\Big\Vert_2$ may be bounded by $O(\sqrt{d(\gamma)}\gamma)$. We hence have
\begin{equation}\label{eqn:bound-1-HD}
    \|\beta-\beta^{*}_H(\gamma)\|_2 = O(\gamma \sqrt{d(\gamma)}).
\end{equation}

Next, we bound $\|\hat\beta(\gamma) - \beta^{*}_H(\gamma)\|_2$. As before, consider the component-wise Taylor series:
\begin{equation}\label{eqn:TSE-HD}
\frac{\hat{V}_{\gamma}^{(1)}}{\hat{D}_{\gamma}}- \frac{E\hat{V}_{\gamma}^{(1)}}{E\hat{D}_{\gamma}} = \frac{1}{E\hat{D}_{\gamma}} (\hat{V}_{\gamma}^{(1)} -E\hat{V}_{\gamma}) - \frac{E\hat{V}_{\gamma}^{(1)}}{(E\hat{D}_{\gamma})^{2}}(\hat{D}_{\gamma} - E\hat{D}_{\gamma}) + R((\Hat{D}_{\gamma},\Hat{V}_{\gamma}^{(1)}), (E\Hat{D}_{\gamma},E\Hat{V}_{\gamma}^{(1)})). 
\end{equation}
First, consider the square of $(\hat{D}_{\gamma} - E\hat{D}_{\gamma})$. This may be handled similar to before (see \eqref{var} onward), to give an error of $O_{\mathcal{P}}(\gamma^{\delta+\zeta-1})+O_{\mathcal{P}}(\gamma^\zeta)$. Similarly, $(\hat{V}^{(1)}_{\gamma} - E\hat{V}^{(1)}_{\gamma})^2$ is $\frac{1}{d(\gamma)}\left( O_{\mathcal{P}}(\gamma^{\delta+\zeta-1}) + O_{\mathcal{P}}(\gamma^\zeta)\right)$, and from \eqref{eqn:HD-Vgamma}, $E\hat V^{(1)}_\gamma=O(\sqrt{d^{-1}(\gamma)})$. Now, observe that the first component of $\hat\beta(\gamma) - \beta^{*}_H(\gamma)$ is $\sqrt{d(\gamma)}$ times the LHS of \eqref{eqn:TSE-HD}. Hence, we have 
\[
\Big\vert\hat\beta(\gamma) - \beta^{*}_H(\gamma)\Big\vert^2 =d^{2}(\gamma)( O_{\mathcal{P}}(\gamma^{\delta+\zeta-1}) + O_{\mathcal{P}}(\gamma^\zeta)),
\]
where $ \vert\cdot\vert$ denotes the Euclidean-1 norm. Recall the basic inequality, for all $X\in\Re^d$, $\|x\|_2\leq \vert x\vert$. Then,
\begin{proposition}\label{prop:HD}
Consider a default generating model given by \eqref{eqn:High-dimensional-defprob}, and let $\hat{\beta}(\gamma)$ be given by \eqref{eqn:High-dim-beta}. Then $\|\hat\beta (\gamma) -\beta\|_{2}^{2}=(O_{\mathcal{P}}(\gamma^{\delta+\zeta-1})+O_{\mathcal{P}}(\gamma^{\zeta}))d^2(\gamma)$.
\end{proposition}
\begin{remark}
Proposition~\ref{prop:HD} shows that for the error to be vanishing, one requires $d(\gamma)=o(\gamma^{-\frac{1}{2} (\delta+\zeta-1)})+ o(\gamma^{-\frac{1}{2}\zeta})$. In practice, one typically uses few (less than 20) covariates for parameter estimation (see Duan et al. 2012 and Duan and Fulop, 2013). Hence, $d(\gamma)=O(\log\gamma)$ appears to be a reasonable choice for the dimensionality of the problem, and the error vanishes. The intercept $\alpha(\gamma)$ is estimated by 
\[
\hat\alpha(\gamma) =\log\left( \frac{\sum_{i=1}^{m(\gamma)}\sum_{t=0}^{T(\gamma)-1}\mathrm{e}^{\frac{1}{\sqrt{d(\gamma)}}\hat{\beta}^{\intercal}(\gamma) V_{i,t}}\mathbb{I}(\tau_{i} \geq t)}{\sum_{i=1}^{m(\gamma)}\sum_{t=0}^{T(\gamma)-1}D_{i,t+1}}\right),
\]
where $\hat\beta(\gamma)$ is now defined by \eqref{eqn:High-dim-beta}. Similar to the proof of Theorem~\ref{theorem:1}, \eqref{eqnlemma2}, one can show that $\|\hat\alpha(\gamma)-\alpha(\gamma)\|_2=O_{\mathcal{P}}(\gamma^{\frac{1}{2}(\delta+\zeta-1)})+ O_{\mathcal{P}}(\gamma^{\frac{1}{2}\zeta})$, although we do not prove this here. Since solving the MLE in the high dimensional set-up can be especially challenging for the intensity model, the proposed estimator provides substantial computational benefits in this setting. 
\end{remark}
\ \\
In practice, in addition to a large number of covariates each of whose effect on  default is small, there may be a small number of covariates which have a significant impact on the default of a firm. We now modify the default generating model in \eqref{eqn:High-dimensional-defprob} to accommodate these. For $D<\infty$, let $V_{i,t} = (V_{i,t}^{H},V_{i,t}^{L}) \in\Re^{d(\gamma)+ D}$, $\beta_1\in\Re^{d(\gamma)}$, $\beta_2\in\Re^D$, and let 
\begin{equation}\label{eqn:Mixed-Cov}
    p(\gamma,V_{i,t}) = \exp\left(\frac{\beta_1^\intercal V_{i,t}^{H}}{\sqrt{d(\gamma)}} + \beta_2^\intercal V_{i,t}^{L} - \alpha(\gamma)\right)(1+H(\gamma,V_{i,t}))
\end{equation}
denote the conditional default probability. Here the covariates $V_{i,t}^{L}$ have a significant effect on the default. Let 
\begin{equation}\label{eqn:HD-Beta-2}
    \hat\beta_2(\gamma) = \frac{1}{\sum_{i,t}D_{i,t+1}} \left({\sqrt{d(\gamma)}}\sum_{i,t}V_{i,t}^{H} ,  \sum_{i,t}V_{i,t}^{L}\right).
\end{equation}
We obtain the following corollary to Proposition~\ref{prop:HD}:
\begin{corollary}
Consider a default generating model given by \eqref{eqn:Mixed-Cov}, and let $\hat\beta_2(\gamma)$, given by \eqref{eqn:HD-Beta-2}. Then,
\[
\|\hat\beta_2(\gamma) - (\beta_1,\beta_2)\|_2^2 = (O_{\mathcal{P}}(\gamma^{\delta+\zeta-1})+O_{\mathcal{P}}(\gamma^{\zeta}))d^2(\gamma).
\]
\end{corollary}
%


\begin{table}\centering
  \caption{High dimensional Approximate MLE
  \label{Reg-HD}}
  {\begin{tabular}{@{\extracolsep{20pt}} p{4cm} p{2cm} p{2cm} p{2cm} p{2cm}}
\\[-1.8ex]\hline
\hline \\[-1.8ex]
Dimensionality $d(\gamma)$ & 8& 12&	18& 24\\
\hline \\[-1.8ex]
$RMSE_\beta$& $0.335$& $0.494$&	$0.742$& $1.03 $ \\
$RMSE_\alpha$& $0.092 $& $0.093 $ & $0.099$& $0.104 $ \\
\hline \\[-1.8ex]
\end{tabular}}    
\begin{tablenotes}
\item \footnotesize {Defaults have been simulated according to \eqref{eqn:High-dimensional-defprob}. Here, the average default probability is kept at 1\%, number of firms of data is 5000, and the number of time periods of data is 200. Dimensionality of covariates is varied across experiments. To keep average default probability constant (to the first order) as the number of covariates vary, we select $\beta$ to be the all $1$s vector of size $d(\gamma)$, and  $\alpha=-7.5$ across all experiments.}
\end{tablenotes}

\end{table}
\section{Proof Outline for Proposition~\ref{lemma:MAP-Reg}}\label{sec:E.C.RegProof}
 In order to prove Proposition~\ref{lemma:MAP-Reg}, we first identify the limiting behaviour of the regularised MLE, \eqref{eqn:MAP-FOC-1} and \eqref{eqn:MAP-FOC-2}. Recall that the regularised MLE solves the $\mathbf{0}$ of
\begin{align} 
     \frac{1}{\tilde{p}(\gamma) T(\gamma)m(\gamma)}\sum_{i \leq m(\gamma), t=1}^{ \tau_i(\gamma)-1}  \frac{V_{i,t} q^{\prime}(V_{i,t})}
{q(V_{i,t})(1-q(V_{i,t}))}
   D_{i,t+1}
&=  \frac{1}{\tilde{p}(\gamma) T(\gamma)m(\gamma)} \sum_{i \leq m(\gamma), t=1}^{ \tau_{i}(\gamma)-1}V_{i,t}\frac{q^{\prime}(V_{i,t})}{1-q(V_{i,t})}+ \nabla_\beta u(\beta)\label{eqn:MAP-FOC-1-EC}\\
\frac{1}{\tilde{p}(\gamma) T(\gamma)m(\gamma)}\sum_{i \leq m(\gamma), t=1}^{ \tau_i(\gamma)-1}  \frac{ q^{\prime}(V_{i,t})}
{q(V_{i,t})(1-q(V_{i,t}))}
   D_{i,t+1}
&=  \frac{1}{\tilde{p}(\gamma) T(\gamma)m(\gamma)} \sum_{i \leq m(\gamma), t=1}^{ \tau_{i}(\gamma)-1}\frac{q^{\prime}(V_{i,t})}{1-q(V_{i,t})}\label{eqn:MAP-FOC-2-EC}.
\end{align}
Observe that 
\begin{equation*}
\frac{q^{\prime}(V_{i,t}) }{q(V_{i,t})(1-q(V_{i,t}))} = 1+O_{\mathcal{P}}(\gamma)    
\end{equation*}
and 
\begin{equation*}
\frac{q^{\prime}(V_{i,t})}{1-q(V_{i,t})} = \exp(\mathbf{\hat\beta}^\intercal V_{i,t}-\hat\alpha(\gamma)) (1+o_{\mathcal{P}}(1)).    
\end{equation*}
Next, observe that with $\hat{p}(\gamma)$ as the empirically observed default probability, we have 
\begin{lemma}\label{lemma:Emp--def--prob}
\[\frac{\hat{p}(\gamma)}{\tilde{p}(\gamma)}-1 = O_{\mathcal{P}}(\gamma^{\frac{1}{2}(\delta+\zeta-1)}) + O_{\mathcal{P}}(\gamma^{\frac{1}{2}\zeta}).\]
\end{lemma}
Recall that
\begin{align*}
    \hat V^R_\gamma&\triangleq\frac{1}{\hat{p}(\gamma)T(\gamma)m(\gamma)} \sum_{i,t}V_{i,t}D_{i,t+1}, \\
    \hat D^R_\gamma&\triangleq\frac{1}{\hat{p}(\gamma)T(\gamma)m(\gamma)} \sum_{i,t}D_{i,t+1}.
\end{align*}
Now, by WLLN and Lemma~\ref{lemma:Emp--def--prob},
\begin{align*}
    \hat V^R_\gamma &= \Sigma\hat\beta\exp\left(\frac{\|\hat\beta\|_2^2}{2} -(\alpha(\gamma)-\hat{\alpha}(\gamma))\right) + \nabla_\beta u(\hat\beta) +o_{\mathcal{P}}(1)\\
    \hat{D}^R_\gamma&= \exp\left(\frac{\|\hat\beta\|_2^2}{2}  -(\alpha(\gamma)-\hat{\alpha}(\gamma))\right)+o_{\mathcal{P}}(1).
\end{align*}

 Then,
\[\hat{V}^R_{\gamma} = \Sigma\beta\hat{D}^R_{\gamma} - \nabla_{\beta}u(\beta) + o_{\mathcal{P}}(1),\]

\textit{For simplicity, we will assume henceforth, that ridge regularisation is used, that is $u(\beta)=-\frac{\|\beta\|^2}{2}$, and that the covariance matrix of the covariates $\Sigma=\mathbf{I}$.} Let $\hat\beta_R(\gamma)$ and $\beta^*_R(\gamma)$ denote the approximate and true regularised MLE respectively. Notice that here, the empirical default probability, $\hat{p}(\gamma) = \frac{1}{T(\gamma)m(\gamma)}\sum_{i,t}D_{i,t+1}$, and hence $\hat D_\gamma =1$. 
Now, for the ridge regularised MLE, in the limit 
\begin{equation}\label{eqn:MAP-lt}
    \Sigma\beta  = \Sigma\hat\beta_\infty +\hat\beta_\infty.
\end{equation}
where $\hat\beta_\infty$ is the asymptotic solution to the regularised MLE. Further,
\[
\hat{V}_{\gamma}^{R} = \hat\beta(\gamma),
\]
where $\hat\beta(\gamma)$ is as defined in \eqref{eqn:hatbeta} with $\Sigma=\mathbf{I}$. Now, use the triangle inequality to get
\[
\|\hat{\beta}_R(\gamma)-{\beta}^*_R(\gamma)\|_2 \leq \|\hat{\beta}_R(\gamma)-\hat{\beta}_\infty\|_2 + \|\hat{\beta}_\infty-{\beta}^*_R(\gamma)\|_2.
\]
We proceed in two steps:\\
\textbf{Step 1:}
Here, we argue that $\|\hat{\beta}_R(\gamma)-\hat{\beta}_\infty\|_2 = O_\mathcal{P}(\gamma^{\frac{1}{2}(\delta+\zeta-1)})+O_{\mathcal{P}}(\gamma^{\frac{1}{2}\zeta})$. This proof works similar to that of Theorem~\ref{theorem:1}. First, recall that from \eqref{eqn:approx-L2reg}, here, $$\hat{\beta}_R(\gamma) = \frac {\hat{V}_\gamma}{2}.$$
Now, following the proof of Lemmas \ref{1Dbound} and \ref{lemma:keylem}, we get the required result.\\
\textbf{Step 2:} Here, we show that $\|\beta^*_R(\gamma)-\beta_\infty\|_2=O_{\mathcal{P}}(\gamma^{\frac{1}{2} (\delta+\zeta-1)})$. Following the proof of Theorem~\ref{thm:CLT} gives the desired convergence rate.\qed
\begin{remark}\label{rem:reg-int}
One can derive similar limits for the regularised intercept, $\alpha_R^*(\gamma)$. Recall that in the limit \eqref{eqn:MAP-FOC-2-EC} point-wise converges to $\mathrm{e}^{\frac{1}{2}(\|\mathbf{u}\|^2-\|\beta\|^2_2)-(v-\alpha)}-1$. Then, the regularised MLE for $\alpha$ satisfies
\[
\frac{1}{2}(\|\beta_R^*(\gamma)\|_2^2 - \|\beta\|_2^2) -(\alpha(\gamma) -\alpha_R^*(\gamma)) \to 0.
\]
Hence,
\[
(\hat\alpha_R(\gamma) -\alpha(\gamma)) \to -\frac{3}{8}\|\beta\|_2^2.
\]
\end{remark}
In particular, $\alpha(\gamma) - \hat\alpha_R(\gamma) = O_{\mathcal{P}}(1)$.
\section{Proofs of Intermidiate lemmas} \label{A4}

\textbf{Proof of Lemma~\ref{lemma:expVCV} :}\\
As is well known, if $X \in \Re^{d}$ is a Gaussian random vector with a covariance matrix $\Sigma = (\sigma_{i,j})$. Then for any $\mathbf{r} \in \Re^{d}$,
\begin{equation}\label{eqn:fact}
E\exp(\mathbf{r}^{\intercal}X ) = \exp\left(\frac{1}{2} \mathbf{r}^{\intercal} \Sigma \mathbf{r} \right)   =  \exp\left(\frac{1}{2} \sum_{i,j}r_{i}r_{j}\sigma_{i,j} \right).
\end{equation}
Let $Y= \left(
\begin{array}{c}
Y_{j_{1}}, Y_{j_{2}}, \ldots Y_{j_{k}}
\end{array}
\right)^{\intercal}$
and $\mathbf{e}_{k}$ be the all 1's vector of length $k$. Let $\Sigma_{Y} = \{\sigma_{Y}(m,n)\}$ be the variance covariance matrix of $Y$, where $\sigma_{Y}({m,n}) = EY_{j_{m}}Y_{j_{n}}$.
Note that we may re-write $\sum_{r=1}^{k}{Y_{j_{r}}}= \mathbf{e}_{k}^{\intercal} Y$.
Then, by (\ref{eqn:fact}),
\begin{equation}\label{eqn:int1}
E\left(\exp(\beta\mathbf{e}_{k}^{\intercal}Y)\right) = \exp\left(\frac{\beta^{2}}{2}\sum_{m,n}\sigma_{Y}({m,n})\right).
\end{equation}
The first part of Lemma~\ref{lemma:expVCV} now follows from definition of $\sigma_{i,j}$. Now, let $j_{1} < j_{2} \cdots <j_{k}$. Then,
Let $\mathbf{e}^{*} =(\beta\mathbf{e}_{k-1}, \beta_{k})$,
where $\beta\mathbf{e}_{k-1}$ denotes a $k-1$ dimensional column vector with each entry
equalling $\beta$.
Let $\mathcal{S}_{2}$ be the set of all pairwise combinations of $j_{1},j_{2},\ldots, j_{k-1}$. From (\ref{eqn:fact}),
\begin{equation}\label{eqn:int}
E\left(\exp(\mathbf{e^{*}}^{\intercal}Y)\right) =  \exp\left(\left(\frac{\beta^{2}}{2}(k-1)\sigma^{2}+\beta^{2}\sum_{i_{m},i_n\in \mathcal{S}_{2} : i_m>i_n} \sigma_{i_{m},i_{n}} +\beta_{k}\beta \sum_{r=1}^{k-1} \sigma_{j_{k},j_{r}} + \frac{1}{2}\beta_{k}^{2}\sigma^{2} \right)\right).
\end{equation}
Differentiating both sides of (\ref{eqn:int}) with respect to $\beta_{k}$, interchanging the derivative and the expectation using the dominated convergence theorem, and then setting $\beta_{k}$ to $\beta$, we get the desired result.
\qed\\

\noindent\textbf{Proof of Lemma~\ref{rem:Vectors}}
 Fix $\lambda \in \Re^{d}$ and consider the array of random variables $L_{n,k}= \lambda^{\intercal} X_{n,k}$. Note that since $X_{n,k}$ is a square integrable martingale array, so is $M_{n,k}$. Further,
 \[
 \sum_{i=1}^{k_n} E(L_{n,i}^{2}\vert\mathcal{F}_{n,i-1}) = \lambda^{\intercal}\left(\sum_{i=1}^{k_n} E(X_{n,i}X_{n,i}^{\intercal}\vert\mathcal{F}_{n,i-1})\right)\lambda = \lambda^{\intercal} M_{n,k_n} \lambda.
 \]
 Hence, by (\ref{eq:Vectorcv}) and the mapping theorem (see Billingsley, 1999), 
\begin{equation}\label{eqn:ssecvectors}
    \sum_{i=1}^{k_n} E(L_{n,i}^{2}\vert\mathcal{F}_{n,i-1}) \xrightarrow[]{\mathcal{P}} \lambda^{\intercal}\Lambda\lambda. 
\end{equation}
Further, since $|\lambda^{\intercal}X_{n,k}| \leq \|\lambda\|_{2}\|X_{n,k}\|_{2}$,
\begin{align*}
    \mathbb{I}\left( \vert\lambda^{\intercal}X_{n,k}\vert^{2} > \kappa \right) \leq \mathbb{I}\left( \|\lambda\|^{2}\|X_{n,k}\|^{2}> \kappa \right). 
\end{align*}
Then by (\ref{eqn:vectorCLT}),  
\[
0\leq E\left(\vert\lambda^{\intercal}X_{n,k}\vert^{2} \mathbb{I}\left( \vert\lambda^{\intercal}X_{n,k}\vert > \kappa \right)\vert\mathcal{F}_{n,k-1}\right) \leq \|\lambda\|^{2}E\left(\| X_{n,k}\|^{2} \mathbb{I}\left(\|X_{n,k}\|^{2}> \frac{\kappa}{ \|\lambda\|^{2}} \right)\vert\mathcal{F}_{n,k-1}\right) \xrightarrow[]{\mathcal{P}} 0,
\]
where the last inequality follows from the conditions of the lemma. Define $\mathbf{S}_{n,k} = \sum_{i=1}^{k}X_{n,k}$. Since $\lambda$ was arbitrary, by Theorem~\ref{thm:MgleCLT}, $\forall \lambda\in\Re^{d}$, $\lambda^{\intercal}\mathbf{S}_{n,k_{n}}$ converges in distribution to a $\mathrm{N}(0, \lambda^{\intercal}\Lambda\lambda)$ random variable. That is, for each $\lambda \in\Re^{d}/\{0\}$, $\lambda^{\intercal}\mathbf{S}_{n,k_{n}} \xrightarrow[]{\mathcal{D}} \lambda^{\intercal}\mathbf{V}$, where $\mathbf{V}$ is $\mathrm{N}(0,\Lambda)$. Since all linear combinations of  $\mathbf{S}_{n,k_n}$ converge to those of $\mathbf{V}$, the  Cramer-Wold theorem implies  that $\mathbf{S}_{n,k_n} \xrightarrow[]{D}\mathrm{N}(0,\Lambda)$. \qed\\

\noindent \textbf{Proof of Lemma~\ref{lemma:LindebergCLT}}
 Using the conditional Cauchy-Schwarz inequality, each term of the summation in (\ref{eqn:Lindeberg}) can be upper bounded by
\begin{equation}\label{eqn:ConditionalCauchySchwarz}
    \left(E\left(\|\Gamma_{k}(\gamma)\|^{4}_{2}\right)\vert \mathcal{F}_{k-1}(\gamma)\right)^{\frac{1}{2}}P(\|\Gamma_{k}(\gamma) \|_{2} > \kappa \vert \mathcal{F}_{k-1}(\gamma))^{\frac{1}{2}}.
\end{equation}
First consider the second term, $P(\|\Gamma_{k}(\gamma)\|_{2}  > \kappa \vert \mathcal{F}_{k-1}(\gamma))$. We use Chebyshev's inequality to get
\[
P(\|\Gamma_{k}(\gamma)\|_{2}  > \kappa \vert \mathcal{F}_{k-1}(\gamma)) \leq \frac{1}{\kappa^{2}} E\left((\|\Gamma_{k}(\gamma)\|^{2}_{2} \vert \mathcal{F}_{k-1}(\gamma))\right).
\]
Using the definition of $\Gamma_{k}(\gamma)$ given in (\ref{eqn:MartingaleCLT}), it is easy to see that for the appropriate $i$ and $t$,
\begin{equation}\label{eqn:CondVar}
E(\|\Gamma_{k}(\gamma)\|^{2}_{2}\vert\mathcal{F}_{k-1}(\gamma))=\frac{1}{\gamma^{-(\zeta+\delta)}}\left( (\|V_{i,t}\|^{2}_{2}\exp(\beta^{\intercal}V_{i,t}))+o_{\mathcal{P}}(\gamma) \right),  
\end{equation}
where the $o_{\mathcal{P}}(\gamma)$ term is a result of the smaller order terms of $q^{\prime}(V_{i,t})$ given by \eqref{eqn:Prob-der}. We now evaluate $\left(E\left(\|\Gamma_{k}(\gamma)\|^{4}_{2}\right)\vert \mathcal{F}_{k-1}(\gamma)\right)$. Note that from a similar argument,
\begin{equation}\label{eqn:higherpower}
\|\Gamma_{k}(\gamma)\|^{4}_{2} =  \frac{1}{\gamma^{-2(\delta+\zeta-1)}}\left(\|V_{i,t}\|^{4}_{2}D_{i,t+1} + o_{\mathcal{P}}(\gamma)\right) ,  
\end{equation}
and hence,
\begin{equation}\label{eqn:Variancehigherpower}
E\left(\|\Gamma_{k}(\gamma)\|^{4}_{2}\vert\mathcal{F}_{k-1}(\gamma) \right) =\frac{1}{\gamma^{-2(\delta+\zeta-\frac{1}{2})}}\left( \|V_{i,t}\|^{4}_{2}\exp(\beta^\intercal V_{i,t}) + o_{\mathcal{P}}(\gamma)\right). 
\end{equation}
Then, using the elementary identity $\sqrt{a+b}\leq \sqrt{a}+\sqrt{b}$, the square root of the RHS of (\ref{eqn:Variancehigherpower}) can be upper bounded by
\begin{equation}\label{eqn:Condexpbound}
\frac{1}{\gamma^{-(\delta+\zeta-\frac{1}{2})}}\left(\|V_{i,t}\|^{2}_{2}\exp(\frac{\beta^\intercal}{2} V_{i,t}) + o_{\mathcal{P}}(\gamma^{\frac{1}{2}})\right).
\end{equation}
Combining (\ref{eqn:CondVar}) and (\ref{eqn:Condexpbound}), we get that
\begin{align}
E\left(\|\Gamma_{k}(\gamma)\|^{2}_{2}\mathbb{I}(\|\Gamma_{k}(\gamma)\|_{2} > \kappa)\big\vert\mathcal{F}_{k-1}(\gamma)\right) \leq \frac{\kappa^{-1}}{\gamma^{-(\delta+\zeta-\frac{1}{2})}}\frac{1}{\gamma^{-\frac{1}{2}\left(\delta+\zeta\right)}} \left(\|V_{i,t}\|^{3}_{2}\exp(\beta^{\intercal}V_{i,t}) + o_{\mathcal{P}}(\gamma^{\frac{1}{2}})\right).
\end{align}
We note that by the law of large numbers,
\[
\sum_{i=1}^{m(\gamma)}\sum_{t=0}^{T(\gamma)-1}\frac{1}{\gamma^{-(\delta+\zeta)}}  \left(\|V_{i,t}\|^{3}_{2}\exp(\beta^{\intercal}V_{i,t}) + o_{\mathcal{P}}(\gamma^{\frac{1}{2}})\right) \xrightarrow[]{\mathcal{P}} C,
\]
as $ \gamma \to 0$ for an appropriate choice of constant $C$. Since $\delta+\zeta-1 >0$, we have that $\forall\kappa>0$,
\[
\sum_{i=1}^{m(\gamma)}\sum_{t=0}^{T(\gamma)-1}\frac{\kappa^{-1}}{\gamma^{-(\delta+\zeta)}}\frac{1}{\gamma^{-\frac{1}{2}(\delta+\zeta-1)}}  \left(\|V_{i,t}\|^{3}_{2}\exp(\beta^{\intercal}V_{i,t}) + o_{\mathcal{P}}(\gamma^{\frac{1}{2}})\right)\xrightarrow[]{\mathcal{P}} 0 ,
\]
as  $ \gamma \to 0$, which completes the proof.\qed\\
\textbf{Proof of Lemma~\ref{lemma:Emp--def--prob} }
Observe that 
\[
\frac{\hat{p}(\gamma)}{\tilde{p}(\gamma)} -1= \frac{\gamma}{\tilde{p}(\gamma)} \left( \frac{1}{\gamma T(\gamma)m(\gamma)}\sum_{i,t}D_{i,t+1} - E(\exp(\beta^\intercal V_{i,t}))\right).
\]
Recall that $\tilde{p}(\gamma) = \gamma\exp\left(\frac{1}{2}\beta^\intercal\Sigma\beta\right) + o(\gamma)$.
Now Lemma~\ref{lemma:Emp--def--prob} follows, proceeding as in the proof of Theorem~\ref{theorem:1}.\qed\\

\textbf{Proof of Proposition~\ref{prop:multiclass}:}\\
We provide an outline of the proof of Proposition~\ref{prop:multiclass}. The detailed steps are similar to those in the proof of Theorem~\ref{theorem:1}, and thus will be omitted. First, notice that using arguments similar to the proof of Theorem~\ref{theorem:1} from the single class setting, 
 \begin{align*}
     \|\hat{e}(\gamma) - e(\gamma) \|^2 &= O_{\mathcal{P}}(\gamma^{\delta+\zeta-1})+O_{\mathcal{P}}(\gamma^{\zeta}),\\
     \|\hat{g}_k(\gamma) - g_k(\gamma) \|^2 &= O_{\mathcal{P}}(\gamma^{\delta+\zeta-1})+O_{\mathcal{P}}(\gamma^{\zeta}).
 \end{align*}
 Similarly, with $\hat{f}_k(\gamma) = \frac{\sum_{i,t}D_{i,k,t+1}}{\sum_{i,k,t}D_{i,k,t+1}}$, $|f_{k}-\Hat{f}_k(\gamma)|^2 = O_{\mathcal{P}}(\gamma^{\delta+\zeta-1})+O_{\mathcal{P}}(\gamma^{\zeta})$. \\
Now,
\[\beta = \left(\Sigma_{YY} - \sum_{k=1}^{K} f_k \Sigma_{XY_K}\Sigma^{-1}_{X_kX_k}\Sigma_{YX_k}\right)^{-1} \left(e(\gamma)- \sum_{k} f_k\Sigma_{YX_k}\Sigma_{X_kX_k} g_{k}(\gamma)\right) + O(\gamma),
\] 
where the $O(\gamma)$ is the result of an analysis similar to Lemma~\ref{1Dbound}. Recall that
\[
\hat{\beta}(\gamma) = \left(\Sigma_{YY} - \sum_{k=1}^{K} \hat{f}_k(\gamma) \Sigma_{XY_K}\Sigma^{-1}_{X_kX_k}\Sigma_{YX_k}\right)^{-1} \left(\hat{e}(\gamma)- \sum_{k} \hat{f}_k(\gamma)\Sigma_{YX_k}\Sigma_{X_kX_k} \hat{g}_{k}(\gamma)\right)
\]
From the above statements, we see that 
\[
\|\beta -\hat{\beta}(\gamma)\|^{2} = O_{\mathcal{P}}(\gamma^{\delta+\zeta-1})+O_{\mathcal{P}}(\gamma^{\zeta})
\]
Now, 
\[
\hat{\eta}_k = \Sigma_{X_kX_k}(\hat{g}_k(\gamma) - \Sigma_{XY_k}\hat{\beta}(\gamma)).
\]
By linearity,
\[
\|\eta_k -\hat{\eta}_k(\gamma)\|^{2} = O_{\mathcal{P}}(\gamma^{\delta+\zeta-1})+O_{\mathcal{P}}(\gamma^{\zeta}).
\]
\qed 
\clearpage

\section{Supplementary Results}\label{EC:Supnum}
\subsection{Additional Numerical Experiments}
\textbf{Robustness to Stationarity}
To test the proposed estimator to deviations from the stationary assumption made in the analysis, the above experiments are repeated by adding a common cyclic drift to the covariates. That is, the covariates evolve according to the equations

\begin{align}
    Y_{t,k}  &= 0.3\cdot Y_{t-1,k} + N_{t,k}(0,1) +\sin{t}, \ k\in[d_1] \label{eqn:driftevo}\\
    X_{i,t,k} &= N_{i,t,k}(0,1) +\sin{t},\ k\in[d_2]\nonumber, 
\end{align}

 In Table~\ref{table:drift}, we find that the performance of the proposed estimator is not significantly affected by addition of a cyclic drift of average value 0. This suggests that for mild deviations from stationarity, the proposed estimator is not significantly worse than the MLE.\\
 \noindent \textbf{Multiple classes of firms: } To illustrate the effectiveness of our estimator in the multi-class setting, we implement a simulation with two classes of firms. We assume that each class has equal number of firms. Here, $\alpha_1=-8$, and $\alpha_2=-8.5$. $\beta$ is kept constant across both classes. We perform parameter using MLE and using the proposed estimator as per Remark~\ref{rem:Multiclass}. Here too, we observe that the error in parameter estimation for the proposed estimator and the MLE is similar. Results are tabulated in Table~\ref{table:Multiclass}.
\subsection{Proposed estimator as initial seed for MLE}
\begin{figure}[htbp!] 
\begin{center}
\includegraphics[height=2.2in]{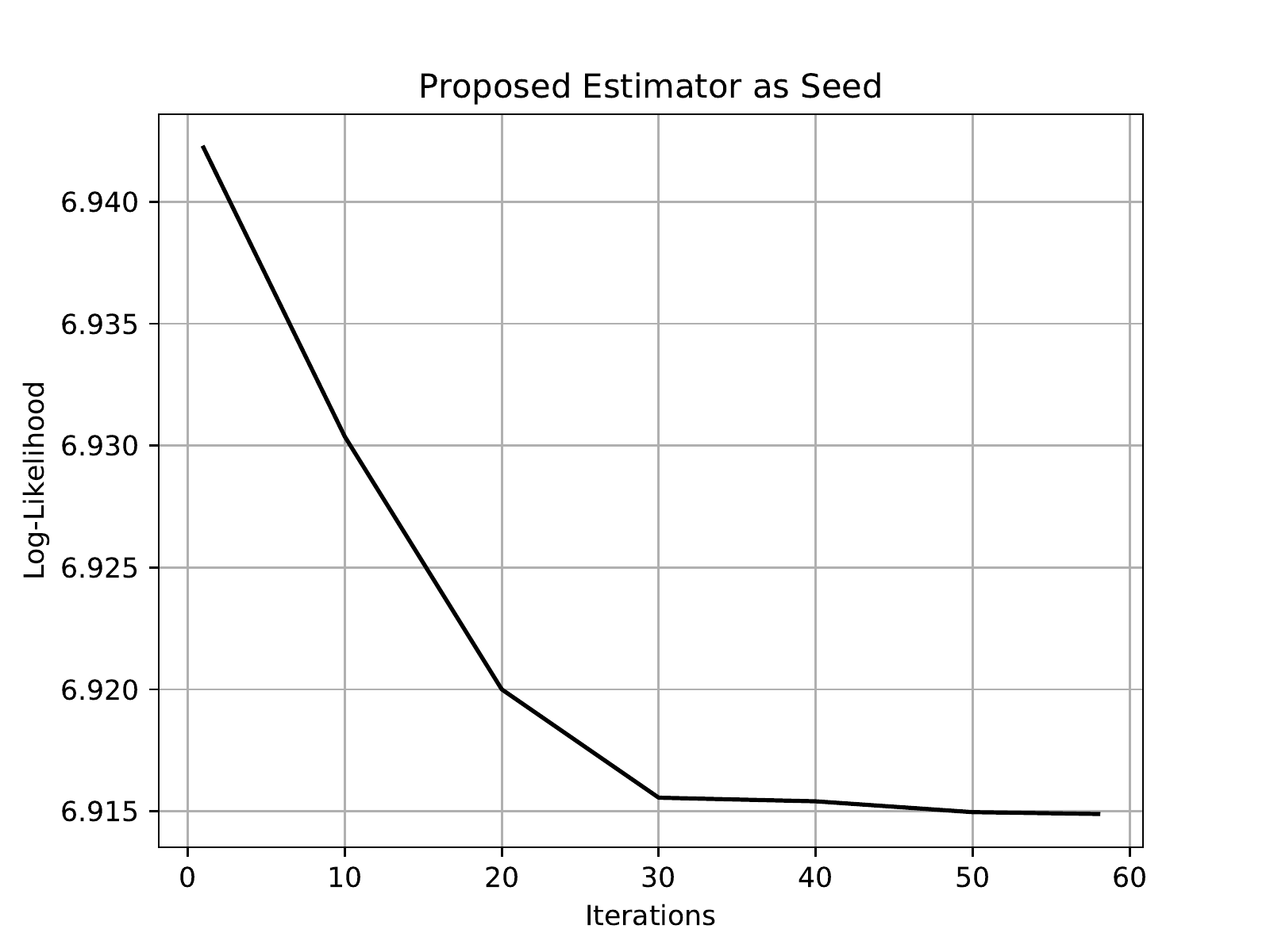}
\includegraphics[height=2.2in]{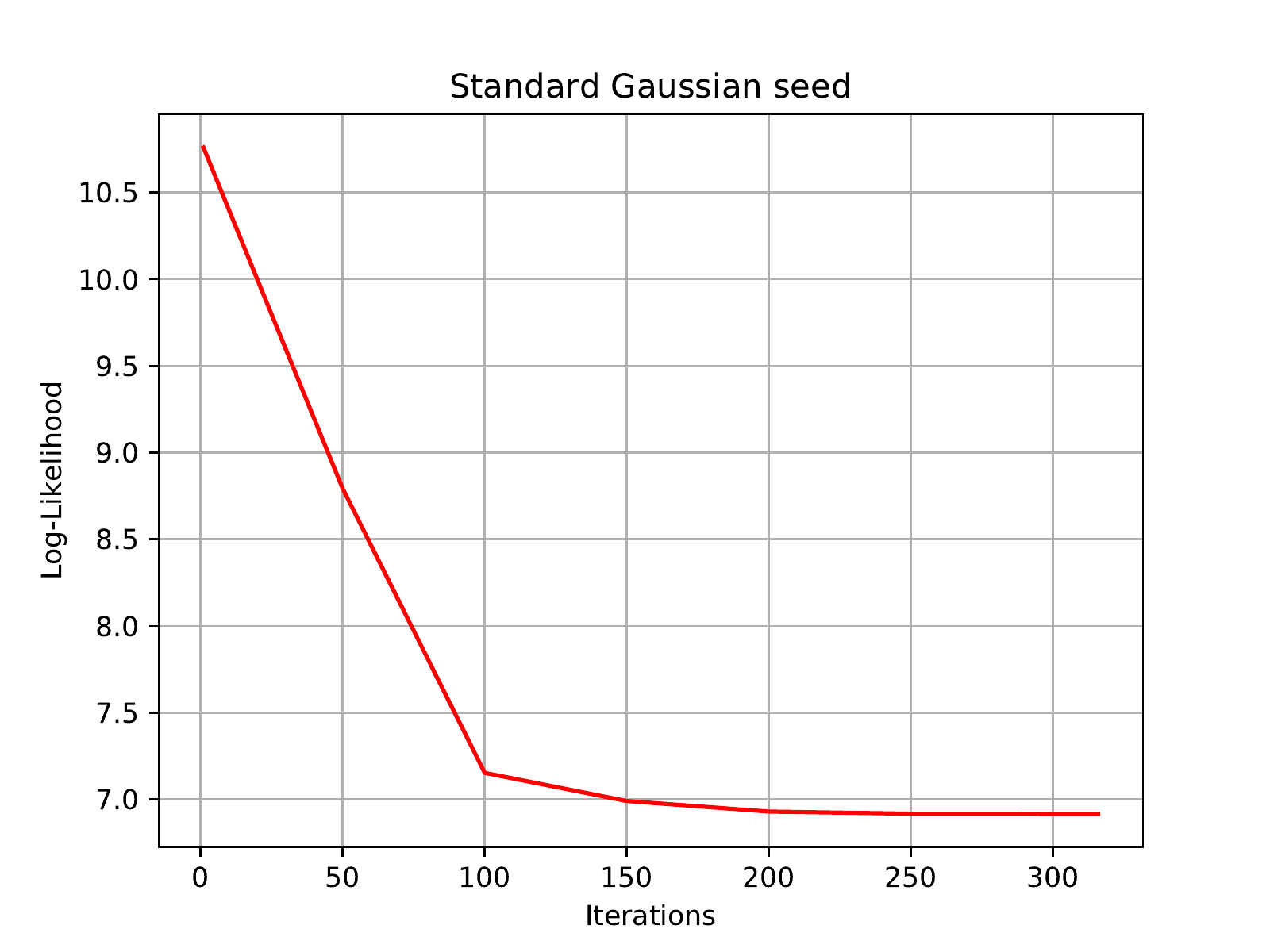}
\includegraphics[height=2.2in]{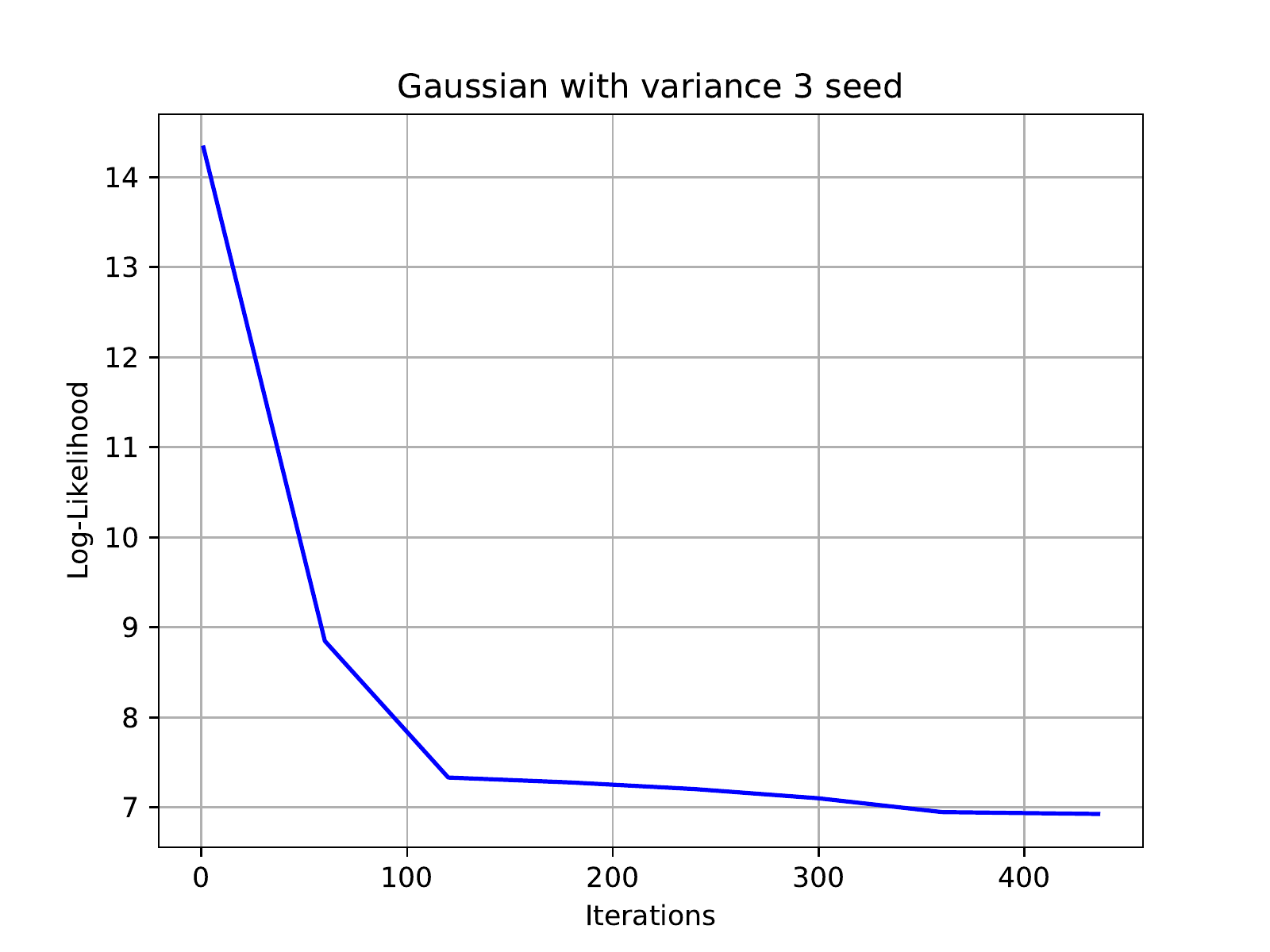}
\caption{Iteration Reduction using Proposed Estimator.} \label{fig:ItrRed}
\end{center}
{Figure~\ref{fig:ItrRed} plots the log-likelihood as a function of number of iterations. Observe that the MLE converges in roughly 60 iterations when the proposed estimator is used as an initial seed, 320 iterations when the seed is selected using a $N(\beta,\mathbf{I})$  random variable, and 450 iterations when the seed is selected using a $N(\beta,3\mathbf{I})$ random variable. The proposed estimator thus gives a reduction in the computational effort required for MLE of 5 and 7 times, respectively over the other two seeds.}
\end{figure}

Figure~\ref{fig:ItrRed} gives an illustration of the log-likelihood as a function of the number of iterations. Here, the number of firms are selected to be $N=10,000$, time periods are chosen to be $T=200$ months, covariates follow the evolution given by \eqref{eqn:Size-large}. We plot the negative of the log-likelihood as a function of number iterations of MLE. We consider two cases: when the initial seed is chosen at random, and  when the proposed estimator is used as a seed. When the proposed estimator is used as an initial seed, the distance between the initial and maximum log-likelihood is $O(\gamma^{\frac{1}{2}(\delta+\zeta-1)}) + O(\gamma^{\frac{1}{2}\zeta})$. Selecting a seed using a standard Gaussian yields a difference of of $O(1)$. This suggests that using the proposed estimator as an initial seed reduces the number of iterations to convergence up to a factor of $\gamma^{-\frac{1}{2}(\delta+\zeta-1)} + \gamma^{-\frac{1}{2}\zeta}$ (we declare that the MLE has converged at the first time when the difference in log-likelihood function computed by successive  iterations of the algorithm drops below $10^{-4}$).
\subsection{Estimation of Covariance matrix from data}
Here, we derive error bounds for estimation of the covariance matrix from data. Recall that our estimate of the covariance matrix was given by $\hat{\Sigma} = \frac{1}{T(\gamma)m(\gamma)}\sum_{i,t} V_{i,t}V_{i,t}^{\intercal}$. Recall also that $\Sigma_{YY}$, $\Sigma_{XY}$  and $\Sigma_{XX}$ denoted the covariance between the common covariates, the cross covariance and the covariance between the idiosyncratic covariates respectively. Define $\hat{\Sigma}_{YY} = \frac{1}{T(\gamma)}\sum_{t\leq T(\gamma)} Y_{t}Y_{t}^{\intercal}$. Now, from the component-wise central limit theorem, it is easy to see that 
\[
\|\Sigma_{YY}-\hat{\Sigma}_{YY} \| = O_{\mathcal{P}}(\gamma^{\frac{1}{2}\zeta}).
\]
It can similarly be seen that $\|\Sigma_{XY} - \hat\Sigma_{XY}\|  = O_{\mathcal{P}}(\gamma^{\frac{1}{2}(\zeta+\delta)})$ and $\|\Sigma_{XX} -\hat{\Sigma}_{XX}\|  = O_{\mathcal{P}}(\gamma^{\frac{1}{2}(\delta+\zeta)})$, and hence the overall error in estimation of covariance is $\|\Sigma-\hat{\Sigma}\|  = O_{\mathcal{P}}(\gamma^{\frac{1}{2}\zeta})$.

\begin{table}
  \caption{Comparison of RMSE: Model correctly specified\label{table:logit_correct_small}}
{\begin{tabular}{@{\extracolsep{5pt}} ccccccc}
\\[-1.8ex]\hline
\hline \\[-1.8ex]
 Time in months & No. of firms & RMSE($\beta_{prop}$) & RMSE($\beta_{ML}$) & RMSE($\alpha_{prop}$) & RMSE($\alpha_{ML}$) \\
\hline \\[-1.8ex]
$200$& $5000$ &$0.2103$         &$0.1638$    &$0.1811$  &$0.1067$\\
$200$& $7000$ &$0.1855$ 		&$0.1343$    &$0.1435$ 	&$0.0900$\\
$200$& $10000$ &$0.1743$ 		&$0.1192$    &$0.1406$	&$0.0763$\\
$200$& $13000$ &$0.1690$  		&$0.1091$    &$0.1398$	&$0.0640$\\
\\[-1.8ex]\hline
\hline \\[-1.8ex]
$200$& $5000$ &$0.2103$		&$0.1638$ &$0.1811$	    &$0.1067$\\
$400$& $5000$ &$0.1441$		&$0.1093$ &$0.0852$ 	&$0.0631$\\
$600$& $5000$ &$0.1230$		&$0.0851$ &$0.7600$	    &$0.0545$\\
$800$& $5000$ &$0.1012$		&$0.0877$ &$0.0651$	    &$0.0541$\\
 \\[-1.8ex]\hline
 \\[-1.8ex]
\hline \\[-1.8ex]
 Time in months & No. of firms & RMSE($\beta_{prop}$) & RMSE($\beta_{ML}$) & RMSE($\alpha_{prop}$) & RMSE($\alpha_{ML}$) \\
\hline \\[-1.8ex]
$200$& $5000$& $0.1845
$&		$0.1228$& $0.1745$&	 $0.0918$\\
$200$& $7000$& $ 0.1584$&		$ 0.1004$&  $ 0.1134$&	 $0.0678$\\
$200$& $10000$& $ 0.1563$&		$0.0905$& $0.1092$&	 $0.0718$\\
$200$& $13000$& $0.1559$&		$0.0802$& $0.0917$&	 $0.0635$\\
\\[-1.8ex] \hline
\hline \\[-1.8ex]
$200$& $5000$& $0.1845
$&		$0.1228$& $0.1745$&	 $0.0918$\\
$400$& $5000$& $0.1305$&		$0.0875$& $0.0744$&	    $0.0711$\\
$600$& $5000$& $0.1294$&		$0.0749$& $0.0681$&	    $0.0645$\\
$800$& $5000$& $0.1278$&		$0.0733$& $0.0657$&  	$0.0639$\\
\hline \\[-1.8ex]
\end{tabular}}
\begin{tablenotes}
\footnotesize\item {True parameters: $\alpha=8.5, \beta=(-0.2, 0.5, 0.5,0.2,-1,0.3,-0.2, 0.5, 0.5,0.2,-0.5,0.3)$, when default probability is 1\% and $\alpha=7.2$ when default probability is 3\%, all else kept the same.
RMSE of the proposed estimator is only slightly larger than that of MLE except when the no. of companies is large. The first set of readings above shows the RMSE when the default probability is kept at approximately 1\% per year, while the second set shows the RMSE when the default probability is 3\% per year. Note that estimation errors are smaller when the default probability is 3\%. To see this, recall that from Theorem~\ref{thm:CLT}, with $P_{def}$ as the average default probability, the estimation error for the MLE is roughly $\frac{1}{ \sqrt{T\cdot m \cdot P_{def}}}$. Hence, $T$ and $m$  kept the same, the estimation error reduces as default probability increases.}
\end{tablenotes}
 
\end{table}

\begin{table}
\centering
  \caption{Comparison of predictive power: Model Correctly specified
  \label{table:Gaussian}}
{\begin{tabular}{@{\extracolsep{5pt}} p{2cm}p{4cm}p{4cm}}
\\[-1.8ex]\hline
  Decile  & Proposed Estimator&MLE \\
\\[-1.8ex]\hline
$1$ &$ 0.9539$&$0.9618 $\\
$2$ &$0.9680$&$0.9719 $\\
$3$ &$0.9727$&$0.9762$\\
$4$ &$0.9776$&$0.9804$\\
$5$ &$0.9804$&$0.9825 $\\
$6$ &$0.9866$&$0.9878 $\\
$7$ &$0.9892$&$0.9921 $\\
$8$ &$1 $&$1$\\
$9$ &$1$&$1$\\

\hline \\[-1.8ex]
\end{tabular}}

\begin{tablenotes}
\footnotesize\item{True parameters are selected so that default probability is approximately 1\% per year.  Number of firms $N$ and time periods $T$ in above experiments are set to $10000$  and $200$ respectively, and the covariates are generated according to (\ref{eqn:Size-large}). Parameter estimation is done according to MLE and with the proposed estimator using transformed data.}
\end{tablenotes}

\end{table}

\begin{table}\centering
  \caption{Approximate MLE with regularisation
  \label{Reg-full}}
  {\begin{tabular}{@{\extracolsep{20pt}} p{4cm} p{2cm} p{2cm} p{2cm}}
\\[-1.8ex]\hline
\hline \\[-1.8ex]
Number of firms& $5000$& $10000$ & $13000$\\
\hline \\[-1.8ex]
RMSE (M)& $0.0909 $&	$0.0704$& $0.0602$\\
RMSE (P)& $ 0.1003$ &$ 0.0941$ &$ 0.0932$\\
\hline \\[-1.8ex]
\end{tabular}}    
\begin{tablenotes}
\footnotesize\item {Defaults are generated according to the intensity model, and ridge regularisation is used for MLE. Parameters are identical to those used in Table \ref{table:logit_correct_small}. Number of time periods of data is kept at 200.  RMSE (P) denotes the error between parameters calculated using \eqref{eqn:approx-L2reg}, setting $\mathbf{Z}=\mathbf{I}$ and $\frac{\beta}{2}$, the asymptotic solution of \eqref{eqn:MAP-FOC-1} and \eqref{eqn:MAP-FOC-2}. Similarly, RMSE (M) denotes error between $\frac{\beta}{2}$ and the regularised MLE.}
\end{tablenotes}

\end{table}

\begin{table}
\centering
  \caption{Comparison of predictive power: Covariates transformed to Gaussian
  \label{table:Non_Gaussian}}
{\begin{tabular}{@{\extracolsep{5pt}} p{2cm}p{4cm}p{4cm}}
\\[-1.8ex]\hline
  Decile & Proposed Estimator&MLE \\
\\[-1.8ex]\hline
$1$ &$0.8612$&$0.8838 $\\
$2$ &$0.8813$&$0.9092 $\\
$3$ &$0.8924$&$0.9224$\\
$4$ &$0.9042$&$0.9352$\\
$5$ &$0.9244$&$0.9571$\\
$6$ &$0.9476$&$0.9734$\\
$7$ &$0.9682$&$0.9892$\\
$8$ &$0.9834 $&$1$\\
$9$ &$1$&$1$\\

\hline \\[-1.8ex]
\end{tabular}}
\begin{tablenotes}
\footnotesize\item {True parameters are selected so that default probability is approximately 1\% per year.  Number of firms $N$ and time periods $T$ in above experiments are set to $10000$  and $200$ respectively, and the covariates are generated as per 5) in Section 4.1. Parameter estimation is done according to MLE and with the proposed estimator using transformed data. In the latter case, the non-Gaussian covariate is transformed by taking its logarithm, to have an approximately Gaussian distribution.}
\end{tablenotes}

\end{table}

\begin{table}
\centering
  \caption{Comparison of RMSE: Missing covariate with small and large coefficients \label{table:logit_incorrect_small}}
 { 
\begin{tabular}{@{\extracolsep{5pt}} ccccccc}
\\[-1.8ex]\hline

\hline \\[-1.8ex]
 $\beta_3$ & No. of firms  & RMSE($\beta_{prop}$) & RMSE($\beta_{ML}$) & RMSE($\alpha_{prop}$) & RMSE($\alpha_{ML}$) \\
\hline \\[-1.8ex]
$0.5$& $5000$&	$0.0851$&		$0.0812$&	$0.1222$&  $0.1220$\\
$0.5$& $7000$& 	$0.0730$&		$0.0710$&	$0.1243$&  $0.1252$\\
$0.5$& $10000$& 	$0.0678$&		$0.0619$&	$0.1254$&  $0.1243$\\
$0.5$& $13000$& 	$0.0634$&		$0.0627$&	$0.1176$&  $0.1187$\\

\\[-1.8ex] \hline\hline\\[-1.8ex]

$2$& $5000$& $0.3176$&		$0.2991$&	$1.834$& $1.844$\\
$2$& $7000$& $0.2804$&		$0.2945$&	$1.910$& $1.909$\\
$2$& $10000$& $0.3071$&		$0.3257$&	$1.885$& $1.881$\\
$2$& $13000$& $0.2722$&		$0.2859$&	$1.943$& $1.951$\\
\hline \\[-1.8ex]
\hline \\[-1.8ex]
 $\beta_{3}$ & No. of firms & RMSE($\beta_{prop}$) & RMSE($\beta_{ML}$) & RMSE($\alpha_{prop}$) & RMSE($\alpha_{ML}$) \\
\hline \\[-1.8ex]
$0.5$& $5000$& $0.0568$&		$0.0523$&  $0.1276$&	$0.1304$\\
$0.5$& $7000$& $0.0631$&		$0.0614$&  $0.1221$&	$0.1191$\\
$0.5$& $10000$& $0.0586$&		$0.0690$&  $0.1243$&	$0.1232$\\
$0.5$& $13000$& $0.0573$&		$0.0525$&  $0.1276$&    $0.1195$\\
\\[-1.8ex]\hline
\hline \\[-1.8ex]
$2$& $5000$& $0.3149$&		$0.3436$&	$ 1.868$&  $1.860$\\
$2$& $7000$& $0.3482$&		$0.3354$&	$1.911$&  $1.921$\\
$2$& $10000$& $0.2825$&		$0.2912$&	$1.885$&  $1.862$\\
$2$& $13000$& $0.2614$&		$0.2629$&   $1.823$&  $1.815$\\
\hline \\[-1.8ex]
\\[-1.8ex]\hline
\end{tabular}
}
\begin{tablenotes}
\footnotesize\item {True parameters: $(\alpha=7.5, \beta_{1}=-0.2, \beta_{2}=0.5)$,
  $\beta_3$ as specified above. Time period in above experiments is set to 200.
   Both the proposed estimator and MLE
estimate  parameters $(\alpha, \beta_{1}, \beta_{2})$ only.
  The RMSE of the two methods is nearly identical. It worsens as model misspecification increases, that is as value of $\beta_3$ increases. The first set of readings shows the RMSE when the default probability is kept at approximately 1\% per year, while the second set shows the RMSE when the default probability is 3\% per year.}
\end{tablenotes}

\end{table}

\begin{table}[!htbp] \centering
  \caption{Number of Defaults by Calendar Year
  \label{1}}
{ 
\begin{tabular}{@{\extracolsep{5pt}} ccccccc}
\\[-1.8ex]\hline
\hline \\[-1.8ex]
  S No. & Year & Number of Defaults & Total Active Firms & Default Percentage \\
\hline \\[-1.8ex]
$1$& $1992$&	$4$ &   $3839$ & $0.10$\\
$2$& $1993$&	$12$ &  $4926$&	$0.24$  \\
$3$& $1994$&	$9$  &  $5723$&	$0.16$  \\
$4$& $1995$&	$8$  &  $6294$&	$0.13$  \\
$5$& $1996$&	$9$  &  $6701$&	$0.13$   \\
$6$& $1997$&	$45$ &  $7135$&	$0.63$  \\
$7$& $1998$&	$74$ &  $7215$&	$1.02$  \\
$8$& $1999$&	$68$ &  $6845$&	$0.99$  \\
$9$& $2000$&	$98$ &  $6640$&	$1.47$ \\
$10$& $2001$&	$154$ & $6249$	  &  $2.46$  \\
$11$& $2002$&	$102$ & $5716$&	$1.78$  \\
$12$& $2003$&	$76$  & $5307$&	$1.43$ \\
$13$& $2004$&	$49$  & $5101$&	$0.57$  \\
$14$& $2005$&	$33$& $5080$	&$0.65$    \\
$15$& $2006$&	$15$& $5050$	&$0.30$    \\
$16$& $2007$&	$25$& $4980$&	$0.50$    \\
$17$& $2008$&  $56$ & $4899$& $1.14$    \\
$18$& $2009$&	$84$ & $4628$	&$1.81$   \\
$19$& $2010$&	$26$ &$4443$&	$0.58$   \\
$20$& $2011$&	$31$ &  $4364$&	$0.71$  \\
$21$& $2012$&	$32$ & $4259$&	$0.75$   \\
$22$& $2013$&	$17$ & $4184$	&$0.41$   \\
$23$& $2014$&	$23$ & $4247$&	$0.54$   \\
$24$& $2015$&	$38$  & $4363$&	$0.87$  \\
$25$& $2016$&	$44$  & $4309$&	$1.02$  \\
$26$& $2017$&	$11$  & $4216$ & $0.53$\\
\hline \\[-1.8ex]
\end{tabular}}
\begin{tablenotes}
\item\footnotesize {The default percentages are between $0.13\%$ and $2.5\%$. These values are used to calculate the effect of contagion- the higher the default percentage in a given period, the more likely are defaults in subsequent periods. }
\end{tablenotes}

\end{table}
\begin{table}[!htbp]
\caption{Transformations of covariates
  \label{transforms}}
{\begin{tabular}{@{\extracolsep{15pt}} m{10cm} m{3cm} }
\\[-1.8ex]\hline
\hline \\[-1.8ex]
 Covariate & Transformation ($y=f(x)$) \\
\hline \\[-1.8ex]
Stock Index Return, DTD Level, DTD Trend, Cash/TA Level, Cash/TA Trend, NI/TA Level, NI/TA Trend & $\log({x+1})$\\
Three Month Treasury Rate  & $\log{x}$\\
Size Level & $(x-2)^{3}$\\
M/B, Sigma & $\log{(\log({x+1}))}$\\
Contagion & $\sqrt{x}$\\
\hline \\[-1.8ex]
\end{tabular}}
\begin{tablenotes}
\footnotesize\item {The covariates are transformed so that marginally, each has an approximately Gaussian distribution. Transformations used are all of power or logarithmic nature. Size Trend is not transformed as it is already marginally Gaussian.}
\end{tablenotes}

\end{table}

\begin{table}[!htbp]\centering
\caption{Estimated Parameters
\label{table:parameters}}
{\begin{tabular}{@{\extracolsep{10pt}} cccccccccc}
\\[-1.8ex]\hline
\hline\\[-1.8ex]
 Index Return& 3-month Rate&DTD(L)& DTD(T)&Cash/TA (L)& Cash/TA (T)\\
 \hline \\[-1.8ex]
$-0.305$& $-0.269$& $-1.195$& $-0.063$& $-0.754$& $-0.057$\\
\\[-1.8ex]\hline
\hline \\[-1.8ex]
NI/TA (L)&NI/TA (T) & Size(L)&Size(T)& M/B& Sigma\\
 \hline \\[-1.8ex]
$-0.144$& $-0.327$& $0.095$& $-2.261$& $0.867$& $0.482$\\
\hline \\[-1.8ex]
\\[-1.8ex]
\\[-1.8ex]
\\[-1.8ex]\hline
\hline \\[-1.8ex]
 Index Return& 3-month Rate&DTD(L)& DTD(T)&Cash/TA (L)& Cash/TA (T)\\
 \hline \\[-1.8ex]
$-0.262$& $-0.260$& $-1.190$& $-0.063$& $-0.754$& $-0.059$\\
\\[-1.8ex]\hline
\hline\\[-1.8ex]
NI/TA (L)&NI/TA (T) & Size(L)&Size(T)& M/B& Sigma\\
 \hline \\[-1.8ex]
$-0.144$& $-0.326$& $0.094$& $-2.256$& $0.847$& $0.474$ \\
\hline \\[-1.8ex]
\end{tabular}
}
\begin{tablenotes}
\footnotesize\item {The notations (L) and (T) correspond to Level and Trend respectively. The first set of readings correspond to the parameters estimated when contagion is not considered. The value of $\alpha$ in this case is $-11.158$. The second set of readings are the estimated parameters when the effect of contagion is considered. The values of $\alpha$, and contagion parameter are $-11.163$ and $0.328$ respectively. }
\end{tablenotes}

\end{table}

\begin{table}[!htbp]
\centering
\caption{Combined Power table (Raw data)\label{2}}

{\begin{tabular}
{@{\extracolsep{10pt}} m{1cm} m{1cm} m{1cm} m{3cm}}

\\[-1.8ex]\hline
\hline \\[-1.8ex]
 Decile & DI &Logit& Proposed estimator\\
\hline \\[-1.8ex]
$1$& 	 $0.808$&	$0.774$& $0.744$\\
$2$&	$0.903$&	$0.875$& $0.862$\\
$3$&	$0.935$&	$0.916$& $0.911$\\
$4$&	$0.960$&	$0.950$& $0.937$\\
$5$&	$0.974$&	$0.972$& $0.956$\\
$6$&	$0.980$&	$0.976$& $0.969$\\
$7$&	$0.990$&	$0.987$& $0.980$\\
$8$&	$1$&	$1$&  $0.989$\\
$9$&	$1$&	$1$&	$1$\\
\hline \\[-1.8ex]    
\end{tabular}
}
\begin{tablenotes}
\footnotesize\item  { We compute the accuracy tables for all three methods.  Here, raw data is used, where most of the covariates are non-Gaussian. It is seen that both logit and DI calibration outperform the proposed estimator.}
\end{tablenotes}
 
\end{table}

\begin{table}[!htbp]
\centering
  \caption{Combined Power table (Gaussian Transformed data) \label{2-Gauss}}
{\begin{tabular}{@{\extracolsep{10pt}} m{1cm} m{1cm} m{1cm} m{3cm}}
\\[-1.8ex]\hline
\hline \\[-1.8ex]
 Decile & DI &Logit& Proposed estimator\\
\hline \\[-1.8ex]
$1$& 	 $ 0.846$&	$0.798$&	 $0.837$\\
$2$&	$0.923$&	$0.893$&     $0.906$\\
$3$&	$0.951$&	$0.933$&     $0.946$\\
$4$&	$0.967$&	$0.955$&     $0.972$\\
$5$&	$0.975$&	$0.974$&     $0.980$\\
$6$&	$0.990$&	$0.976$&     $0.990$\\
$7$&	$1$&	$0.985$&	     $1$\\
$8$&	$1$&	$1$&	  $1$\\
$9$&	$1$&	$1$&	$1$\\
\hline \\[-1.8ex]    
\end{tabular}
}
\begin{tablenotes}
\footnotesize\item {The set-up of Table \ref{2} is repeated, but covariates are transformed to be individually Gaussian. This leads to a significant (about 9\%) improvement in the performance of the proposed estimator.}
\end{tablenotes}
  
\end{table}

\begin{table}[!htbp]\centering
  \caption{Combined Power table (Gaussian data with contagion)\label{2-Guass+cont}}
{\begin{tabular}{@{\extracolsep{10pt}} m{1cm} m{1cm} m{1cm} m{3cm}}
\\[-1.8ex]\hline
\hline \\[-1.8ex]
 Decile & DI &Logit& Proposed estimator\\
\hline \\[-1.8ex]
$1$& 	 $0.856$&	$0.812$    & $0.845$\\
$2$&	$0.934$&	$0.904$   & $0.910$\\
$3$&	$0.959$&	$0.944$   &$0.953$\\
$4$&	$0.968$&	$0.962$   &$0.980$\\
$5$&	$0.978$&	$0.977$   &$0.990$\\
$6$&	$0.986$&	$0.984$    &$1$\\
$7$&	$1$&	$1$      &$1$\\
$8$&	$1$&	$1$         &$1$\\
$9$&	$1$&	$1$         &$1$\\
\hline \\[-1.8ex]    
\end{tabular}
\begin{tablenotes}
\footnotesize\item {Finally, a contagion factor is added and calibration is performed using all three methods. In all three this leads to an improvement in performance. Here as well as in Table~\ref{2-Gauss}, the proposed estimator is better than logit, and marginally worse than the DI method.}
\end{tablenotes}
}
  
\end{table}

\begin{table}\centering
  \caption{Comparison of RMSE: Covariates with a drift
  \label{table:drift}}
  \centering
{\begin{tabular}{@{\extracolsep{5pt}} cccccc}
\\[-1.8ex]\hline
  No. of firms & RMSE($\beta_{prop}$) & RMSE($\beta_{ML}$) & RMSE($\alpha_{prop}$) & RMSE($\alpha_{ML}$) \\
\\[-1.8ex]\hline
$5000$& $0.2217$&	$0.1766$&	$0.1631$&	$0.1154$\\
$7000$& $0.2187$&	$0.1719$&	$0.1563$&	$0.1067$\\
$10000$& $0.1780$&	$0.1318$&	$0.1004$&	$ 0.0687$\\
\hline \\[-1.8ex]
\end{tabular}}
\begin{tablenotes}
\footnotesize\item{True parameters: $\alpha=8.5, \beta=(-0.2, 0.5, 0.5,0.2,-1,0.3,-0.2, 0.5, 0.5,0.2,-0.5,0.3)$ are selected so that default probability is approximately 1\% per year. Time period $T$ in above experiments is set to 200 months and the covariates are generated from a non stationary distribution with a sinusoidal drift as in (\ref{eqn:driftevo}). The modeller is assumed to know all covariates. It is seen that the proposed estimator still performs almost as well as MLE.}
\end{tablenotes}

\end{table}

\begin{table}
\centering
  \caption{Comparison of RMSE: Multiple classes of firms
  \label{table:Multiclass}}
{\begin{tabular}{@{\extracolsep{5pt}} m{2cm}m{4cm}m{5cm}}
\\[-1.8ex]\hline
  No. of firms & RMSE(P) & RMSE(M) \\
\\[-1.8ex]\hline
$5000$& $0.1851$&	$0.1532$\\
$7000$& $0.1616$&	$0.1278$\\
$10000$& $0.1325$&	$0.1051$\\
\hline \\[-1.8ex]
\end{tabular}}
\begin{tablenotes}
\footnotesize\item{True parameters: $\alpha_1=8.5, \alpha_2=8, \beta=(-0.2, 0.5, 0.5,0.2,-1,0.3,-0.2, 0.5, 0.5,0.2,-0.5,0.3)$ are selected so that default probability is approximately 1\% per year for class 1 and 2\% per year for class 2. Time period $T$ in above experiments is set to 200 months and the covariates are generated according to \eqref{eqn:Size-large}. It is seen that the proposed estimator still performs almost as well as MLE. Here, RMSE(P) denotes the RMSE of the proposed estimator, and RMSE(M) denotes the RMSE of MLE.}
\end{tablenotes}
\end{table}

\end{document}